\documentstyle [12pt] {article}
\def \vt { \vartheta }
\def\L {\Lambda }
\def\l {\lambda } 
\def \t {\theta }
\def\a {\alpha }
\def\dh {\partial }
\def \d {\delta }
\def \D {\Delta }

\def \g {\gamma }
\def \G {\Gamma }
\def \O {\Omega }
\def \o {\omega }
\def \b {\beta }

\def \s {\sigma }
\def \e {\epsilon }
\def \ud { {1 \over 2} } 
\input psfig

\if@twoside\oddsidemargin25mm
\else\oddsidemargin25mm\evensidemargin25mm\marginparwidth25mm\fi%
\begin{document}
\begin{titlepage}

\title{$\ $\ \\$\ $\ \\ 
SYSTEMATICS OF STRING LOOP
THRESHOLD CORRECTIONS IN 
ORBIFOLD MODELS \thanks
{\it Supported by the 
Laboratoire de la Direction des Sciences
de la Mati\`ere du Commissariat \`a l'Energie Atomique }}
\vskip 2cm
\author{  {M. Chemtob\thanks  
{\it e-mail: chemtob at spht.saclay.cea.fr } 
} \\ \\ \\
{\em Service de Physique Th\'eorique
CE-Saclay 
F-91191 Gif-sur-Yvette Cedex France\thanks
{\it Laboratoire de la Direction des Sciences
de la Mati\`ere du Commissariat \`a l'Energie Atomique } 
}}

\maketitle 
\vskip 2cm

\begin{abstract}
The string theory one-loop threshold corrections are 
studied in  a background field  approach  due to 
Kiritsis and Kounnas which uses space-time curvature as an infrared regulator.
We review the conformal field theory aspects of the method for the 
special case of the semiwormhole space-time solution.
The comparison between the string 
and effective field theories vacuum functionals  is made for 
the  low derivative order, as well as for 
certain higher-derivative, gauge and gravitational  interactions. 
We study the dependence of string loop renormalization  corrections 
on the infrared cut-off.  Numerical applications 
are  considered for  a sample of four-dimensional 
abelian orbifold models   with a view 
to deduce the  systematic trends 
of the  moduli-independent threshold corrections.
The implications on the  
perturbative string theory unification  are  examined.
We present numerical results for the gauge interactions 
coupling constants as well as  for 
the quadratic order gravitational ($R^2$) and the 
quartic order gauge  ($F^4$) interactions.
\end{abstract}


\rightline {hep-th/97xxxxx}

\rightline  {T97/019}

\end{titlepage}

\section{ INTRODUCTION  }
\label{sec:intro}

With a  single 
dimensionful free parameter (the Regge slope $\a ' $)
and a handful of 
dynamical parameters (the moduli fields vacuum 
expectation values)
string theory must  strive at  providing 
an unified description for all the  known elementary  particles and  their 
interactions. Weakly coupled  solutions,  in spite of the  
runaway dilaton vacuum  sickness, 
have the important  advantage   of 
calculability.
Both the gauge symmetry bosons and matter particles
are then  manifest as 
elementary massless
excitations, while the effective action  can be 
constructed  controllably
using the string theory  world sheet 
and  space-time loop expansions. 
The excellent  good contact achieved  by 
the solvable  perturbative string theory  models 
with particle physics phenomenology 
is surely an  encouraging sign.

As is well known,
the matching of  loop contributions to the  scattering 
amplitudes of a string theory to those of
its low energy, $\alpha ' \to 0$ limit,
field theory descendant, induces
finite contributions
to the local interactions  coupling constants in the  effective action. 
These so-called heavy threshold corrections, which 
reflect the  decoupling of massive string  modes,
are expected to relax the
restrictive  unification constraints 
imposed on the various 
coupling constants 
at tree level.  In particular, 
the departures from universal
values of the gauge interactions 
coupling constants
could indeed  be large enough to  have a phenomenological 
impact on the issue of the  high 
energy extrapolation of the standard model of the electroweak 
gauge interactions.

Three main features 
distinguish the heavy threshold corrections in string  theories
from their  grand unification  theories analogs:
(i) The  heavy modes decoupling in string theory 
involves summations over
infinite towers of massive
excitations;
(ii) For weakly coupled string solutions, where the string mass scale 
$\a'^{-\ud }$ lies  close to the Planck scale, it is necessary to care
about the back-reaction effects of gauge interactions on 
gravitational interactions and conversely;  
(iii) String theories have finite ultraviolet  behavior but   are 
subject to infrared divergences associated with 
vacuum  tadpoles of the massless modes.  

The  first item in the above list suggests that, 
unless  some cancellation mechanism is at work, 
nothing prevents threshold corrections from 
attaining  large sizes. The second item  underscores the 
importance of having a description  of the world sheet theory 
consistent  with conformal and modular 
invariance.
As to  the last item, 
it clearly points out to  the need of implementing 
a consistent infrared  cut-off regularization procedure. 

Little    attention was 
given  so far to the issue of the infrared  regularization scheme dependence.
The original works
\cite{minahan,kapl,dkl} mainly focussed on the non-constant, 
moduli or gauge group factors dependent, parts   where
the infrared sensitivity cancels away to some extent.
The  approach initiated in \cite{kapl,dkl}, and developed  further
in \cite{anton,anton92,anton93,deren,ibalu,carov,mayr93,bailin94,kapl1},  
has served an important purpose in testing the 
string theory  dualities \cite{cardoso,harvey}. It has also been 
applied in several phenomenological studies
\cite{antell,nano,dienes,iba91,mayr2,chem,dienes1}.
Recently, a
more  complete approach 
was presented by Kiritsis and Kounnas \cite{kiko}.
The idea is to use a  curved
space-time as an infrared cut-off regulator,
observing that such a regularization scheme  can be
consistently and workably realized
for both  string and field theories. 

Curved gravitational and gauge backgrounds are defined as the 
solutions  of the  
string theory equations of motion, or of the perturbatively 
equivalent equations expressing the 
cancellation of the world sheet 
conformal Weyl anomaly.
The search for exact solutions of the classical theory
has been actively pursued in recent years, using 
the techniques  of unitary coset models \cite{coset} or of
solution-generating duality  transformations \cite{dualite}. 
Solutions of the quantum theory, exact to  all orders of $\a '$, 
have also been discussed in cosmology \cite{basf} or 
particle physics \cite {bars} applications.
(We have 
cited  a very small  fraction of the extensive 
literature on this subject.)
From  the standpoint where  a curved 
space-time is viewed primarily as a technical device,
a convenient class of solutions is provided by 
the models  with ${\cal N}=4$  
world sheet supersymmetry
\cite{koun3}. 
Solvable, perturbatively stable solutions, 
which depend on a free parameter 
associated to the space-time curvature, 
can be found here  by assembling together suitable direct products 
of compact or non-compact 
 WZW (Wess-Zumino-Witten) current algebra  
 sigma-models \cite{koun3,koun1,koun2}. 
 The simplest solution of this kind,  
the so-called  semiwormhole space-time 
solution \cite{calhast}, 
is associated with the conformal theory, 
$W_k^{(4)}= SU(2)_k\times U(1)$, and has
the asymptotic  (large level $k$) geometry, 
$S^3\times R^+$.
A curvature  regularized heterotic 
string theory can then be constructed by substituting  
for the  world sheet coordinate and spin fields 
of the  uncompactified  four-dimensional Minkowski space-time, 
conformal (left-moving sector)  and superconformal  
(right-moving sector) blocks of appropriate central charge 
and world sheet supersymmetry. 

An   important consequence in this approach  of the solvability of the 
space-time conformal block  
is the existence of marginal deformations of the theory
which  represent conformally  invariant  perturbations by
constant gravitational or gauge backgrounds.
This makes the 
approach well-suited  for 
studies of the higher-derivative interactions in the effective action.
String loop corrections to the 
effective actions of ten-dimensional  
string  theories  were discussed
some time ago   \cite{mizrachi,sakai}.   
Undertaking the  
analogous programme for    four-dimensional string theories
is a challenging task because 
each compactification comes with its particular gauge symmetry 
group  and matter content. Moreover, the non-renormalization constraints 
are less restrictive in four dimensions.

Since the initial proposal of the  curvature 
regularization  approach, further developments were  
reported by Kiritsis and Kounnas \cite{kiko1}
and  Petropoulos and Rizos \cite{peri,petro},
independently and  in collaboration \cite{kkpr1,kkpr2,kkpr3}. 
In the present work we shall focus on 
similar issues.
Our principal goal will be to perform a  quantitative study 
of threshold corrections to the gauge  and gravitational 
interactions, including certain  higher-derivative terms, 
based on four-dimensional  orbifold models. 
While  some overlap between  
our presentation and that of the above authors 
is  unavoidable, in reviewing their approach,
we shall  try to bring out 
the essential points and  emphasize
certain complementary  aspects.

Our discussion of the  higher-derivative interactions focusses on  the
quadratic terms in the curvature  tensor field  $(R^2)$
and the quartic terms  in the  gauge fields $(F^4)$. 
Although of academic interest for particle physics phenomenology, 
because of the enormous suppression by one power 
of  $\alpha '$ relative to the conventional linear gravity term,  
the quadratic gravitational interactions have important  implications 
on the consistency of quantized gravity 
\cite{stelle,gb,ovrut} 
and as a mechanism  to trigger 
supersymmetry  breaking \cite{hindawi}.   The 
supergravity completion of quadratic gravity is
discussed in \cite{theisen,cecotti} and the
constraints on its structure from 
mixed sigma-model or Kahler and
gravitational anomalies
in  \cite{cardoso1}.
The studies  of string loop corrections to the topological 
Gauss-Bonnet gravitational interaction, initiated in 
\cite{anton92}, have been  pursued further in  
\cite{cardoso,harvey1,forger}.

The quartic gauge interactions  are affected at low energies 
by an even larger
factor $\a '^2$ suppression relative to the minimal quadratic interactions.
Nevertheless, their contributions 
could have  some phenomenological impact at high energies, particularly 
in the event that nature would  have chosen the so-called 
weak scale string solutions \cite{lykken}, characterized by
a tension parameter close to  the Fermi scale.
For the class of four-dimensional heterotic string  vacua with ${\cal N}=4$ 
supersymmetry and an unbroken gauge symmetry, such as those
obtained by toroidal 
compactification  from the  ten-dimensional theory, the exact structure of the
one-loop $F^4$ interactions  
can be determined  \cite{grosswi,lerche}, thanks  
to the supersymmetry 
relationship with the Green-Schwarz anomaly canceling countertem, $B X_8$. 
These  properties of the quartic gauge interactions 
have  been recently exploited to  test the strong-weak coupling 
duality map between the four-dimensional toroidally compactified 
heterotic and type I theories
\cite{baki}.

Regarding the numerical applications, 
our motivations remain essentially unchanged with respect to
our previous work \cite{chem}. We 
shall mainly focus  on 
the  constant,  moduli-independent component  of 
threshold corrections since this remains the poorly understood 
part. We calculate threshold corrections  for 
a  sample of representative orbifold models
covering a range of gauge symmetry groups and  matter fields content, with a  
view to uncover possible
systematic trends. The updated results for the gauge coupling constants
reported in the present work  
include the gravitational back-reaction effects. 

The present paper includes  three additional  sections.
In Section \ref{sec:secin}, we review 
the approach of Kiritsis and Kounnas \cite{kiko}, as applied
to the semiwormhole space-time solution.
The following items are discussed: 
Conformal field theory aspects  of the semiwormhole solution 
and identification of certain of its zero-modes deformations;
Matching of the 
string vacuum functional with the field theory  effective action;
Effective action renormalization;  Extension to 
the D-term auxiliary fields.
Proceeding next  to the phenomenological part of the paper,
we  present  in Section \ref{sec:secgauge} 
numerical results for the moduli-independent  
threshold corrections in the gauge  coupling constants, the 
quadratic gravitational interactions 
and the quartic order gauge interactions. A brief discussion 
of the moduli-dependent threshold corrections is given.
Section \ref{sec:seconc} summarizes our 
conclusions. Appendix \ref{sec:appa} provides  technical help for the partition 
function expansion in powers of the background fields;
Appendix \ref{sec:appb} for the action of zero-modes operators;
Appendix \ref{sec:appc} for the approximate 
evaluation of  certain modular integrals 

\section{\bf  Infrared Regularization approach of Kiritsis and Kounnas}
\label{sec:secin}

\subsection{\bf Conformal field theory aspects}
\subsubsection{\bf Partition function}
Consider the familiar way of constructing  four-dimensional 
heterotic string solutions,  consistently  with  
the two-dimensional  world sheet Weyl conformal symmetry. One  assembles
the coordinate and spinor  fields describing the 
(uncompactified,  compactified, 
gauge) target space 
into conformal (left-moving sector) 
and superconformal (right-moving sector)  blocks 
whose total central charges, 
$c=26, \bar c=15$, cancel
those of the conformal and  superconformal ghosts systems.
An infrared regularized theory can then  be defined by 
replacing  the groups  of 
free  fields coordinates for the uncompactified Minkowski  space-time 
$R^4$ by  those of interacting conformal and superconformal theories  
corresponding to a (compact or non-compact) curved  space-time background.
The background  gravitational and gauge fields  
are required to obey the conformal  symmetry 
equations of motion  and to 
depend on at least one free (curvature) parameter which monitors 
the decompactification limit.

A very  convenient space-time background
is that of the semiwormhole solution. This is 
the simplest choice  amongst a
large  family of 
solutions  \cite{koun1,koun2,koun3} having 
a world-sheet ${\cal N}=4$ superconformal algebra.  The 
semiwormhole background 
is described by an $ SU(2)_k\times U(1)_Q$ 
non-linear sigma-model given by  the direct product of 
a level $k$ WZW   model for the spatial coordinates,
times a  non-compact (Liouville or Feigin-Fuchs) 
model with
background charge $Q$
for the time coordinate, $X^0(z,\bar z)=X^0(z)+\bar X^0(\bar z)$.  
The level $ k $ is a free discrete integral parameter representing the 
space-time curvature,  or mass gap, such that
$k\to \infty $ retrieves the 
decompactification limit. To fit the requisite central 
charge, $ c= {3k\over k+2}+1+3Q^2=4$,  one sets the background charge at 
$Q=(2/(k+2))^\ud $.
The left-moving sector
$SU(2)_k$ current algebra
is described by 
three current generators, $ I_i (z),
[i=1,2,3]$, obeying   the  operator product expansion (OPE),
\begin{equation}
I_i(z)I_j(w) = {1 \over 2}  {k\d_{ij}
\over (z-w)^2 }+{i\e_{ijk}I_k(z)\over z-w}
+\cdots, 
\label{eq:ope}
\end{equation}
where $z , w$  are world sheet complex coordinates and 
we use the  so-called field theory 
normalization convention  for
the highest root vector squared length,  $\psi^2=1$.
The right-moving sector includes, in addition
to the $SU(2)_k\times U(1)_Q$ current algebra 
with generators $\bar I_i(\bar z)$ and time coordinate $\bar X^0(\bar z)$,
four free fermion fields $\psi_a (\bar z), [a=0,1,2,3]$,
which build up an affine $SO(4)_1 \simeq SU(2)_{H^+}\times SU(2)_{H^-} $
level 1 algebra, with generators,    
\begin{equation}
S^\pm _i=\ud (\pm \psi_0\psi_i +
\ud \e_{ijk}\psi_j \psi_k)
= [{1\over \sqrt 2} \bar \dh H^\pm ,
e^{ i\sqrt 2 H^\pm } e^{-i\sqrt 2 H^\pm}], \quad [i=3,(1+i2), (1-i2)]
\label{eq:gen}
\end{equation}
where the  bosonic fields counterparts  of the
fermion fields,
$H^\pm  (\bar z)$, take
values on circles $S^1$ of (dimensionless) radii set at the 
self-dual value,  $r={ R\over \sqrt {\a '}} =1$.
The combined algebraic system of bosonic and fermionic operators
$(\bar I_i, S^\pm _i$)  
can be embedded in an ${\cal N}=4, \bar c=6$ superconformal algebra,
whose  (stress tensor, supersymmetry, $SU(2)$ group) generators
$(T, G^a, S_i)$ are  constructed by forming 
suitable products of 
the elementary  field operators, $(\bar I_i, S^+_i,  S_i^-)$
\cite{koun1,koun2,ferrara}.

Aside from the global symmetry under 
$SU(2)_{H^+ } $, the
semiwormhole world sheet theory 
is symmetric under
the diagonal vector subgroup,
$ SU(2)_N  $ of $SU(2)_{H^-} \times SU(2)_k$,  with generators 
$ N_i = I_i+S^-_i \quad [i=1,2,3]$.   The unbroken 
discrete symmetries of its  ${\cal N}=4$ superconformal algebra  include 
the $ Z_2^+ $ parity described by $e^{2\pi  i S^+ }$, where $S^+$ is the 
representation spin, 
and, for $k$ even,   the  $Z_2$ automorphism of  $SU(2)_k$, denoted $Z_2^-$,
which acts on  the generators, $I_i, \ S^-_i$, 
of $SU(2)_k $ and $SU(2)_{H^-}$.
Both of  these parities play an essential 
r\^ole with respect to the space-time 
supersymmetry.
To expose the $Z_2^- $ parity, it is convenient to introduce 
the auxiliary  functions, 
\begin{equation}
Z_k\bigg [ {\a \atop \b }\bigg ] (\tau , \bar \tau ) 
= \sum_{l=0}^k e^{2i\pi \b l}\chi_{l,k}(\tau )
\bar \chi_{(1-4\a )l +2\a k, k} (\bar \tau ),  
\end{equation}
where $\tau =\tau_1 +i\tau_2, \bar \tau = \tau_1-i\tau_2$ denote  
the world sheet torus modular  parameter and
$\a , \b $ are spin structure labels taking the values $(0,\ud )$. 
The affine Lie group character  and theta functions, defined generically as,
$$ \chi_{\l ,k} (\tau , w ) = Tr(q^{L_0}
e^{-2\pi i w I_3})
, \quad  \vt _{n , k} (\tau , w)
= \sum_{\g\in  (M+{\l \over k })} q^{\g^2 /2 } e^{-2\pi i k w \g  } ,$$
where $M$ is the group long root lattice
and $\l $  the representation 
highest  weight, are,
for the  unitary representations  of $SU(2)_k$, with 
$l= 2j \in [0, \cdots , k] $  twice the spin of the  representation, 
given by the familar formulas
\cite{gepner}, 
\begin{eqnarray}
\chi_{l,k}(\tau , w)&=&{C_{l+1,k+2} (\tau , w)\over C_{1,2} 
(\tau , w)}, \quad 
 C_{L,K} (\tau , w)=
\vt_{L, K}(\tau , w)-\vt_{-L, K}(\tau ,w); \nonumber \\
\vt_{L,K}(\tau , w)&=&\sum_{n\in Z} q^{(n+{L\over 2K})^2}
e^{-2\pi i K  (n+{ L\over 2K} ) w}, 
\label{eq:char}
\end{eqnarray}
where the summation index $n$ is twice the spin projection 
and $q=e^{2\pi i\tau }.$ 
The  $SL(2,Z)$ modular group transformation  laws of $Z_k[{\g \atop \d }]$
are similar to those of 
$(\vt [{\g \atop \d }]/\eta )^2$, 
\begin{equation}
 S (\tau \to -{1\over \tau } ): \quad  Z_k
 \big [{\g \atop \d } \big ] \to e^{4\pi i k\a \b }
Z_k \big [ {\d \atop \g }\big ]; \quad T (\tau \to \tau +1):
\quad  Z_k\big [ {\g \atop \d }\big ]  
\to e^{-2\pi ik  \a^2 } Z_k \big [ {\g \atop \d  +\g }\big ] .
\label{eq:trsf}
\end{equation}
For $k$ even, not multiple of 4, 
the $SO(3)_{k/2} 
\sim  SU(2)_k/Z_2$ orbifold model partition function
is constructed by means 
of the projection 
\cite{gepner}, 
\begin{eqnarray}
Z_{SO(3)_{k/2}} (\tau , \bar \tau )&=& (1+S+TS)Z_U (\tau , \bar \tau )
-Z_{SU(2)_k}(\tau, \bar \tau ), \nonumber \\
&=&\bigg [(1+S+TS)\sum_{l= 0 (even) }^k \chi_{l,k} (\tau )\bar \chi_{l,k} (\bar \tau ) - 
\sum_{l=0 }^k \chi_{l,k}(\tau ) \bar \chi_{l,k} (\bar \tau )\bigg ], \nonumber \\ &=&
 \ud \sum_{\g , \d } 
e^{-2i \pi \g \d } Z_k\bigg [{\g \atop 
\d } \bigg ] (\tau , \bar \tau ),
\label{eq:so3}
\end{eqnarray}
$Z_U $ denoting the 
$Z_2$-singlet partition function of the untwisted sector
and $Z_{SU(2)_k}$ that of
the covering group.

We shall need to consider the class of compactified heterotic string models,
$W_k^{(4)} \times K$  with an internal space Kahler  manifold  $K$ 
allowing for ${\cal N} $ conserved supercharges.
One must distinguish here the cases of 
maximal space-time supersymmetry,  ${\cal N}=4$,  from the  non-maximal cases, 
${\cal N}\le 2$. 
The maximal case, corresponding, say, to
$K=T^6 $ or $ K= W_{k'}^{(4)}\times T^2$, allows in  principle for
four conserved supersymmetry charges,
even under $Z_2^+$,  
\begin{equation}
 \t_\pm (\bar z) =e^{{i\over \sqrt 2}
 (H^+\pm {H'}^{+})} , 
 \quad \tilde \t_\pm (\bar z)=e^{{i\over \sqrt 2} 
 (H^-\pm {H'}^{-})}  ,
\end{equation}
where ${H'}^\pm $ are bosonic fields analogs of $H^\pm $ 
for the internal space fermions.
However, since  the  two charges, $\tilde \t_\pm (\bar z)$,  have
non-local OPE with the  $W_k^{(4)}$ superconformal
algebra generators and are hence unphysical, 
these models have only ${\cal N}=2$ supersymmetry.  
The halving of the space-time  supersymmetry  
(from ${\cal N}=4$ to ${\cal N}=2$)
takes place because, in addition to the 
conventional GSO (Gliozzi-Scherk-Olive) modular invariant 
projection on string  states odd 
with respect to  $e^{2\pi i (S^++S^{'+})}$,
one must also project  with respect to 
the  discrete $Z^-_2$ symmetry. 
The construction  of partition functions for the maximal 
supersymmetry case is discussed in 
\cite{ferrara,kiko1}.
For the  non-maximal case, corresponding to a choice of 
internal space  which preserves  supersymmetry  ${\cal N}=2$ 
($K=T^4/Z_2\oplus T^2)$
or ${\cal N}=1$ 
($K=T^6/G$), the charges $\tilde \t^\pm $ are  
not conserved. Since the  $Z_2^-$ symmetry group then cannot be embedded in $K$,
the projection  need  only involve the conventional 
$Z_2^+$ symmetry \cite{kiko1}. 
In particular, for both  ${\cal N}=1,2$  compactifications,
the $SO(3)_{k/2}$ partition function factorizes out in the 
expression of the full partition function. Thus, 
for models  obtained through the substitution,
$R^4\times K \to W_k^{(4)} \times K$, the string one-loop  amplitudes 
are derived from those associated with 
a flat space-time  by inserting 
the correction factor,
\begin{eqnarray}
Z_W(\tau , \bar \tau )&\equiv  &{\G (SU(2)_k)\over V(SU(2)_k)}
\equiv -{X'(\mu )\over 2\pi V(\mu ) }
 \equiv { (\sqrt {\tau_2} \eta \bar \eta )^3 \over V(\mu )} \ud 
\sum_{\a \b =0,\ud }e^{-2i\pi \a \b } 
Z\bigg [ {\a \atop \b }\bigg ] (\tau , \bar \tau ), \nonumber \\
&=& {(\sqrt {\tau_2} \eta \bar \eta )^3\over V(\mu )}
\bigg [\sum_{l= 0 (even) }^k \chi_{l,k} (\tau )\bar \chi_{l,k} (\bar \tau ) + 
\sum_{l=1 (odd)}^k \chi_{l,k}(\tau ) \bar \chi_{k-l,k} (\bar \tau )\bigg ],
\label{eq:zw}
\end{eqnarray}
where the derivative is defined as 
$X'(\mu ) ={\dh \over \dh \mu^2 } X(\mu )$,
and the normalization factor
$V(SU(2)_k)= {1\over 8\pi \mu^3} =
{ (k+2)^{3\over 2} \over 8\pi }$
corresponds to  the (string theory corrected) volume of the 
group space manifold, $ SU(2)_k \sim S^3$,
with a  dimensionless  
radial  scale parameter, $ r=(k+2)^{\ud } ={1\over \mu } $.
The prefactor, $(\sqrt {\tau_2} \eta \bar \eta )^3$,  
accounts for the  determinant of the free 
bosonic spatial coordinates. Furthermore,
describing the momentum modes in the 
time coordinate $U(1)_Q$ model by 
the continuous series of unitary representations, 
$e^{\sqrt {2\over \a '} \b X^0(z, \bar z)}, \ [\b = ip_0-{Q\over 2} ]$
will yield for $X^0(z,\bar z)$ 
the same determinantal factor, $1/(\sqrt {\tau_2} \eta 
\bar \eta )$, as for a
free bosonic coordinate.

The following two representations of the function
$X(\mu )$, defining the semiwormhole partition function, eq.(\ref{eq:zw}), 
will prove useful later: 
\begin{eqnarray}
X(\mu ) &=&{1\over 2\mu } \sum_{(m,n) \in Z^2}
e^{i\pi (m+n+mn)} e^{-{\pi \vert m-n\tau \vert ^2\over 4\mu^2 \tau_2}}
=\sqrt {\tau_2 } \sum_{(m,n)\in Z^2} e^{i\pi n}
q^{[2\mu (m-{n+1\over 2})+{n\over 2\mu} ]^2}
\bar q^{[2\mu (m-{n+1\over 2})-{n\over 2\mu }]^2},\nonumber \\
 X(\mu )& =& [Z_T(\mu )-Z_T (2\mu )],  \quad 
Z_T(\mu )= \sqrt {\tau_2 }\sum_{(m,n)\in Z^2} q^{{1\over 4}(m\mu +n  /\mu )^2} 
\bar q^{{1\over 4}(m\mu -n /\mu )^2}.
\label{eq:xmu}
\end{eqnarray}
These formulas were obtained in \cite{kiko} and can be 
derived  by use of the familiar  conformal field theory 
methodology \cite{cft}.   
The representation  given by the first line in (\ref{eq:xmu}) 
is well-suited to studying the 
decompactification limit,
$k \to \infty $. At  fixed $\tau_2$, one 
directly infers that $\lim_{\mu \to 0} Z_W= 1 
+O(e^{-{1\over \mu^2}}, e^{-{\tau_2\over \mu^2}}) $.
The representation in the second line   involves the partition function 
$Z_T(\mu = {1\over r} )$ for the lattice $\Gamma (1,1)$
of radius $R=r\sqrt {\a '}=\sqrt {\a '}/ \mu $.
This is a modular invariant function of $\tau $
obeying the duality property, $Z_T(\mu )= Z_T({1\over \mu })$.
The second representation in eq.(\ref{eq:xmu})  can be directly used 
in the limit $\tau_2 \to \infty$ to derive 
an exponentially convergent  asymptotic expansion,
\begin{equation}
lim_{\tau_2 \to \infty } Z_W (\tau , \bar \tau ) 
=-{1\over 2\pi V(\mu ) } {\dh X(\mu )\over \dh \mu^2}
= {(\tau_2)^{3\over 2}\over V(\mu )}  \bigg [ \bigg (e^{-\pi \mu^2 \tau_2} -{1\over \mu^4} 
e^{-{\pi \tau_2\over \mu^2 }} +\cdots \bigg ) 
-4\bigg (\mu \to 2\mu \bigg  ) \bigg ]  .
\label{eq:xlim}
\end{equation}
valid for fixed $\mu $.
We observe that the limits $\mu \to 0 $
and $\tau_2 \to \infty $ do not 
commute, reflecting the non-uniform convergence 
of the sums over momentum 
and winding integers, $m $ and $n $, respectively.
Indeed,  whereas $Z_W \to 1$ if  the limit $\mu \to 0$ is taken first, 
taking the limit  $\tau_2 \to 
\infty $  prior to 
$\mu \to 0 $,  yields $ Z_W\to 0$.

\subsubsection{\bf Marginal deformations}
Consider the regularized zero-modes conformal generators for the 
heterotic string  semiwormhole  solution  $W_k^{(4)}$ with an orbifold 
six-dimensional internal space $K$:
\begin{equation}
 L_0= {(j+\ud )^2\over k+2} 
+\sum_a {J_a^2\over k_a}+N+E_0-1, \quad 
 \bar L_0= {(j+\ud )^2\over k+2} +{\bar Q^2\over 2} 
 +\sum_{i=1}^3 {\bar Q_i^2\over 2}+\bar N +E_0-\ud ,
\label{eq:cfw}
\end{equation}
where the $SU(2)_k\times U(1)_Q$ conformal weights
contribute additively as,
$ {j(j+1)\over k+2} +{Q^2\over 8} ={j(j+1)+1/4\over k+2},$
$E_0$ is the  vacuum  energy 
shift from the internal space degrees of freedom and 
$N , \bar N $ are the oscillator number operators.
We have denoted the zero-modes 
charges for the  Cartan subalgebra of the fermionic 
$ SO(8) $, level 1,  affine algebra (two-dimensional transverse 
space-time and six-dimensional internal space) by 
$( \bar Q, \bar Q_1, \bar Q_2, \bar Q_3) $
and those for the Cartan subalgebra of the unbroken
gauge symmetry group, $\prod_a G_a$, of levels $k_a$, by  $J_a$. 
The normalization conventions are such that, 
$ J_a(1) J_b(0) = \d_{ab } {k_a\over 2} +\cdots , \quad 
 \bar Q_a(1) \bar Q_b(0) = \d_{ab } +\cdots .  $
 No confusion should arise  from using  the same symbols to denote
the current densities (functions 
of $z , \bar z$) and their associated zero-modes.

Let us focus on the space-time $(I_3; \bar I_3, \bar Q)$ and 
gauge $J_a $  operators and 
rewrite the conformal generators succinctly as,
\begin{equation}
\bar L_0= {\bar Q^2 \over 2} + {\bar I_3^2\over k}
+\cdots ,  \quad 
  L_0=\sum_a { J_a^2\over k_a} + {I_3^2\over k}
+\cdots ,
\label{eq:cfwa}
\end{equation}
where the dots in (\ref{eq:cfwa}) stand  
for all the  remaining contributions
implicit in (\ref{eq:cfw}).
Observing that the conserved
charges,   $ \bar I_3  $ and $ \bar I_3+ \bar Q $  
tend, in the decompactification 
limit,  $k \to \infty $,  to the space orbital 
and angular momentum helicity operators,  
one is led to describe  the perturbations   due to
finite gauge and gravitational  background 
fields in terms of the 
conformal  vertex operators,
\begin{equation}
V_F^a= {1\over 2\pi } \int d^2\s 
{F^a \over \sqrt {k_a (k+2)} }J_a(z)
[\bar I_3(\bar z) +\bar Q (\bar z)]
, \quad 
 V_R= {1\over 2\pi } \int d^2 \s {R\over \sqrt {k(k+2)}} 
I_3(z) [\bar I_3(\bar z) +\bar Q (\bar z)],
\label{eq:vert}
\end{equation}
where $\bar Q (\bar z)= {i\over 2}\e_{3ij} \psi_i \psi_j  
=i\psi_1\psi_2 $ is the spatial  helicity  current density, having 
$\bar Q = \bar Q_0 $ as the  zero-mode component of its Laurent 
series expansion, $\bar Q (\bar z)=\sum_{n\in Z} \bar Q_n \bar z^{-n-1}$.
Added to the world sheet action $S_0$, the extra  action  
$\d S = V_R+\sum_a  V_F^a$  
is a conformal weight  $(1,1)$
marginal perturbation leaving the
conformal symmetry of the model intact.
While the case of greatest practical interest of 
conformal operators with   constant 
field strength parameters exists 
only for non-flat 
theories, the flat space-time limit is useful to set the constant 
normalization factors in eq.(\ref{eq:vert}).  
In the sigma-model  classical limit of large $k$, where the generators 
can be expanded with respect to the space coordinates,
$X^a(z, \bar z), \ [a=1,2,3]$ as 
$\bar I_a \equiv  -i\sqrt {k\over 2} Tr(\tau_a g^{-1} \bar \dh g) 
=\sqrt {k\over 2} (\bar \dh X_a -\ud \e_{abc} X^b \bar \dh X^c +\cdots )$, 
using, $g(z,\bar z)= e^{{i\over 2} \vec \tau \cdot \vec X }$, we find that  
$V_F^a$ reproduces the vertex operator for an uniform  
electromagnetic field $F_{\mu \nu }$,  
$$V(A^a_\mu  )= {g_{st}\over 2\pi } \int d^2 \s A_\mu ^a 
(\bar \dh X^\mu +\cdots ) {J^a \over \sqrt {k_a} } , \quad [A^a_\mu (X) =-\ud  
F^a_{\mu \nu } X^\nu ]$$
with the identification, $ F^a= \sqrt 2 g_{st} F_{12}^a$ where 
$g_{st} $ is the string theory  coupling constant. The fermionic 
terms, indicated by dots, 
are reconstructed by supersymmetry. A similar statement holds for $V_R$
in relation with a constant curvature gravitational field,
$(R_{ab})_{\mu \nu }.$
With parameters  $F^a  $ and $ R$ independent of 
$z, \bar z$, the perturbed  action
depends then  solely on the 
zero-modes operators,
\begin{equation}
 \d S= -2\pi (2\tau_2) 
 \bigg [{F^a\over \sqrt {k_a(k+2)}} J^a 
 (\bar Q +\bar I_3)
 +{R\over \sqrt {k(k+2)}}  I_3
 (\bar Q +\bar I_3) \bigg  ],
\label{eq:pert}
\end{equation}
where we have accounted for the change of 
coordinate variables  from
the real  (Euclidean metric)  orthogonal set, 
$\s =(\s_1, \s_2) \in [0,1]^2 $, such that $ds^2=d\s_1^2 +d\s_2^2 $, to the 
complex set, 
$ z=(\s + \tau t )/2, \quad 
 \bar z=(\s +\bar  \tau  t )/2 $,  using $\int d^2 \s = \int d^2z \sqrt {h} 
 =2\tau_2 \int dz d\bar z$.
The zero-modes conformal generators of the perturbed theory, 
$(L'_0,\  \bar L'_0)$,
can  now  be identified   by  comparing 
the Lagrangian and
Hamiltonian representations
of the world sheet  one-loop  functional integral,
$$ Z= \int [DX][D\psi]  e^{S_0+\d S}  
=\ud  \int_F {d^2\tau \over \tau_2} Tr\bigg (e^{-2\pi \tau_2( L_0+\bar L'_0)}
e^{2i \pi  \tau_1(L_0-\bar L'_0)}\bigg ).$$
One can then  describe the conformal perturbations $\d S$ as 
deformations of the Cartan subalgebra  tori for the conserved
fermionic and gauge  symmetry groups by 
defining an associated extended Narain orthogonal coset  moduli space, 
with a lattice of conserved charges, $\G (r+1,\bar r +1)$, where $r$ is the 
rank of the gauge group and $\bar r$ that  of the $SO(2\bar r)$ 
group of conserved fermionic charges. The one unit additions here refer to the 
$I_3, \bar I_3$ charges.
The vertex operators  parameters  $F^a , R$ 
provide us with a local  description of the moduli space of deformations. 
A description of the global structure, 
incorporating the back-reaction effects, is developed  by
acting on the  zero-modes  lattice  with the 
orthogonal  group, $SO(r+1, \bar r +1,R)$. 
The transformations
which reproduce the perturbations
in  eq.(\ref{eq:pert}), decompose into three factors:
(i) The  right-moving sector  rotation of 
angle $\t ' = \cos^{-1}({k\over k+2})^\ud $
which introduces the 
total angular momentum  projection and its 
orthogonal complement,
\begin{equation}
({\bar I_3 \over \sqrt k}, {\bar Q\over \sqrt 2})
\to ({\bar I_{3\t '} \over \sqrt k}, {\bar Q_{\t '} \over \sqrt 2}) \equiv 
\bigg ({\bar I_3 +\bar Q \over (k+2)^\ud}, 
{-2 \bar I_3 +k \bar Q\over 
 (2k(k+2))^\ud } \bigg );
\label{eq:ort}
\end{equation}
(ii) The left-moving sector   rotation of angle $\t $
which mixes the space-time and gauge group 
charges,
\begin{equation}
({I_3\over  \sqrt k }, {J^a\over \sqrt k_a}) \to 
({I_{3\t } \over  \sqrt k }, {J^a_\t \over \sqrt k_a} )\equiv 
(\cos \t {I_3 \over \sqrt k} 
+\sin \t  {J_a\over \sqrt k_a},
\sin  \t {I_3 \over \sqrt k} 
+\cos \t  {J_a\over \sqrt k_a});
\label{eq:ort1}
\end{equation}
(iii) The Lorentz  boost of  hyperbolic angle $\psi /2$
which mixes the rotated  left sector  and right sector generators,
\begin{equation}
\pmatrix {{\bar I_{3\t ' }\over \sqrt k} \cr {I_{3\t } \over \sqrt k }\cr }  
\to \pmatrix {{\bar I_{3\t '\psi }\over \sqrt k} 
\cr {I_{3\t \psi } \over \sqrt k }\cr } = 
 \pmatrix {\cosh { \psi \over 2} & \sinh { \psi \over 2} \cr 
\sinh {\psi \over 2 } & \cosh {\psi \over 2 }\cr }
\pmatrix {{\bar I_{3\t ' }\over \sqrt k} \cr {I_{3\t } 
\over \sqrt k }\cr }.
\label{eq:lor}
\end{equation}
The induced conformal generators increments,
\begin{eqnarray}
\d \bar 
L_0&=&[{\bar Q_{\t ' }^2\over 2 }+{\bar I_{3\t '\psi  }^2 \over k } +\cdots 
] - [{\bar Q^2\over 2 }+{\bar I_3^2 \over k } +\cdots ],    \nonumber \\
&=&{\cosh \psi -1\over 2} 
 \bigg [{(\bar Q+\bar I_3)^2\over k+2} 
 +(\cos \t {I_3\over \sqrt k}+\sin \t {J^a\over 
 \sqrt {k_a}})^2 \bigg ] \nonumber \\
&+&\sinh \psi {\bar Q +\bar I_3 \over (k+2)^\ud }
 (\cos \t {I_3\over \sqrt k}+\sin \t {J^a\over \sqrt {k_a}}),\nonumber \\ 
\label{eq:parm0}
\end{eqnarray}
obey level matching, 
$\d L_0\equiv L'_0
-L_0=\bar L'_0-\bar L_0 \equiv \d \bar L_0 $, by construction. 
Comparing the dependence  of $\d L_0 , \ \d \bar L_0$ 
for infinitesimal  values of the parameters, 
$R, F^a$, with that of 
the  perturbed action $\d S$, eq.(\ref{eq:pert}), using the formal 
identification, $\d S \simeq  - 4\pi \tau_2 \d \bar L_0 $, 
imposes the following connection formulas
between the two sets of parameters $(R,F^a)$ and $(\t , \psi )$: 
\begin{eqnarray}
 F^a&=&
\sinh \psi \sin \t ,
 \quad R= \sinh \psi \cos  \t ; \nonumber \\
{ \cos \t \choose \sin \t }&=& 
{1\over  ( F^{a2}+ R^2)^\ud } 
{R  \choose F^a} , 
\quad \sinh \psi = (  F^{a2}+ R^2)^\ud . 
\label{eq:parm}
\end{eqnarray}
Using these relations, the conformal operators increments  can
also  be expressed in the alternate forms,
\begin{eqnarray}
\d \bar L_0= \d L_0 &=&
\ud C_-\bigg [ {(\bar Q +\bar I_3)^2\over k+2} 
+{(R {I_3\over \sqrt k} +F^a {J^a\over \sqrt k_a })^2 
\over (F^{a2}+R^2 )}\bigg ]
+{(\bar Q +\bar I_3)\over \sqrt { k+2}} 
(R {I_3\over \sqrt k} +F^a {J^a\over \sqrt k_a }), \nonumber \\
&=&\ud \bigg [ \sqrt {C_+} {(\bar Q +\bar I_3)\over \sqrt {k+2} } 
+{1\over \sqrt {C_+} }     
(R {I_3\over \sqrt k} +F^a {J^a\over \sqrt k_a })
\bigg ]^2 -
{(\bar Q +\bar I_3)^2\over  k+2}, 
\label{eq:incr}
\end{eqnarray}
where $C_\pm = \pm  1 +(1+F^{a2}+R^2)^\ud $. 
The additional terms,  of 
$O(F^{a2}, R^2)$  and beyond,
with respect to  those in 
eq.(\ref{eq:pert}),
which are given by the term  involving the factor 
$C_-$ in  the first line of (\ref{eq:incr}), 
are associated with the back-reaction corrections.

The deformed theory  can be represented  in still another 
parametrization by means of the generalized  gravitational background fields 
(metric tensor $G_{\mu \nu }(X)$, 
2-form    $ B_{\mu \nu }(X)$, dilaton $\Phi (X) $) 
and gauge background fields ($A_\mu ^I= A_\mu^a\hat J_a^I $)
which appear as 
coupling constants in the space-time and gauge sectors of the  world sheet 
sigma-model action,
\begin{eqnarray}
S&=& -{1 \over 4\pi  \alpha ' } \int \int  d^2\sigma \bigg \{
\sqrt h h^{\alpha \beta } \bigg [  (\partial_\alpha X^\mu 
\partial_\beta X^\nu +i\bar \psi_R^\mu 
\rho_\alpha \nabla_\beta \psi_R^\nu ) G_{\mu \nu } (X)+ 
\dh_\a F^I \dh_\b F^I \bigg ]\nonumber \\
&+&i\epsilon^{\alpha \beta } \bigg (\partial_\alpha X^\mu 
\partial_\beta X^\nu  B_{\mu \nu }(X) 
+\partial_\alpha X_L ^\mu \partial_\beta F_I A_\mu^I (X) \bigg )
-\alpha' \sqrt h R^{(2)} \Phi (X) \bigg \},
\label{eq:wsa}
\end{eqnarray}
using  familiar notations for the world sheet 
metric and antisymmetric  tensors, covariant derivative 
and curvature, $h, \ \e , \  \nabla_\b = \dh_\b X^\mu D_\mu , \ R^{(2)}$.
To deduce the semiwormhole solution background fields, it is convenient  
to write the $SU(2)_k$ WZW action
in the realization \cite{hassan}, 
$ g(z,\bar z)= e^{{i \over 2} \g (z,\bar z)  \s_2} 
e^{{i\over 2} \b (z, \bar z)  \s_1} 
e^{{i\over 2} \a (z, \bar z) \s_2} , $  with  the spatial coordinates angles, 
$\g  \in [0,2\pi ], \b \in [0,\pi ],\a 
\in [0,2\pi ] $.
Including the action for the  $U(1)$ 
gauge coordinate fields, $\phi (z) = \sqrt {2\over k_a} F^I(z)$, 
such that $J_a(z) =\dh \phi (z)$,
and that for the non-compact  time coordinate field,
$X^0(z,\bar z)$,  
the total action reads:   
\begin{eqnarray}
S_0&=&{k\over 8\pi }\int d^2 z 
\bigg (\dh \a\bar \dh \a +  \dh \b \bar \dh \b + \dh \g \bar \dh \g 
+2\cos \b \dh \a \bar \dh \g +
{2k_a\over k} 
\dh \phi \bar \dh \phi \bigg )  \nonumber \\
&-&{1\over 2\pi } \int d^2 z  \bigg ({1\over \a '} \dh X^0 \bar \dh X^0 
+ {1\over 2  \sqrt {2\a '} } \sqrt {h} R^{(2)} Q X^0 \bigg ).
\label{eq:wzw}
\end{eqnarray}
One can now represent the marginal perturbations   associated to 
a finite constant magnetic field $H$ and an infinitesimal 
constant gravitational  field, $\d \l $,  in
terms of the Cartan subalgebra generators,
$I_3=\dh \g +\cos \b \dh \a , \
\bar I_3=\bar \dh \a +\cos \b \bar \dh \g $,
by adding to $S_0$ the extra action, 
\begin{eqnarray}
 \d S &=& {\sqrt {k k_a} \over 2\pi } 
H\int d^2 z\bar I_3 (\bar z)  J^a(z)+\d \l  {k\over 8\pi }\int 
d^2 z\bar I_3 (\bar z)I_3 (z),    \nonumber \\ 
&=&{\sqrt {k k_a}H\over 2\pi } \int  d^2 z (\bar \dh \a +\cos \b \bar \dh \g ) 
\dh \phi  \nonumber \\ 
&+&{k\d \l \over 8\pi } \int  d^2z
(\bar \dh \a +\cos \b \bar \dh \g )
(\dh \g +\cos \b \dh \a ).
\label{eq:def1}
\end{eqnarray}
Since $\a , \g $ are Killing (isometry)  coordinates of the
semiwormhole  manifold, this has 
an orthogonal coset moduli space of vacua described by
$M \in O(2,2,Z)\setminus O(2,2,R)/O(2,R)\times O(2,R),$ where 
$M $ is a $4\times 4$ matrix constructed from the metric and torsion tensors 
in the basis of Killing coordinates, $\t_\pm = (\g \pm \a )/2 $, which
transforms  under $\O \in O(2,2,R)$, to leading 
$O({1\over k})$,  as $M\to  M'=\O M \O ^T$, along with 
the dilaton shift, $\Phi \to \Phi '=\Phi +{1\over 4} \ln det (G'G^{-1})$.
The case of a  finite   parameter  $\l $ can then  be described by means of
the so-called solution-generating transformation method
\cite{hassan,kigi}. Starting with the unperturbed  background 
fields, denoted  $\hat G, \hat B$, one performs first the particular non-linear 
transformation,   $\hat M\to \Omega M \Omega^T $,
$$\hat M=\pmatrix {\hat G^{-1} & \hat G^{-1}  \hat B\cr 
\hat B^T \hat G^{-1} & \hat G +\hat B^T \hat G^{-1} \hat B }, \quad 
\Omega =\ud \pmatrix {R+S & R-S \cr  R-S & R+S }, \quad [R, \ S \in O(2,R)]$$
specialized to the case, $R =S^T$, with $R$  a two-dimensional rotation of angle $a$,
followed by the  variables  rescalings, $\t_ +\to \t_+/\cos a , \ 
\t_-\to \t_- /(\cos a -k \sin a) $ 
and a constant shift of $B_{\t^- \t^+ } \to B_{\t^- \t^+ } +
\cos a (k\cos a +\sin a)$, 
corresponding to a total derivative term.
The deformed WZW model background fields 
depend on the rotation angle parameter $a$ through the  two parameters 
\cite{hassan}: 
$\l_+= \cos^2 a +{k\over 2} (k\sin^2 a -\sin 2a ), \ \l_- ={k\over 2} 
(k\sin^2 a -\sin  2a) $. 
Expressing the perturbed  action, $S_0 +\d S $,
so as to  achieve a matching with 
the generic  form of the world sheet action  in (\ref{eq:wsa}), 
gives us the  following background fields,
solutions of the classical  string equations of motion:   
\begin{eqnarray}
G_{tt}&=&1, \quad G_{\b \b } ={k\over 4} ,\quad  
G_{\b \a }=0, \quad G_{\b \g } = 0,    
\quad G_{\a \a }= {k\over 4} \bigg [{\l_+ -\l_- \cos \b \over \d } -{8H^2\cos^2\b 
\over \d^2}\bigg ],   \nonumber \\
G_{\g \g }&=& {k\over 4} \bigg [{\l_+ -\l_- \cos \b \over \d } -{8H^2
\over \d^2}\bigg ], \quad 
G_{\a \g }= {k\over 4}\bigg [ {\l_+ \cos \b -\l_-  \over \d } -{8H^2\cos \b 
\over \d^2}\bigg ];   \nonumber \\
B_{\b \a }& =& B_{\b  \g }=0, \quad 
B_{\a \g }= {k\over 4}\bigg [ {\l_+ \cos \b +\l_-
\over \d }\bigg ]; \nonumber \\
\Phi - \Phi_0 &=&-{1\over 4} \ln det {G\over G_0}= 
-\ud \ln \d  + \ln (g_X );  \nonumber \\
{\sqrt {k_a} A_\a \over \cos \b } &=&\sqrt {k_a} 
A_\g =2\sqrt {2k\over k_a\a '} {H\over \d }.
\quad \bigg [\d = \l_++\l_-\cos \b 
\bigg ]
\label{eq:bgf}
\end{eqnarray}
Because of the invariance under a uniform constant rescaling of 
$\l_+, \ \l_-, \ H$, one can  group $\l_\pm $ in a single 
parameter, $\l \in [0, \infty ]$, which is  defined  as,
$ \l_\pm  = \l \pm  {1\over \l }$. 
The unperturbed space-time case, $H=0, \l =1$, is described by  the 
gravitational background fields,
$ G_{tt}=1, 
G_{\a \a } = G_{\b \b }=G_{\g \g }= {k\over 4},
\  B_{\a \g }= 
G_{\a \g } ={k\over 4} \cos \b ; \ 
\Phi_0  =-{QX^0\over \sqrt {2\a '}} =
-{X^0\over \sqrt {\a ' (k+2)}}$. 
The spatial coordinates are to be identified as,
$[\a ,\b , \g ]=[{X^m \over 
\sqrt {\a ' k}}], [m = 1,2,3]$. 
The dilaton field  is
normalized with respect to the flat space-time  limit, 
$k \to \infty,  \l \to 1$, such that
$e^{<\Phi >}=g_{st}={g_X\over \sqrt 2} $, the string theory loop 
expansion parameter, identifies with 
the four-dimensional gauge coupling constant
in the string theory normalization.
(The relationship between the 
field  and string theories normalized coupling constants and gauge potentials,
characterized by a highest root  of squared length,  
$\psi^2=1$ and $\psi^2=2$, respectively,  is described by: 
$ g_{ft}=\sqrt 2 g_{st}, 
A_{st}=A_{ft}\sqrt 2$,  leaving the product 
$ Ag $ invariant.)

The comparison of the  dependence of the  conformal weights on the 
background  fields, eq.(\ref{eq:bgf}), with the corresponding 
dependence of the mass spectrum for  particle propagation in the  same
background  fields, based on the mass-shell conditions, 
${\a '\over 4} M^2 = L_0$,  can be 
used \cite{kiko1} to establish  
the connection formulas between  the sigma-model 
parameters, $[H, \lambda ]$ and the  vertex operators parameters
$[F^a, R]$:
\begin{equation}
 H={F^a\over \sqrt 2 C_+}, \quad  
\l_+ = \sqrt 2 C_+ , \quad 
\l_- = \sqrt 2 {R\over  C_+}.
\end{equation}
This yields the explicit formulas:
\begin{eqnarray}
H &=&{F^a/\sqrt 2 \over 1+(1+F^{a2}+R^2)^\ud }
={F^a\over \sqrt 2}[1-{1\over 4} 
F^{a2} -{1\over 4} R^2 +{1\over 4 } F^{a2} R^2 +{1\over 8} 
R^4 +O(F^{a2}R^4, R^6)], \nonumber \\ 
\l &=&\bigg [{1+R+(1+F^{a2}+R^2)^\ud \over  
1-R+(1+F^{a2}+R^2)^\ud }\bigg ]^\ud 
= 1+\ud (1  -{1\over 4} F^{a2} )R \nonumber \\
&+&{1\over 8}(1 -{1\over 2}F^{a2} )
R^2 +O(F^{a2}R^3, R^3),
\label{eq:parametres}
\end{eqnarray}
where the second equations give the power expansions 
for small deformation  parameters.  

\subsection{\bf 
String theory vacuum functional and field theory effective action }

The one-loop 
partition function  for the deformed
string theory, 
\begin{equation}
Z(F^a,R)= \ud \int_F {d^2 
\tau \over \tau_2 } Tr (q^{L'_0} \bar q^{\bar L'_0} ) ,
\label{eq:defp}
\end{equation}
defines $Z(F^a,R)$ as 
the generating 
functional of one-loop vacuum-to-vacuum transition  
amplitudes with external lines 
insertions of the background fields 
$F^a, R$.
For four-dimensional  heterotic string orbifold models, one can write the 
explicit formula, 
\begin{eqnarray}
Z(F^a, R)&=&{ V^{(4)} \over (4\pi )^2}
\int_F{d^2\tau \over \tau_2} 
{1\over 2\pi^2\tau_2^2} {Z_W  \over \vert 
\eta \vert^4 } {\sum  } '  Z_0(q,\bar q)Z_G (q)
e^{-4\pi \tau_2 \d L_0}; \nonumber \\
{\sum } ' &= &\ud \sum_{\bar \a ,\bar \b } 
(-1)^{2\bar \a +2\bar \b } 
{1\over \vert G \vert } \sum_{g,h} \chi(g,h)\e (g,h)
\ud \sum_{ \a , \b }\ud \sum_{\a' \b'}
\eta(g,h; \a , \b ; \a' , \b' ), \label{eq:part30}  \\
Z_0(q,\bar q)& =& {\bar \vt [{\bar \a \atop \bar \b }] \over\bar  \eta }
\prod_{i=1}^3{\bar \vt [{\bar \a +g_i\atop \bar \b +h_i}] \over\bar  \eta }
\prod_{i=1}^3\bigg \vert {\eta \over \vt [{\ud  +g_i\atop \ud  +h_i}] }
\bigg \vert ^2, \quad  
Z_G(q)= 
\prod_{I=1}^8{ \vt [{\a +g_I\atop \b +h_I}] \over \eta }
\prod_{I'=1}^8{\vt [{\a' +g_{I'}\atop \b' +h_{I'}}] \over \eta }.
\label{eq:part3}
\end{eqnarray}
The factor ${1\over \tau_2}$ in eq.(\ref{eq:defp}) 
arises from the ghosts contributions.
The internal and gauge spaces 
determinantal factors in the partition function, eq.(\ref{eq:part3}),
are denoted 
by  $Z_0$ and $Z_G$, respectively. 
The summations over the  right and left sectors spin structures, 
$(\bar \a , \bar \b )$ and $    (\a , \b , \a', \b')$ and  the 
left and right  sectors  orbifold spatial and gauge  twists, 
$(g_i, h_i), (g_I,h_I, g_{I'},h_{I'})$ are  represented by the 
primed summation symbol. The sums  include 
the twisted subsectors  degeneracy factors $\chi (g, h) $, 
the discrete torsion phase factor, $\e (g,h)$,  and the phase factors, 
$\eta (g,h;\a , \b ; \a', \b' )$, which effect the 
extended GSO  orbifold projection. 
More information  concerning these factors, the  definition of 
Dedekind eta-function $\eta (\tau )$ and
Jacobi theta-functions $\vt  [ {\a \atop \b }] (\tau )$, especially the 
phase conventions, is provided in \cite{chem}.
The space-time volume  $V^{(4)}$ appears through the integration
over the flat limit $D=4$  Minkowski space-time translations 
zero-modes,
\begin{equation}
 {\a '} ^{2}V^{(D)}\int {d^Dp\over  (2\pi )^D } 
e^{-\pi \tau_2 \a' p^2}= {V^{(D)}\over 
(4\pi^2\tau_2 )^{D/2}}.
\label{eq:plat}
\end{equation}
The dependence on the background fields parameters  can be exposed by 
expanding the exponential  factor inside the trace,
   $$Z(F^a,R)=\sum_{(m,n)\in Z^2}Z_{m,n}F^m R^n .$$
The power expansions  of $Z(F^a,R)$ 
out to quartic orders in $F^a, R$
are provided in  Appendix
\ref{sec:appa}.
The first few terms, relevant for our purposes here, read:
\begin{eqnarray}
 Z_{0,0} & =& {V^{(4)}\over 2 (2\pi )^4}  
 \int_F {d^2\tau \over \tau_2^3} 
{Z_W \over \eta^2 \bar \eta ^2} 
{\sum } '  Z_0  (q, \bar q) Z_G (q) , \nonumber \\
 (4\pi )^2 [Z_{2,0} F^{a2}; Z_{0,2}R^2 ]& =& 4V^{(4)}
 \int_F {d^2\tau \over \tau_2} 
{Z_W \over \eta^2 \bar \eta ^2} 
{\sum } '  Z_0  (q, \bar q) Z_G (q) \nonumber \\
& \times &\bigg [  { F^{a2}\over k_a (k+2)}
 \bigg ([(\bar Q +\bar I_3)^2 -{k+2\over 8\pi \tau_2 }]
 [J^{a2}-{k_a \over 8\pi \tau_2} ] 
-{k_a (k+2)\over  (8 \pi \tau_2)^2 } \bigg  ) \nonumber \\
& ;&{R^2\over k(k+2) }
\bigg ([ (\bar Q+\bar I_3)^2 -{k+2\over 8\pi \tau_2} ]
[ I_3^2-{k\over 8\pi \tau_2}]
-{k(k+2)\over (8\pi \tau_2 )^2} \bigg )\bigg ], \nonumber   \\
 (4\pi )^2Z_{4,0}F^{a4}& = & {16\pi^2 V^{(4)} F^{a4} \over 3k_a^2(k+2)^2} 
 \int_F d^2\tau \tau_2 
 {Z_W\over \eta^2\bar \eta^2}{\sum } ' 
 Z_0 (q,\bar q)  Z_G (q)  \nonumber \\
&\times &\bigg [ (\bar Q +\bar I_3 )^4 J_a^4 +
6[k_a(\bar Q +\bar I_3)^2+(k+2) J_a^2]  
[{k_a(k+2) \over 256\pi^3 \tau_2^3}  
-{1\over 8\pi \tau_2} (\bar Q +\bar I_3 )^2 
J_a^2 ]  \nonumber  \\
&+&{3\over 64 \pi^2\tau_2^2} [k_a (\bar Q +\bar I_3)^2 
+(k+2) J_a^2]^2 \bigg ].
\label{eq:string}
\end{eqnarray}
The basic assumption of the background field approach 
concerns the equivalence of the low energy string theory  limit to an 
effective point field theory.  Substituting 
in the corresponding  effective action, denoted  $S_{EFF}$, the expressions 
for the gauge and gravitational  fields for the semiwormhole background,
we expect that $S_{EFF}$ will take the same functional form as $Z$ with 
respect to $F^a $ and $R$.
Specifically, we shall proceed as follows:
First, we  write a  
general ansatz for the effective action, 
at tree and one-loop levels, as a function of the 
bosonic components of the
gravitational   and gauge fields, 
$  G_{\mu \nu }, B_{\mu \nu }, \Phi , A^a_\mu $, of universal character.
The motivations for including the 2-form and dilaton fields are inspired 
partly from consideration of the underlying four-dimensional
string theory, partly from  a possible
embedding   in a ten-dimensional theory, where these fields are part of the 
gravitational supermultiplet.
The structure of $S_{EFF}$ is strongly constrained by the requirements of 
gauge symmetry,  global (holomorphic couplings) and local
${\cal N}=1$ supersymmetry, and of the combined
axionic, $\d B= d\L $, and Peccei-Quinn symmetries
acting on the 2-form and dilaton fields, 
which are bosonic partners in the 
dilaton four-dimensional 
chiral superfield. 
Next, we substitute in $S_{EFF}$ 
the semiwormhole 
background fields solutions. Finally, 
we make a term by term identification of 
powers of $F^a , R$ between the string theory 
functional, $Z $, and the corresponding field theory 
functional formed by adding to $S_{EFF}$  the contributions from the 
one-loop massless modes.  The matching  equations for the coefficients 
of $F^a, \ R$ are further analyzed as functional relations with respect to
the infrared cut-off, ${1\over k}$. 

The ${\cal N}=1$ supersymmetric 
four-dimensional  effective bosonic action, including the tree and 
one-loop levels terms, 
up to quadratic (quartic)  order  in derivatives of the  gravitational
(gauge) fields, but omitting momentarily matter fields, 
takes the general form:
\begin{eqnarray}
S_{EFF}&=&{1\over \a '} \int d^4X\sqrt G \bigg [ {1\over \a '}Z_\Lambda 
+(e^{-2\Phi } + Z_R)     R + 
(e^{-2\Phi } + Z_\Phi ) 4(D_\mu \Phi )^2 \nonumber \\
&-&(e^{-2\Phi } + Z_H){1 \over 12} 
(H_{\mu \nu \rho })^2  
-{\a ' k_a \over 8  }    \sum_a( e^{-2\Phi }+Z_{F^a}) 
(F_{\mu \nu }^a)^2  \nonumber \\
 &+& \a '\bigg ((e^{-2\Phi } + Z_{1R}) \ t_1 \ (R_{\mu \nu \rho \s })^2 
+ (e^{-2\Phi } + Z_{2R})\  t_2\  (R_{\mu \nu })^2 
+ (e^{-2\Phi } + Z_{3R}) \ t_3 \ R^2 \bigg )  \nonumber \\
&-&{{\a '} ^{2}k_a^{3/2} \over 8}  \sum_a (e^{-2\Phi } + Z_{0F^a})\  r \ 
 (F_{\mu \nu }^aF_{\nu \rho }^a
 F_{\rho \mu }^a)  \nonumber \\
&-&{{\a '} ^{3}\over 16} \sum_a k_a^2 \bigg ( (e^{-2\Phi }+ Z_{1F^a})\ s_1 \ 
(F_{\mu \nu }^a)^2 (F_{\rho \sigma })^2
 +(e^{-2\Phi }+ Z_{2F^a})\ s_2 \ F_{\mu \nu }^aF_{\nu \rho }^a
 F_{\rho \s  }^aF_{\s  \mu }^a\bigg ) \bigg ] +\cdots , 
\label{eq:effa}
\end{eqnarray}
where the saturation of  space-time indices employs the familiar  
convention, $(F_{\mu \nu })^2 = F_{\mu \nu }F^{\mu \nu }, \cdots $, 
using the metric tensor  $G_{\mu \nu }$ to raise and lower indices.
In terms of the  differential forms notations, with 
$A=A_\mu dX^\mu  ,  \o = \o_\mu dX^\mu   $
as the gauge and spin connections 1-forms,   matrices in the 
gauge and tangent space-time  $SO(4)$  group, $A=A^aT_a, 
\ \o =\o_{ab} J_{ab}$, 
the field strength 2-forms and space covariant derivative are: 
$F=dA+A^2, R=d\o+\o^2,  D = d+\o ,$ where $R= \ud R_{ab\mu \nu } dX^\mu \wedge 
dX^\nu $.
The alternate  tensorial  notation
will also be  used  for the 
curvature scalar, $R=R^\mu _{\ \mu } $,  and the Riemann and  Ricci curvature   tensors, 
$  R_{\mu \nu \rho \s }, \ R_{\mu \nu }= R^\l _{\ \mu \l \nu } $. 
The tree and one-loop  level  terms in eq.(\ref{eq:effa})
can be recognized by the specific coupling 
with the dilaton field,
$e^{(2g-2) \Phi },$ with the values  $g=0,1$ for the 
genus parameter of the world sheet surface.
We have accounted for the fact that a cosmological 
constant term,  $Z_\L $, is 
only present at one-loop level. 
The modified 3-form field strength, $H$,  associated with 
the 2-form Neveu-Schwarz field, includes the gauge and gravitational 
second Chern-Simons 
3-form  terms $(\o_{3Y}, \ \o_{3L}$, such that $d\o_{3Y}=Tr(F\wedge F), 
\ d\o_{3L}=Tr(R\wedge R)$) in the form familiar from 
ten-dimensional string theory, 
\begin{eqnarray}
H_{\mu \nu \rho }&=&\dh_{[\mu } B_{\nu \rho ]}-{\a '\over 4}  
((\o_{3Y} )_{\mu \nu \rho }
-(\o_{3L} )_{\mu \nu \rho }), \ 
(\omega_{3Y})_{\mu \nu \rho }= Tr(A_{[\mu } 
F_{\nu \rho ] } -{1\over 3} 
A_{[\mu  } A_{\nu ]} A_{\rho ]}) , \nonumber \\
(\omega_{3L})_{\mu \nu \rho }&=& 
Tr(\o _{[\mu } R_{\nu \rho ]} -{1\over 3} 
\o_{[\mu  } \o_{\nu ]} \o_{\rho ]}) , \ 
\dh_{[\mu } B_{\nu \rho ]} = \dh_\mu B_{\nu \rho }
- \dh_\nu B_{\mu \rho }- \dh_\rho B_{\nu \mu }.
\label{eq:form}
\end{eqnarray}
While the structure of $H$ in the ten-dimensional case  is motivated by 
considerations of supergravity and anomalies cancellation, 
the analogous structure for the four-dimensional case rather relies 
on  the fact that the string S-matrix elements for the 3-point functions, 
$BGG, BAA, \cdots $,  are insensitive to the internal 
space sector. Moreover, since 
the vertex $B A A$ 
is not renormalized by string  loop 
effects \cite{mayr}, no internal 
renormalization constant is needed 
in the definition of  $H$. 

The familiar unification relations
\cite{ginsparg} for the tree level 
coupling constants of the
(Einstein-Hilbert) gravitational and  (Yang-Mills) 
gauge interactions,
$${2\kappa^2\over \a '}={16\pi G_N\over \a '} 
=g^2_{st}= {g_X^2\over 2}={k_a g_a^2\over 2} ,$$
have been explicitly incorporated in (\ref{eq:effa}). 
The renormalization constant $Z_{F^a}$ in (\ref{eq:effa}) 
identifies then  with the one-loop 
contribution to the inverse squared gauge coupling constant ${1\over g_a^2}$.
For a  short-cut  derivation of
the above tree level relations, one can  apply a
dimensional reduction argument starting with  the 
ten-dimensional heterotic string effective action \cite{gross,tseytlin}. 
This is a valid procedure in the heterotic string  for
the  gravitational interactions and for those 
gauge  interactions which arise from 
the gauge space 
(as opposed to the internal space) sector. The internal space  
coordinates contribute then through a volume factor which can be absorbed
by transforming  the ten-dimensional into the four-dimensional dilaton field.

We shall also need information about the 
higher-derivative gauge and gravitational
interactions at tree level.  Unfortunately, 
no systematic studies seem to exist 
for the  gravitational coupling constants of naive dimension-4, 
$t_{1,2,3}$, and even less for 
the gauge coupling constants  of dimension-6, $\  r, $ or 
dimension-8, $\  s_{1,2}$.  
The cubic gauge  interaction term  $F^{a3} $ in eq.(\ref{eq:effa})  has 
been included for completeness purposes only, since its 
projection onto the Cartan subalgebra vanishes
by virtue of the antisymmetry with respect to the space-time indices.
We have considered only the subset of 
higher-derivatives operators with a 
maximal numbers of field strength factors,
and disregarded the independent  interactions involving 
covariant derivatives, 
such as the naive dimension-6 interactions,  $ D_\mu F^{a \mu \nu } 
D^\rho F^a_{\rho \nu } \cdots  $, which can be expressed in terms of 
quartic order  fermionic couplings by use of the equations of motion.  
We have also omitted writing a large number  of dimension-4 
generalized gravitational interactions, involving the 2-form 
and dilaton fields, given schematically as \cite{gross}, 
$\d S_{EFF}=\int \sqrt G [(D H)^2+ RHH+ R (D\Phi )^2 +
(D\Phi )^2(D^2\Phi ) +H (DH)(D\Phi )+\cdots  ].$

At this point we should recall  that a subset of the 
coupling constants in $S_{EFF}$ are 
ambiguous due to 
the freedom of 
fields redefinition.
These inherent  limitations of 
the first-quantized on-shell formalism of string theory
afflict the description of the four-dimensional effective action in the same 
manner as they do in the  ten-dimensional case,
at tree level  \cite{gross,tseytlin}   as well as  at
one-loop  level \cite{forger}. 
Thus, the metric tensor redefinitions, $\d G_{\mu \nu } =\a ' (b_1 R_{\mu \nu }
+b_3 R G_{\mu \nu } +\cdots ) ,$ with field independent constant  coefficients, 
$b_i$, leave the structure of 
$S_{EFF}$ unaffected, except for the following 
shifts in  the quadratic interactions: 
$\d t_2= -b_1, \ \d t_3= \ud [b_1 +(D-2)b_3]$. More generally, the 
consideration of both metric and dilaton fields redefinitions is known 
to leave one with only two so-called essential gravitational 
constants  at 
order $\a '$ \cite{tseytlin}, 
$S_{EFF}\sim  \int \sqrt {G} e^{-2\Phi } [(R_{\mu \nu \rho \sigma })^2 +\rho_1 (\dh_\mu \Phi )^4]$.
For the gravitational interactions, a convenient physically motivated choice 
is to set  the  values of the two, so-called a priori  ambiguous
coupling constants,   $t_2, \ t_3$ in eq.(\ref{eq:effa}), 
in such a way that  the tree level 
quadratic gravitation interactions become 
proportional to the so-called  Gauss-Bonnet, ($GB$),   topological term, 
\begin{eqnarray}
(GB)&=& (R_{\mu \nu \rho \s })^2
-4(R_{\mu \nu })^2 +R^2 
= (C_{\mu \nu \rho \s })^2
-2(R_{\mu \nu })^2 +{2\over 3} R^2, \nonumber \\ 
\bigg [ \chi (M)&=& {1\over 32\pi^2 }
\int_Md^4X \sqrt {G} (GB)={1\over 16\pi^2} 
\int_M  Tr (R\wedge R^d  )\bigg ]
\label{eq:gb}
\end{eqnarray}
where $ C_{\mu \nu \rho \s }$  is the conformal invariant
Weyl curvature   tensor and  $R^d _{ab} = \ud \e_{abcd} R_{cd} $
is the dual curvature 2-form. 
The formula in the second line summarizes 
the Gauss-Bonnet theorem, with  $\chi (M) $ denoting the
Euler characteristic 
of the four-dimensional manifold $M$.
In order to fix now the absolute size of the 
quadratic gravitation terms,   one can apply 
a dimensional  reduction   procedure
starting with  the known 
results for the  tree level ten-dimensional 
action \cite{gross}. This yields us  the results, 
$ t_1={1\over 8},t_2=-\ud , t_3={1\over 8}.$

The local and global supersymmetry transformations in 
$S_{EFF}$ provide useful information  on certain additional 
bosonic interactions. Thus, the necessity of  quadratic 
gravitational interactions  arises from the fact that these 
are related by ten-dimensional supergravity to the gravitational 
Chern-Simons term  in $H$. More directly, the four-dimensional supersymmetry 
constraints on the quadratic derivative order interactions, impose 
the following supersymmetry completion in the tree level action,
\begin{eqnarray}
\d S_{EFF}= \int d^4 X \sqrt {G} 
{1\over 4}\bigg [R(S) [ (F_{\mu \nu }^a)^2+(GB)] 
+I(S)[F^a_{\mu \nu } \tilde {F}^{a\mu \nu  } +  R_{\mu \nu \rho \sigma } 
R^{d \mu \nu \rho \sigma }]\bigg ] ,
\end{eqnarray}
where $S(X)=\ud [ e^{-2\Phi (X)}+i{a(X)\over 8\pi ^2} ]  $
is the dilaton chiral superfield combining the dilaton with the real scalar 
field dual to the 2-form,  $dB =\star da$,
such that $<a>$ is the $\theta $-vacuum angle. 
The one-loop contributions
are strongly constrained by consideration of supersymmetry in combination with 
the duality  symmetries \cite{ibalu,cardoso1}.
Of course, a  fixed $(GB)$ interaction 
at tree level does not imply that the same combination
should also occur in the one-loop interaction, 
excluding there   the so-called naked 
$(R_{\mu \nu \rho \sigma })^2$  or
$(C_{\mu \nu \rho \sigma } )^2$ terms.

An  analogous situation  arises with the 
cubic and quartic order gauge interactions.
The terms appearing in eq.(\ref{eq:effa}) comprise 
the set of independent space-time structures, consistent
with the use of equations of motion and neglect of fermionic terms. 
However, the decomposition with respect to the 
gauge symmetry group dependence 
in eq.(\ref{eq:effa}), where we 
have ignored the charged generators and the cross terms between the 
Cartan subalgebra generators, will bring
more independent terms   depending on the gauge group.
Also, the dimensional reduction prescription 
lacks generality  here
since compactification partially  breaks the ten-dimensional
gauge symmetry. 
Nevertheless, for orientation purposes, let us 
rewrite in our present notations 
the gauge interactions for the ten-dimensional, $E_8\times E_8$ heterotic 
string theory, as given in  
\cite{gross},
\begin{eqnarray}
L_{10} & =& \sqrt {G_{10}} e^{-2\Phi_{10} }
\bigg [ \sum_a -{1\over 8}tr( F^a F^a) \nonumber \\
&+& { {\a '}^2\over 2^{11}} 
\sum_{a , b} \bigg ( 8tr( F^a F^a F^bF^b) + 4 tr(F^a F^b F^aF^b) \nonumber \\
&-&tr(F^aF^a)tr(F^bF^b) -2tr(F^aF^b)tr(F^aF^b) \bigg ) \bigg ],
\label{eq:het10}
\end{eqnarray}
where the group indices $a ,b $ run  over all the group (charged and 
uncharged) generators and we use an operator notation where 
the trace symbol refers  to a
sum over  the space-time indices, 
$tr(F^a F^b)= F^a_{\mu \nu } F^b_{\nu \mu }, \ \cdots $
Identifying the part of the 
dimensionally reduced interaction diagonal in 
the Cartan subalgebra generators with 
eq.(\ref{eq:effa}), gives us the following qualitative
estimates for the  four-dimensional tree level 
coupling constants: 
$s_1= -{s_2\over 4} = {3\over 2^7} ={3\over 128}$.

The dependence on the 
three independent constants, 
$Z_{iR^2} t_i$, could possibly  be resolved 
by considering background fields depending on two deformation parameters 
in addition to $R^2$. There is no guarantee, however, that  such a
procedure would  be successfull,  due to
the  field redefinition ambiguities. 
We are led, at a preliminary  stage, 
to restrict consideration  to the specific, but unknown,   linear 
combinations of the higher-derivative  interactions 
which are singled out by the  structure of the string theory 
background fields. Before discussing this 
point, we need to express the effective action $S_{EFF}$ in terms of the 
parameters $F^a, R$. For this purpose, 
we  substitute the solutions (\ref{eq:bgf})   for the 
gravitational  and gauge fields backgrounds  in the 
effective action (\ref{eq:effa}), 
perform the integration over the 
space-time manifold by using, 
\begin{eqnarray}
 \int d^4 X \sqrt G &=&
{k^{3\over 2} \over 4} (1-2H^2)^\ud \int dX^0 \int_0^{2\pi } d\a 
\int_0^{2\pi } d \g  \int_0^{\pi } d\b {\sin \b \over \d } , \nonumber \\
&=&V^{(4)} (1-2H^2)^\ud \int_{-1}^{1} {d(\cos \b ) \over 
\l +{1\over \l } + (\l -{1\over \l })
\cos \b }, \nonumber \\  
 \bigg [ {V^{(4)}\over \int dX^0} &=& {1\over \a'^2} \int d^4 X \sqrt G_0 =
\pi^2 k^{3\over 2} =(2\pi )^3 V(\mu_e)\bigg ]
\label{eq:vol}
\end{eqnarray}
and expand the integrals 
in powers of $R,F^a$. To perform these tasks we have used 
the symbolic calculations ``mathematica" software  package.  The leading terms 
in the power expansion of the one-loop part of the  action read:
\begin{eqnarray}
S_{EFF}^{1-loop} & =& V^{(4)}   \bigg [
Z_\Lambda ( 1-{F^{a2}\over 8}  -{R^2\over 24} ) +
Z_R {4\over k} ({3\over 2} -{F^{a2}\over 16} +{3R^2\over 16} )
+Z_\Phi {4\over k}  (1 -{F^{a2}\over 8} 
+{R^2\over 8} ) \nonumber  \\
&-& Z_H{4\over k} (8+F^{a2} +R^2) 
-Z_{F^a} {1\over kk_a}(F^{a2} ) \nonumber \\ 
&+&{4\over k^2} \bigg ( Z_{3R^2} t_3(9+{3F^{a2}\over 8}+{263R^2\over 24})
+ Z_{2R^2} t_2(3
+{101R^2\over 24}+O(F^{a2})) \nonumber \\
&+& Z_{1R^2} t_1(3
+{141R^2\over 24} +O(F^{a2})\bigg ) 
-{1 \over k^2k_a^2}\bigg (s_1Z_{1F_a^4}+{ s_2\over 2} Z_{2F_a^4}\bigg )
F^{a4} +\cdots \bigg ].
\label{eq:effa1}
\end{eqnarray}
The cubic gauge interactions $F^{a3}$ cancel out, as already pointed out.
The renormalization constants associated to the 
interactions of increasing derivative order are accompanied by
increasing powers
of ${1\over k}$, as follows simply  from the fact  that 
${k \over \a '}$ is the sigma-model loop expansion 
parameter. 
We have   only displayed 
in eq.(\ref{eq:effa1}) the leading powers of ${1\over k}$ for each interaction.
The omitted dimension-4 interactions, $(DH)^2, \cdots $,
enter at $O({1\over k^2})$. 
Higher powers in ${1\over k} $ should also  be  present, since 
the  background fields in  eq.(\ref{eq:bgf}) are solutions  of the 
tree level action $S_{EFF}$  truncated to  the  dimension-$4$ interactions,
or equivalently,  prior to the consideration of sigma-model  loop  
contributions. Accounting for these effects will induce for 
each of the interactions,
correction factors  of the form $ [1 +{1\over k} (a_1 +a_2 R^2 
+a_3 F^{a2} +\cdots ) + O({1\over k^2})] $, which multiply 
the renormalization constants, 
$ Z_\Lambda , Z_R , \cdots $. 
The need for these subleading terms in ${1\over k}$  will appear explicitly
in the following.

While the coefficients of $F^{am} R^n$ in $S_{EFF}$ stand for 
1PI amplitudes with respect to the massless modes, the corresponding 
coefficients in $Z$ stand for the full amplitudes, including massless 
and massive modes. On can match the two expansions,
eqs.(\ref{eq:effa1}) and  (\ref{eq:string}), 
only after adding 
to $S_{EFF}$ (or subtracting from $Z$) the one-loop massless modes 
contributions, which we shall denote, 
$z(F^a, R)= \sum_{(m,n)\in Z^2}  z_{m,n} F^{am} R^n$.
The leading order  constant term yields the functional 
equality, as a function of $k$, 
\begin{equation}
 V^{(4)} \bigg [ Z_\L + {4\over k} ({3\over 2}Z_R 
 +Z_\Phi -8Z_H  +x_1Z_\L ) 
+{12\over k^2} (3 Z_{3R^2} t_3 
+Z_{2R^2} t_2 +Z_{1R^2} t_1 +x_2 Z_\L +y_1) +\cdots  \bigg ]
= Z_{0,0}-z_{0,0}, 
\label{eq:cosmz}
\end{equation}
where the coefficients   $x_{1,2}$ are associated with 
the omitted higher-derivatives interactions  and $y_1$ with
the uncalculated sigma-model loop corrections.
The dependence on $k $ in the string theory functions 
$Z_{n,m}$  appear explicitly through the 
powers of $ k$ and $k+2$ and implicitly through the zero-mode operators 
and the partition function factor $Z_W$.
The  identification for the quadratic $F^{a2}$ term, 
retaining the leading order at  $k\to \infty $,
involves a specific linear   combination 
of  the gauge and gravitational fields one-loop renormalization constants, 
denoted by $g'_{a}$:
\begin{eqnarray}
{(4\pi )^2 \over g_a^{'2 }}& \equiv  &(4\pi )^2[{1 \over g_a^2 } +{k_a\over 4} 
(2Z_\Phi +Z_R +16Z_H)+{kk_a \over 8} Z_\L ] ={1\over V^{(4)} } 
(Z_{2,0}-z_{2,0} ), \nonumber   \\
& =&-{4k\over  (k+2)} 
 \int_F {d^2\tau \over \tau_2} Z_W 
  {\sum } ' { Z_0 Z_G \over \eta^2\bar \eta^2}
\bigg [ {i\over \pi } \bigg (  \dh_{\bar \tau  }\bigg (\ln 
{\bar \vt [{\bar \a \atop \bar \b }]\over \bar \eta }\bigg )  +{k+2\over 6} 
\dh_{\bar \tau } \bigg )
\bigg (J^{a2}-{k_a  \over  8\pi \tau_2} \bigg ) \nonumber \\
&-&{k_a(k+2)\over (8\pi \tau_2)^2 } \bigg ] -{z_{2,0} \over V^{(4)} }.
\label{eq:fa2}
\end{eqnarray}
We have used an explicit representation of the zero-mode operators 
discussed in Appendix \ref{sec:appb}. The action of $(\bar Q +\bar I_3)^2$ 
gives rise to the terms involving the derivative $\dh_{\bar \tau } $.
The free derivative operator, in eq.(\ref{eq:fa2}), 
is understood to act only on the factor $X'(\mu )$.
The discussion greatly  simplifies
in the supersymmetric  case, since
the only  non-vanishing contributions there are those  arising  from the
$ \bar Q^2 $  operator insertions.
The functional relations,  $Z_{0,0}=0, \ z_{0,0}=0$,
in eq.(\ref{eq:cosmz}), imply  then, 
$Z_\L =0$, corresponding to a vanishing one-loop cosmological 
constant. 
If one accepts the fact that $\bar Q^2$ on the right-hand side
of eq.(\ref{eq:fa2}) contributes
to $Z_{F^a}$ only, then
one easily infers  from the $O(k^0)$ term  the equality,
$  (2Z_\Phi + Z_R +16Z_H )=0.$ 
Since a  derivation based on the S-matrix approach
for the heterotic string,
provides  us  with the equalities \cite{kkpr1}:
$Z_R=0 $ (non-renormalized Newton constant) and 
$  Z_H=0$,   one deduces:  $Z_\Phi =0$, so that  the absence of 
wave function renormalization for the  2-form field entails its absence for 
the dilaton field.
These relations are consistent with the matching of the 
$O({1\over k} )$ terms in eq.(\ref{eq:cosmz}).
It also  follows that, for supersymmetric vacua,  the right-hand side of 
eq.(\ref{eq:fa2}) gives us the entire   one-loop 
corrections to the gauge coupling constants, $g_a^{-2}$. 
We shall continue using the primed  coupling constant notation
to remind ourselves of the general case.

The quadratic order gravitational  constants appear first in the 
quadratic order,  $R^2$ terms,
where they are  mixed with  the  renormalization 
constants of the gravitational multiplet, $Z_\Phi , ...$ 
It would be desirable, of course, to be able to separate the various
constants here.  The consideration of higher-order or mixed terms 
$O(F^{a2}R^2 )$ or $ O(R^4)$, not provided 
in eq.(\ref{eq:effa1}), could possibly give us other independent linear 
combinations. However,  these relations would involve still 
higher-order interactions.  
Resolving the dependence on the  three independent coupling 
constants,  $Z_{iR^2} t_i  $,
is probably beyond the possibilities of the present formalism, because of 
the fields redefinition ambiguities. 
Also  for the quartic gauge interactions, 
resolving the dependence on the  two coupling constants 
$s_i$ and unfolding their group theory 
substructure, raises technical complications  beyond the  scope of this work.
As stated above, 
we shall  restrict ourselves in this work  to the specific  linear combinations
arising from the  string theory perturbations, $R$ and $F^a$.
This leads us to introduce two effective  quadratic 
gravitational and quartic  gauge coupling constants, $g_{R^2}$ 
and $g_{F^4_a}$, defined as linear combinations 
of the independent coupling constants  by the equality:
\begin{eqnarray}
S_{EFF}^{1-loop}&=& {4V^{(4)} \over 24k^2}
\bigg (263 Z_{3R^2} t_3+
101 Z_{2R^2} t_2+141 Z_{1R^2} t_1\bigg ) R^2
-{V^{(4)} \over k^2k_a^2}\bigg (s_1Z_{1F_a^4}+{ s_2\over 2} Z_{2F_a^4}\bigg )
F^{a4}, \nonumber  \\
&=& \bigg [V^{(4)} {2 R^2\over k^2} {1\over g_{R^2}^2}
-{4V^{(4)} F^{a4} \over k^2k_a^2} {1\over  g^2_{F_a^4}}
\bigg ]_{1-loop}.
\label{eq:gepe}
\end{eqnarray}
The tree level  value of $g_{R^2}$ may be obtained indirectly  through 
a comparison with the results  from the S-matrix approach 
\cite{anton92,cardoso1,harvey1}.  We find:
$g_{R^2}= g_X$. For the gauge case, the  tree 
level value of $g_{F^4_a}$ is unknown.
If one  could only retain the constant $s_1$, say,  then a comparison with 
eq.(\ref{eq:het10}) would 
give: $ g_{F^{a4}}= g_X$. In the following, we shall parametrize the 
tree level value as, 
$ g_{F^{a4}}^2= g_X^2/s_a$, where $s_a$  are group-dependent
unknown  quantities which are expected, however, to be of order unity.

The  identification between  
eqs.(\ref{eq:string}) and (\ref{eq:effa1}) for  the $R^2$ term 
gives an equation for a linear combination, denoted 
${1\over {g'}_{R^2}^2 }$, 
of the gravitational coupling constant 
and the other renormalization constants, 
\begin{eqnarray}
{(4\pi )^2  \over {g'}_{R^2}^{2} } &\equiv &  (4\pi )^2 [{1\over g_{R^2}^2}
+ {k\over 8} (2Z_\Phi +3Z_R -16Z_H ) -{k^2\over 48}Z_\L ] ={1\over V^{(4)} }
(Z_{0,2}- z_{0,2}) , \nonumber \\
 &=&{2k\over  (k+2)} 
 \int_F {d^2\tau \over \tau_2} Z_W 
  {\sum } ' { Z_0 Z_G \over \eta^2\bar \eta^2}
\bigg [ {i\over \pi } \bigg ( \dh_{\bar \tau  } \bigg (\ln 
{\bar \vt [{\bar \a \atop \bar \b }]\over \bar \eta }\bigg ) +{k+2\over 6} 
\dh_{\bar \tau } \bigg )
  {i\over \pi }\bigg ( 
\tilde E_2 (\tau )-{k+2\over 6}
\dh_\tau   \bigg  ) \nonumber \\  &-& 
{k(k +2)\over (8\pi \tau_2)^2 } \bigg ] - {z_{0,2}\over V^{(4)} } . 
\label{eq:part4}
\end{eqnarray}
The formulas in eqs.(\ref{eq:fa2}, \ref{eq:part4}) 
essentially  agree with
the results reported by \cite{kkpr1}, up to  a few 
minor  modifications due to different conventions. 
The derivatives $\dh_\tau , \bar \dh_{\bar \tau }$  in eq.(\ref{eq:part4})
do not act on any of the factors other than the 
space-time partition function $X'(\mu )$. In the limit $\mu \to 0$, their 
action is simply given by
 $\dh_{\bar \tau } \ln  X'(\mu ) \simeq  
-{1 \over 4i\tau_2 },$ with $\dh_\tau $ obtained  by  complex conjugation. 
For supersymmetric vacua, the 
vanishing of the constants $Z_{\L , \Phi , R, H}$, 
give us in principle the equality,  $g'_{R^2}= g_{R^2}$.
This would seem to imply that no terms 
of $O(k)$ can be present on the right-hand side of eq.(\ref{eq:part4}), 
in contradiction to what is actually found  by a careful
analysis of the $k$ dependence (see eq.(\ref{eq:gravit1}) below). 
The reason for the mismatch in eq.(\ref{eq:part4}) is  
 due to our omission of the higher-derivative interactions 
$R H^2, (D H)^2, \cdots $ in constructing  $S_{EFF}$. 

The identification of eq.(\ref{eq:string})
for the string theory  quartic order $F^{a4}$ term,
with the corresponding field theory term in eq.(\ref{eq:effa1}),
gives us an equation for 
a linear combination of the quartic 
gauge interactions  coupling constant $g^{-2}_{F^{a4} }$ and 
the renormalization constants of the lower-order interactions.
Denoting the uncalculated coefficients as,  $x_\L , x_\Phi , 
\cdots $,  we have: 
\begin{eqnarray}
{(4\pi )^2\over  {g'}^2_{F^{4a}}}  &\equiv & 
(4\pi )^2\bigg [ {1\over  {g}^2_{F^{4a}}} 
+k^2k_a^2 x_\L Z_\L  
+kk_a^2 (x_R Z_R +
x_\Phi Z_\Phi +x_H Z_H+x'_\L Z_\L ) 
\nonumber \\ &+&kk_a  x_F 
(Z_{F^a}+{x''}_\L Z_\L ) +k_a^2(
\sum_{i=1}^3x_{iR^2} Z_{iR^2}+y'_1)
\bigg ] ={1\over V^{(4)} } (Z_{4,0}-z_{4,0}) ,  \nonumber  \\
&=&-{\pi^2 k_a^2\over 3} 
\int_F 
{d^2\tau \over \tau_2} \tau_2^2
{Z_W\over \vert \eta \vert^4} {\sum } ' Z_0Z_G
\bigg [ \zeta \  \bar Q'^4 \bigg ({1\over 4}  (Q^a)^4
-{3k_a\over 2\pi \tau_2}Q_a^2
+{3\over 4\pi^2 \tau_2^2} \bigg ) \nonumber \\
&+&  \xi \  \bar Q'^2 \bigg ({3  Q_a^4\over 8\pi \tau_2}
-{3 \over 8\pi ^2 \tau_2^2}Q_a^2
-{3\over 4\pi ^3  \tau_2^3} \bigg )
+k^2 \bigg ( {3\over 256 \pi ^2 \tau_2^2 } Q_a^4 
+{3\over 64 \pi ^3 \tau_2^3} Q_a^2 \bigg ) \bigg ] -{z_{4,0} \over V^{(4)} }
, \nonumber \\
&=&-{\pi^2\over 3} \int_F 
d^2\tau \tau_2
{Z_W\over \vert \eta \vert^4} {\sum } ' Z_0Z_G 
\bigg [16\bar Q^{'2} J_a^4 (\bar Q^{'2}+{3\over 2\pi \tau_2})
-{12\over \pi \tau_2}J^{a2} \bar Q^{'2} (\bar Q^{'2}+{1\over 4\pi \tau_2}) \nonumber \\
&+&{3\over 4\pi^2 \tau_2^2} \bar Q^{'2} (\bar Q^{'2} -{1\over \pi \tau_2}) 
+O(k)
\bigg ] -{z_{4,0}\over V^{(4)} } ,
\label{eq:jauge4}
\end{eqnarray}
where we use the convenient abbreviations, 
$\bar Q' = \bar Q +\bar I_3$ and $Q_a^2 ={8\over k_a} J_a^2$, 
such that $Tr(Q_a^2)=2$.
In the linear combination, denoted by 
$g'_{F^{4a}}$,  the unspecified coefficients of type, 
$x_\L , x_R, \cdots $, correspond to the easily  calculable higher-order 
terms in the expansion of $e^{-4\pi \tau_2 L_0}$. The coefficients of 
type, $x'_\L , y'_1, \cdots $, arise  from the omitted
higher-derivatives interactions and from the sigma-model loop corrections  
to the background fields which we have not calculated.
It is advisable to refrain from expanding the right-hand side in powers of 
$k$ until one exhibits the $k$-dependence of the modular integrals.  
This is the reason for introducing the auxiliary quantities, 
$\zeta = ({k\over k+2})^2 = 1-{4\over k} +O({1\over k^2}), \quad 
\xi = -{k^2\over 2(k+2)}= 
-{k\over 2}+1 -{2\over k} +O({1\over k^2})$.
The last equation in (\ref{eq:jauge4}) provides a 
useful simplified formula for the loop correction to the $F^{a4}$ interaction, 
where  we retained   the $O(k^0)$ term only, while  dropping
the $O(k) $  and  $O({1\over k^n})$ terms. 
\subsection{\bf Renormalized effective action}
\subsubsection{\bf Gauge interactions  coupling constants}

We proceed now to the  final stage  of the discussion, which consists
in identifying the string vacuum functional integral $Z$ with
the effective action $S_{EFF}$, after subtracting from $Z$ the one-loop
field theory contributions induced by $S_{EFF}$.
For this purpose,  we shall need to  
expose, on one hand, 
the  string theory ``massless modes" contributions  and, on the other hand, 
the field  theory high energy modes contributions.
We use quotes  here to remind ourselves that,
because of the finite 
mass gap, set by $\mu $,
the finite curvature theory has 
no massless modes as such, but instead 
a tower of 
momentum and winding  massive  modes whose masses are sent
to zero at the four-dimensional  decompactification limit.  
The contributions of the  would-be massless modes are 
isolated by taking 
the limit $q\to 0$, inside  the 
integrand of eq.(\ref{eq:fa2}), individually for the different 
terms associated with the power  factors ${1\over \tau_2^n}$, 
for all the factors 
except $Z_W$. 
As long as one works with a  finite infrared cut-off $\mu $, 
the  limit $q\to 0$  can be taken safely.
The various terms in  the partition function reduce, at  
$ q\to 0$,  to appropriate 
supertrace (fermion parity $(-1)^F$    weighted) sums
over the would-be massless modes.
The connection  formulas can be obtained by using 
the results describing the  action of 
the zero-modes operators on the  states or on
the  determinantal  factors, which are detailed in  
 Appendix \ref{sec:appb}.  The $q\to 0$ limit for the angular momentum 
 projection operator is given by:
\begin{equation}
(\bar Q +\bar I_3)^2 \to 
2\bar q {d\over d\bar 
q} \ln {\bar \vt [{\bar \a \atop \bar \b }]\over \bar \eta } =
-{1\over 12}  
+2\sum_{n=1}^\infty {n\bar q^n\over  1-\bar q^n} 
+{i\over \pi} {\dh_{\bar \tau } \bar  \vt [{\bar \a \atop \bar \b } ] \over 
\bar \vt [{\bar \a \atop \bar \b } ] }
 =  (-1)^F (-{1\over 12} +\chi^2)  +O(\bar q),
\label{eq:nul}
\end{equation}
where $\chi $  denotes the space-time helicity.
The  dependence  on $q , \ \bar q $ which appears in the  low $q, \ 
\bar q $ expansions,  may be written schematically as,
\begin{eqnarray}
{\sum }' {Z_0 Z_G\over \vert \eta \vert ^4} (\bar Q +\bar I_3)^2
\pmatrix {J_a^2 \cr   1 \cr E_2(\tau )}
& \to &  (-1)^F ({1\over 12} -\chi^2) \pmatrix {J_a^2 \cr 1  \cr 
1-24q+O(q^2) } \nonumber \\ &\times &
\bigg ({l_{-1}\over q} +l_0 +O(q^{1\over N})\bigg )
\bigg (r_0 +O(\bar q^{1\over N})\bigg ),
\label{eq:nul1}
\end{eqnarray}
where the power ${1\over N}$ of the next-to-leading  order terms is 
determined for orbifold models by the order of the orbifold point 
symmetry group.  The string theory trace sums include the contributions
from the physical, on-shell  modes, $<L_0-\bar L_0>=0$, as 
well as from the non  level-matched modes, $<L_0-\bar L_0>\in Z$. 
The modular invariance constraints are essential to  ensure
the convergence in the projected sums  of the modular 
integrals at $\tau_2 \to \infty $. While the massless modes contributions 
to the $J_a^2 $ and $1$ operators involves then 
the product, $l_0r_0$, those of 
$E_2(\tau )$ enter with the combination, $l_0r_0-24l_{-1}r_0$. 

Since we  keep the infrared cut-off fixed, 
the   low energy theory  must  be defined accordingly  
as a finite curvature field theory with respect to 
the  same set of  space-time background fields as in the string theory.
Fortunately, there
is no need to redo a new
calculation for this case,  since  the result
can be obtained by applying  
the $\alpha ' \to 0$ limit and 
using the familiar correspondence formulas  between the 
string theory  modular integral and 
the field theory heat kernel (Schwinger proper-time)
representations,
\begin{equation}
\mu^2 ={1\over k+2}\to \mu_e^2={1\over k} 
,\quad  \tau =\tau_1 +i\tau_2 \to i \tau_2= i{t_H\over \pi \a' },\quad 
\int_F {d^2\tau \over \tau_2}\to  
\int_{1 \over \L^2}^\infty  {dt_H \over t_H}=
\int_{1\over \pi \a' \L^2}^\infty
{d\tau_2\over \tau_2} ,
\label{eq:ft}
\end{equation}
based on the identification, 
$e^{-t_Hp^2} \sim e^{-\pi \a' \tau_2 p^2}$.
The field theory truncated   space-time partition function  factor
is obtained by removing  the  winding modes terms in the sum representation,
which corresponds  to perform the  substitutions, 
\begin{eqnarray}
Z_T(\mu ) \to   \hat Z_T(\mu_e) & =&
\sqrt {\tau_2} \sum_{m\in Z} e^{-\pi \tau_2 \mu_e^2 m^2} 
=\sqrt {\tau_2} \vartheta_3 (i \tau_2 \mu_e^2) ={1\over \mu_e } 
\vt_3({1\over -i \tau_2 \mu_e^2}),  \nonumber \\
X(\mu )\to \hat  X(\mu_e)&=&\sqrt {\tau_2} 
[\vt_3(i\tau_2 \mu_e^2)- 
\vt_3 (4i\tau_2 \mu_e^2)] .
\label{eq:ft1}
\end{eqnarray}
The ultraviolet finiteness of closed strings,
which follows   from the 
restriction to a fundamental
domain of the modular group,  so that 
$t_H=\pi  \a ' \tau_2 \ge  {\sqrt 3\over 2} 
\pi \a ' $, 
indicates that the parameter  $ {1\over \a '}$
actually plays the r\^ole of a  string theory ultraviolet cut-off. 
In order to separate out  the divergent contributions arising from  the 
effective field theory high energy modes,    this 
must be equiped  with an ultraviolet cut-off, which  will 
be represented by a 
dimensional mass parameter,   $\L $.
A  convenient ultraviolet  regularization within 
the heat kernel formalism is 
by imposing a
lower bound on the  integrals, in the manner exhibited  in 
eq.(\ref{eq:ft}).
The logarithmic and power divergences at  $\L \to \infty  $ will 
appear in close correspondence with the string
theory divergences in the infrared cut-off  at $\mu  \to 0$.
As required by naive dimensional analysis,  the cut-off dependence
must involve  the product 
$\Lambda^2 \alpha ' \to 0$.
Since the limit $\a' \to 0$ must 
precede the $\L \to \infty $
limit, by the very definition of the string theory
effective action,   there is no need to worrying about the  positive 
power divergences in $\L ^2\a '$, which will 
simply cancel away in the limit $\a '\to 0$. 
Special care is needed for the logarithmic dependence on cut-off. 
This  must be absorbed inside the bare coupling constants 
in the process of  defining
the renormalized, cut-off independent  coupling constant. 
A  convenient supersymmetry preserving  renormalization 
is  the so-called 
modified dimensional reduction $(\overline {DR})$ prescription \cite{siegel}.
This  is defined by performing an analytic continuation in the 
space-time  dimension, $D=4-\e $, only for the integration measure,
while evaluating  all algebraic expressions
at  $D=4$. Once the $\overline {DR} $ renormalized scheme 
constant is  defined, the 
conversion to other schemes is straightforward.

For the  gauge interactions case, 
the  relationship  between the $\overline {DR} $ renormalized 
coupling constant, denoted $g_a (p)$, 
and the bare (or unrenormalized)  field theory 
coupling constant $g_a =g_a(\L )$, using, for convenience,  
a Gaussian factor cut-off,  in place of the sharp cut-off in eq.(\ref{eq:ft}),
is described at one-loop order by the formula: 
\begin{eqnarray}
 {(4\pi )^2\over g_{a}^{2} (\L )}-
 {(4\pi )^2\over g_{a}^{2}(p)}& =& b_a \bigg [ 
 \int_0^\infty dt_H {e^{-{t_Hp^2\over \L ^2}}
 \over t_H^{1-\epsilon  }} -{1\over \e }\bigg ]  \nonumber \\
 &=&b_a [({p^2\over \L^2})^{-\e } \G (\e ) -{1\over \e } ]
 = b_a[-\g_E-\ln {p^2\over \L^2}] 
=b_a [-\g_E +\int_{1\over \L^2}^{1\over p^2} 
{dt_H\over t_H}], 
\label{eq:ft3}
\end{eqnarray}
where $b_a$ identifies with the beta-function slope parameter, 
$\b_a (g)= {\dh g_a \over \dh \ln p }= -b_a{ g_a^3\over (4\pi )^2}+\cdots $.
If so required, momentum dependent coupling constants could 
also be introduced for the other low-order interactions, $R, (D\Phi )^2, 
(H)^2$, in  an analogous way. However, 
since no logarithmic divergences will  arise
from these interactions, there is no need in considering the  renormalized 
coupling constants associated with $Z_{\L ,\Phi , R , H}$. 
In order to account for the general case, including the non-supersymmetric
solutions where these  renormalization constants are non vanishing, 
we shall consider the primed coupling constants, $ g'_a,\ 
g'_{R^2}, \  g'_{F^{a4}} $.
It will also prove
unnecessary to consider the tree level values of the $ R, (D \Phi )^2,
(H)^2 $ interactions 
coupling constants since these 
will cancel between the left-hand  and right-hand sides.

Recapitulating our procedure, we consider 
for the low order interactions
coupling constants in $ Z $ and $S_{EFF}$ the sum of 
tree  and one-loop level contributions to the $F^{a2}$ term, rewrite  
the string theory one-loop contributions as, $ Z_{2,0}= \sum_{i=1}^3 I_i= 
\sum_{i=1}^3 (I_i-I_i^0) +\sum_{i=1}^3 I_i^0$,
by subtracting  and adding  the massless modes contributions 
(designated below  by the  quantities $I_i^0$ with a suffix $0$) 
and rewrite 
the field theory  one-loop contributions, denoted $z_{2,0}=\sum_{i=1}^3 L_i^0$, 
after trading the bare coupling constant  for
the $\overline {DR}$ renormalized  coupling constant. 
Equating the total tree and one-loop 
string and field theories unrenormalized coupling  constants as, 
$${(4\pi )^2k_a \over g_X^2 }+\sum_{i=1}^3 I_i= 
{(4\pi )^2k_a \over {g'}_{a } ^2 (\L ) }+\sum_{i=1}^3 L_i^0,$$
the matching equation, with string and field  theory terms placed  on the
left-hand and right-hand sides,  respectively,
is given by:
\begin{eqnarray}
&&{(4\pi )^2k_a \over g_X^2 }+\sum_{i=1}^3 (I_i-I_i^0)+\sum_{i=1}^3 I_i^0=
 {(4\pi )^2\over g_{a }^{'2} (p ) }
 +b_a(-\g_E +2\ln {\L \over p})+\sum_{i=1}^3 L_i^0,
\label{eq:match} \\
 I_1&= &{i\over \pi^2 V(\mu ) } {2k\over (k+2)}
 \int_F {d^2\tau \over \tau_2}
 {X'(\mu ) \over \eta^2\bar \eta^2} 
{\dh \over \dh \bar \tau } \ln 
\bigg [{\bar \vt [{\bar \a \atop \bar \b }]
\over \bar \eta }\bigg ] Z_0  (q, \bar q) 
[J^{a2}-{k_a  \over  8\pi \tau_2} ] Z_G (q)  , \nonumber \\
 I_2&=&{ik\over 3  \pi^2 V(\mu ) } \int_F 
 {d^2 \tau \over \tau_2} {\dh_{\bar \tau } 
 X'(\mu ) \over \eta^2\bar \eta^2} 
Z_0  (q, \bar q) 
[J^{a2} -{k_a \over  8\pi \tau_2} ] Z_G (q), \nonumber \\
 I_3&=&- {k k_a \over 32  \pi^3 V(\mu )}
 \int_F {d^2\tau \over \tau_2^3} 
 {X'(\mu ) \over \eta^2\bar \eta^2} 
Z_0  (q, \bar q) 
Z_G (q).
\label{eq:integrales}
\end{eqnarray}
No confusion should arise from the fact that 
the definition  ${g'}^{-2}_a(p) =g_a^{-2} (p) 
+{k_a\over 4}(2Z_\Phi +Z_R+16 Z_H) +{kk_a\over 8} Z_\L $,
 uses a mixed notation involving the  $\overline {DR}$ renormalized gauge coupling 
constant along  with the unrenormalized   
$ R, (D\Phi)^2, \cdots $ interactions coupling 
constants. 
The would-be  massless string modes contributions
are obtained by taking the limit $\tau_2\to \infty $, 
\begin{eqnarray}
I_1^0&=&-{2k\over \pi (k+2) }\bigg [ J_1 STr
\bigg (({1\over 12} -\chi^2) J^{a2}\bigg ) 
-{k_a\over 8\pi }
J_2Str({1\over 12 }-\chi^2) I )\bigg ] = {k\over \pi (k+2)} [{J_1\over 2}  b_a
+{k_a J_2 \over 8\pi } h ] 
, \nonumber \\
I_2^0&=&-{k\over 12 \pi^2 } \bigg [ J_2 Str(J^{a2}) -{k_a\over 8\pi }
J_3Str(I)\bigg ] = -{k\over 24\pi^2} [J_2 c_a -{k_a J_3 \over 8\pi } z], \nonumber \\
I_3^0& =& -{kk_a\over 32 \pi^3 } STr(I) J_3 = -{k_a k J_3\over 64\pi^3} z , 
\quad \bigg [J_n(\mu )={1\over V(\mu )}\int_F
{d^2\tau \over \tau_2^n} X'(\mu ) \quad  [n=1,2,3]\bigg ]
\label{eq:ii0}
\end{eqnarray}
where the $(-1)^F$ (space-time 
fermion number signature) weighted supertraces
over massless modes are defined as:
\begin{eqnarray}
 &&b_a= -4
 STr (J_a^2 ({1\over 12}-\chi^2)) =
 -{k_a\over 6} [n_S c(R_S) +2 n_F c(R_F) -11n_V c(R_V)], \nonumber  \\ 
 && h=2 STr\bigg (({1\over 12} -\chi^2) I\bigg ) 
 ={1\over 3 } [n_S +2n_F -11 n_V], \nonumber \\
 c_a&=&2
 STr (J_a^2 I), \quad  z=2STr(I).
\label{eq:traces}
\end{eqnarray}
The symbol  $I $ stands for the notation, $STr (I)= \sum (-1)^F l_0r_0 $,
using eq.(\ref{eq:nul1}). 
The normalization of the gauge charges is such that, 
$Tr(J_a^2) ={k_a\over 2} Tr(\hat J_a^2) = {k_a\over 4} c(R)$, 
with $c(R)$ the Dynkin index of the group $G_a$  representation $R$, 
and $n_{S,F,V}$  denote  the numbers of real scalar, chiral or 
Majorana fermion, vector massless modes. 
Note that $b_a $ is the beta-function slope parameter introduced earlier in
 eq.(\ref{eq:ft3}). 
The proper massive threshold corrections  are  isolated in 
the differences   $ \D'_a \equiv \sum_{i=1}^3\d I_i=I_i-I_i^0$,
which are defined by  the same integrals as the
$I_i $, eq.(\ref{eq:integrales}),    with the 
massless limit part  $q\to 0$ of the integrands  subtracted out. 
These subtracted integrals are infrared finite, so one can safely take the 
limit $\mu \to 0$ and therefore   set $Z_W \to 1$.
The quantities corresponding to $I_i^0$ in the field theory case,
which appear as 
$L_i^0$ in eq.(\ref{eq:match}), are defined as: 
\begin{eqnarray}
L_1^0&=&-{2\over \pi }\bigg  [ K_1 Str\big [({1\over 12} 
-\chi^2) J^{a2}\big ] -{k_a\over 8\pi }
K_2Str\bigg (({1\over 12 }- \chi^2) I \bigg )\bigg ] =
{1\over \pi } [{K_1\over 2} b_a +{k_a 
K_2 \over 8\pi } h], \nonumber \\
L_2^0&=&-{2\over 8 \pi^2 } (1+{1\over 3\mu_e^2}) 
\bigg [K_2 Str(J^{a2}I) -{k_a\over 8\pi }
K_3Str(I) \bigg ] = -{1\over 8\pi^2} (1+{k\over 3}) [K_2 c_a- { k_a K_3 \over 
8\pi } z ],  \nonumber \\
L_3^0&=&-{2(k+2) k_a \over 64 \pi^3 } Str(I) K_3 = -{
(k+2) k_a K_3\over 64\pi^3 } z .  \nonumber \\ 
 && \bigg [K_n(\mu_e )={1\over V(\mu_e )}
\int_{1\over \pi \a' \L^2} ^\infty {dt \over t^n}
{\dh \over \dh \mu_e^2} \big [\sqrt t 
\big ( \vt_3(it\mu_e^2) -\vt_3(4it\mu_e^2)\big ) \bigg ]
\label{eq:li0}
\end{eqnarray}
The divergent dependence at  $k\to \infty $  in the 
formulas given by eqs.(\ref{eq:integrales}), 
(\ref{eq:ii0})  and (\ref{eq:li0}) originate from 
the explicit 
power  factors of $k $ or $(k+2)$  and from  the contributions to the
modular integrals  in the cusp  region, 
$\tau_2\to \infty $.   
The ultraviolet divergences at  $\L \to \infty $ in the 
field theory integrals, eq.(\ref{eq:li0}), arise in close correspondence 
with the infrared divergences of the string and field theories 
modular integrals at $\tau_2, \ t 
\to \infty $. 
The dependence on these cut-off parameters can be 
easily isolated through the simple estimates, 
$$ \int_F {d^2\tau \over \tau_2^n} {X'(\mu )\over V(\mu )} \sim (\mu^2)^{n-1},
\quad  \int_{1/\L ^2}^{1\over \mu_e^2}  {dt \over t^n} 
{\hat X'(\mu_e )\over V(\mu_e )}
\sim \bigg ({\L ^2\over \mu_e^2}\bigg )^{-n+1},$$
where for the case  $n=1$, one must substitute for the right-hand sides,
$ \ln \mu  $ and $  \ln {\L \over \mu_e }  $, respectively.
In order to analytically  evaluate the modular integrals  $I_i^0, L_i^0$, 
so as to expose the
dependence on the infrared and ultraviolet cut-offs, it is convenient to use 
an approximate expression for the 
space-time partition  function factor corresponding to a 
truncation  which leaves only  the momentum modes,
analogously to eq.(\ref{eq:ft1}). 
We follow an approximate procedure, 
due  to \cite{kiko}, which is detailed in  Appendix
\ref{sec:appc}. Useful 
formulas
for the integrals $J_n , K_n$,  accurate to 
 $O(e^{-{1\over \mu^2}})$
 and $O(e^{-{\L^2 \over \mu^2}})$, are: 
\begin{eqnarray}
J_1&=&2\pi [\g_E-3+\ln {\pi \sqrt{27} \mu^2\over 8}],\quad  
J_2=-{2\pi^2\over 3} (1+2\mu^2), \quad  
J_3=-\pi (\ln 3 +{28\pi^2\over 15} \mu^4);\nonumber \\
K_1&=& 2\pi [\g_E -2 +\ln { \mu_e^2\over 4 \L^2}], \quad 
K_2=-2\pi^2\L^2 (1+{2\mu_e^2\over 3\L^2}), \quad  K_3=
 -\pi^3\L^4 (1+{28\mu_e^4\over 15\L^4}).
\label{eq:jika}
\end{eqnarray}
Substituting in (\ref{eq:match})  yields the final formula 
for the $\overline {DR}$ renormalized field theory coupling constant, 
\begin{eqnarray}
 {(4\pi )^2\over g_a^{'2} (p)}&\equiv &
(4\pi )^2[{1 \over g_a^2 (p^2)} +{k_a\over 4} 
(2Z_\Phi +Z_R +16Z_H)+{kk_a \over 8} Z_\L ] =
 {(4\pi )^2k_a\over g_{X}^{2}}+b_a \ln \bigg (p^2 {2\pi \sqrt {27} 
 \over  e^{1-\g_E }M_S^2 }\bigg ) +\D_a, \nonumber \\
 \D_a&=& 
 ( -{hk_a\over 12} +{c_a(k+2)\over 36 } ) (1-3\L ^2\a')
 -{zk_a(k+2) \over 96} (-
 {\ln 3\over \pi^2}  +(\L^2\a ')^2) +\D'_a, 
\label{eq:jauge}
\end{eqnarray}
where we have  reinstated the string scale through the substitution,
 $p^2 \to \a' p^2 
 \equiv {4p^2\over M_S^2} , \quad 
 \L^2 \to \L^2 \a', $  while  choosing the following 
 convention for the string mass scale, 
$ \a' \equiv {4\over M_S^2}$ .  
 We  have exhibited the $\L^2 $ dependent terms, 
 although these cancel away in the relevant limit,  $\a '\to 0$, fixed $ \L $. 
The logarithmic dependence on  the infrared and ultraviolet cut-offs, $\mu , 
\mu_e  $ and $ \L $, has canceled away, leaving the familiar running scale dependence
with an improved  string unification scale,
\begin{equation}
 M_X^2 = {2 e^{1-\g_E} \over  \a' \pi \sqrt{27}}
= M_S^2 {e^{1-\g_E} \over 2 \pi \sqrt{27}}.
\label{eq:unif}
\end{equation}
Interestingly, our result for the  effective unification 
scale is equal to  that of Kaplunovsky
\cite{kapl}, although his 
derivation employed  a sharp  infrared cut-off on the modular integral.
The coincidence of the two results 
reflects an infrared insensitivity of the unification  scale.

The positive powers  of $k$ 
present in $\D_a$ arise from  the 
massless supertraces $c_a , z$ and the corresponding massive 
supertraces included inside 
$\D'_a$. 
The  term of $O(k)$ in the matching equation,  (\ref{eq:jauge}),
relates the cosmological constant,  $Z_\L $, 
to the linear terms in $k$ in $\D_a$, arising with the massless traces, 
$c_a, \ z$ and their massive modes counterparts in $\D'_a$.
The implication  here is that 
the potentially divergent string loop divergences can be absorbed 
inside the constant $Z_\L $. 
Equivalently said, the infrared  divergences signal instabilities 
associated with tadpoles (one-point functions) of 
the dilaton and trace of the  graviton fields,   which can be removed 
by considering a  loop corrected effective action $S_{EFF}$ 
with a finite cosmological 
constant term.    This is the familiar
Fischler-Susskind mechanism \cite{susskind} of
cancellation  of string loops divergences by  massless tadpoles 
corrections to the equations of motions.
Since the renormalization constants $ Z_\L , Z_\Phi, \cdots $ 
can be interpreted as
string loop effects corrections to 
the conformal invariance constraints,  it is  natural to
find that the renormalization  constants
accompany  the divergent dependence in the 
infrared parameter $k$.

The supersymmetric case  should be
immune to  a dilaton tadpole instability, as indeed follows from 
the fact that the massless supertraces,
$ c_a, z$ and the corresponding massive ones in $\D'_a$ have cancelling 
contributions from bosons and fermions within each of the  (massless and 
massive) supermultiplets. 
The massless supertraces $b_a, h$ are 
non-vanishing  helicity weighted sums, which induce finite corrections in 
the gauge coupling constants.
\subsubsection{\bf Quadratic gravitational interactions}
The above derivation can be repeated  word by word 
for the quadratic gravitational interactions coupling constant, denoted 
$g_{R^2}= g_{R^2} (\L )$.
Again, one decomposes the string theory one-loop  contribution into 
the sum of three integrals, $I_i^R$, and writes the matching equation as,
\begin{eqnarray}
{(4\pi )^2\over g_{X}^{2} } +\sum_{i=1}^3 I_i^R& =&
{(4\pi )^2\over g_{R^2}^{'2}(\L )  }+\sum_{i=1}^3 L_i^{R0} ,
\nonumber   \\
 {I_1^R\choose I_2^R}&=& {1\over \pi^3 V(\mu ) } {k\over k+2}
 \int_F {d^2\tau \over \tau_2}
 {X'(\mu ) \over \eta^2\bar \eta^2} 
{    \dh_{\bar \tau } \ln 
{\bar \vt [{\bar \a \atop \bar \b }]\over \bar \eta } 
\choose {1\over 6\mu^2} \dh_{\bar \tau }\ln X'   }
 (\tilde E_2 + {i(k+2)\over 24  \tau_2} )
 {\sum } ' Z_0 (q,\bar q) Z_G(q) ,   \nonumber \\
 I^R_3&=& {k^2\over 64  \pi^3 V(\mu ) }
 \int_F {d^2\tau \over \tau_2^3} 
 {X'(\mu ) \over \eta^2\bar \eta^2} {\sum } '  
Z_0  (q, \bar q) 
Z_G (q).  \nonumber  \\
\bigg [\tilde E_2 (\tau )&=& {i\pi \over 12} (E_2(\tau ) -{3\over \pi \tau_2})
=\dh_\tau \ln \eta -{i\over 4\tau_2} = 
{i\pi \over 12} (1-24 \sum_{n>0} {nq^n\over 1-q^n} ) -{i\over 4\tau_2}  \bigg ]
\label{eq:gravit}
\end{eqnarray}
Next, one
separates out  the contributions of would-be massless modes  by 
taking the limit $q\to 0$ in the various 
terms in the integrands  with fixed 
${1\over \tau_2^n}$ powers, for all factors except $Z_W$. The integrals 
$I_i^R$ reduce then to: 
\begin{eqnarray}
  I_1^{R0}=- {k\over 24 \pi (k+2) }[h_RJ_1 +{k-4\over 2\pi }h J_2] , \quad 
  I_2^{R0}= - {k\over 576 \pi^2 } [z_RJ_2 +{k-4\over 2\pi }z J_3] , \quad 
 I_3^{R0}=  {k^2z\over 128 \pi^3 } J_3 ,
\label{eq:gravit0}
\end{eqnarray}
 where $J_i$ are defined in eq.(\ref{eq:ii0}). The corresponding 
one-loop  field theory integrals, entering as   $z_{0,2}=\sum_{i=1}^3L_i^{R0}$, 
 are given by analogous formulas to those for $I_i^{R0}$, with 
 the substitution $J_i\to K_i$, and the insertion of 
an  overall  factor $(1+{k\over 3})$ in $L_2^{R0}$.
The proper massive threshold corrections are isolated
in  the quantity $\D'_R=\sum_{i=1}^3 (I_i^R-I_i^{R0})$, given 
by the same integrals as $\sum_i I_i^R$ with the asymptotic 
$q\to 0$ limit removed out. 
The divergent dependence on the infrared cut-off parameter is
interpreted along similar lines as in the gauge interactions case,
by matching the functional dependence on $k$ 
of the string and field theory amplitudes, including tree and one-loop 
contributions. The logarithmic divergences are handled by
introducing a renormalized coupling constant.
The matching equation  
for the  quantity, ${g'}_{R^2} (p)$, corresponding to the 
$\overline {DR}$ renormalized quadratic gravitation  coupling constant
combined with the lower dimensions unrenormalized
coupling constants,  reads:
\begin{eqnarray}
{(4\pi )^2\over {g'}_{R^2}^2(p)}= 
{(4\pi )^2\over g_X^2(p)} +\sum_{i=1}^3(I_i^{R0}-L_i^{R0}) -b_R (\g_E
+\ln {\L^2\over p^2} ) +\D'_R.
\label{eq:gravit01}
\end{eqnarray}
Substituting the expressions for the massless modes contributions, yields the
final formula:
\begin{eqnarray}
 {(4\pi )^2\over {g'}_{R^2 }^{2} (p)}&\equiv &
(4\pi )^2 [{1\over g_{R^2}^2(p)}
+ {k\over 8} (2Z_\Phi +3Z_R -16Z_H ) -{k^2\over 48}Z_\L ]=
 {(4\pi )^2\over g_{X}^{2}}+b_R \ln \bigg ( p^2 \a'{\pi \sqrt {27} 
 \over  2 e^{1-\g_E }} \bigg )+ \D_R  ,\nonumber \\
\D_R&=& - {1\over 4} \bigg [ -{h\over 18} (k-4) (1-3\L ^2\a ')
 -(k+2) [{z_R(1-3\L^2\a ')\over 216}-{z(k-4)\over 288}  
 (-{\ln 3\over \pi^2}  +(\L^2\a ' )^2) ]  \nonumber \\
 &-&{z (k+2)^2\over 32}
 ({-\ln 3\over \pi^2}  +(\L^2\a ' )^2)
 \bigg ] +\D'_R, 
\label{eq:gravit1}
\end{eqnarray}
where:
\begin{eqnarray}
h_R= -12b_R=  2 STr\bigg ( ({1\over 12}-\chi^2 )
(l_0r_0-24l_{-1}r_0)\bigg ), \nonumber \\
z_R&=& 2 STr ( l_0r_0-24l_{-1}r_0 ), \quad z= STr(l_0r_0).
\end{eqnarray}
The beta-function slope parameter  $b_R$ is related to the conformal anomaly
\cite{cardoso1,duff,davies}.
In the expression of $\D_R$, eq.(\ref{eq:gravit1}), 
the  terms involving the massless supertraces 
$h_R, z_R , z$  originate from 
$I_1^R, I_2^R,  I_3^R $, respectively. Both $ z_R , z$ and their 
massive counterparts in 
$I_{2,3}^R$ vanish for supersymmetric solutions, 
because of  the bosonic and fermionic modes cancellation.
The vanishing of the terms $O(k^2)$ and a subset
of the  $O(k)$  terms in $\D_R$
is consistent with the vanishing of the 
renormalization constants $Z_{\L , R, H,\Phi }$ for that case, as already 
encountered in discussing the quadratic  gauge interactions. However,
since the helicity supertraces, $h_R, \ h $ are finite, in general, the
infrared divergent term of $O(k)$, which originates  in the 
term, ${k+2\over 24 \tau_2 }$ in $I_2^R$ 
seems to  remain unmatched  
by a corresponding term in the   effective action.
We attribute this 
discrepancy to the   higher-derivative interactions,  such as, 
$R H^2, (DH)^2 , ...$ which we have 
discarded from the effective action.
Thus, a subset of these interactions should acquire one-loop 
renormalization corrections in order to ensure a consistent infrared 
finite theory.
\subsubsection{\bf Quartic gauge interactions}

We shall  use an analogous procedure to describe the one-loop corrections in 
the quartic  $F^{a4}$  gauge coupling constants.
The  separation of massless modes, by subtraction of the $\tau_2 \to \infty $
limit, introduces 
the following massless modes  
supertraces: 
\begin{eqnarray}
h&=&-2STr((\bar Q +\bar I_3)^2 I),\quad 
h_2=2STr((\bar Q +\bar I_3)^4 I),  \quad 
b_a=4STr((\bar Q +\bar I_3)^2 J_a^2), \nonumber \\
c_a &=&2STr(J_a^2), \quad 
l_a=4STr(J_a^4),\quad 
d_a=4 STr((\bar Q +\bar I_3)^4 J_a^4), \nonumber \\
e_a&=&2STr((\bar Q +\bar I_3)^4 J_a^2),\quad 
f_a=4 STr((\bar Q +\bar I_3)^2 J_a^4).
\label{eq:jauge41}
\end{eqnarray}
Modular  integrals $J_n, \ [n=-1,0,1,2]$  are introduced 
by using the same defining equation, eq.(\ref{eq:ii0}). 
The field theory one-loop contributions 
are obtained through the substitution $\mu \to \mu_e , $ with 
corresponding integrals, $K_n , \ [n=-1,0,1,2]$. A renormalized 
running coupling  constant  is introduced within a $\overline {DR}$ scheme, using the 
relationship with the bare coupling constant,
$ g^{-2}_{F^{4a} } (p)-g^{-2}_{F^{4a} }(\L ) = (4\pi )^2 
b_{F^{4a} } 
\ln ({p^2e^{\g_E} \over \L ^2})$.
The matching equation including the singular 
dependence at  $ k\to \infty $  reads:
\begin{eqnarray}
{(4\pi )^2\over  {g'}^2_{F^{4a}}}  &\equiv & 
\bigg [{(4\pi )^2\over  {g}^2_{F^{4a}}} +k^2k_a^2 x_\L Z_\L  +kk_a^2 (x_R Z_R +
x_\Phi Z_\Phi +x_H Z_H) +kk_a  x_F Z_F +k_a^2 x_{R^2} Z_{R^2}
\bigg ], \nonumber  \\
&=&{(4\pi )^2 k_a^2s_a\over g^2_{X} } +
{1\over 8} (\zeta h_2-2\xi b_a  +{k^2\over 2} l_a ) 
\ln ({p^2 \pi \sqrt {27} \over 2 e^{1-\g_E}} ) 
+{1\over  (4\pi )^2} ({k^2\over 2} c_a + \xi h )  (J_2-K_2) \nonumber \\
&+& (\xi  f_a-\zeta e_a) (J_0-K_0)+
{2\pi \over 3} \zeta d_a (J_{-1}-K_{-1}) +\G '_a. 
\label{eq:jauge42}
\end{eqnarray}
where $\xi =  -{k\over 2 }+1 -{2\over k} +\cdots ,\quad 
\zeta  = 1 -{4\over k } +\cdots $, as defined previously.
The proper massive one-loop contributions, denoted by $\G '_a$,  are given 
by the same integral as in eq.(\ref{eq:jauge4}) with the asymptotic 
$\tau_2 \to \infty $ limit removed.
To complete the list  of formulas for the $J_n$ integrals 
given in eq.(\ref{eq:jika}), we  quote the results for 
the two other needed integrals,  
valid  up to the same exponential accuracy:
$$ J_{-1} =[-{155\over 8\pi \mu^4} +{3\pi \over 4} -\sqrt {6} \pi^2 \mu^2 
+O(\mu ^4) ],\quad 
J_0 =[-{21\over 2\mu^2} + \pi \sqrt {3} -{3\pi^2 \mu^2\over 2} 
+O(\mu^4)]  .$$
We have used the same approximate representation as for $J_1$
in  eq.(\ref{eq:ap8}).
The corresponding field theory integrals, 
$K_{-1}, \ K_0  $, can also be evaluated by using the 
analog of the approximate  representation in eq.(\ref{eq:ap9}), 
$${K_0 \choose K_{-1}}= 4\mu_e^2 \bigg [ {\dh \over \dh \mu_e } 
{e^{-\mu_e^2/\L ^2} \over \mu_e^3} {1 \choose {1+\mu_e^2/\L ^2\over 
\pi \mu_e^2 } }  -2(\mu_e \to 2\mu_e ) \bigg ],$$
which indicates that the integrals $K_0 , \ K_{-1} $ vanish 
in the limit, $\L^2 \a '\to 0$, finite 
$\mu $.
The field theory dependence on $k$ in eq.(\ref{eq:jauge42}) involves 
several unknown coefficients, $x_\L , \cdots $.
For interactions of increasing derivative order, 
the matching  relations impose non-trivial relations among wider 
subsets of the renormalization constants.
The identification for the  $ O(k^2) , \ O(k) , \ O(k^0)$ 
terms, respectively,  yields the following formal equations: 
\begin{eqnarray}
Z_\L &\simeq  &\bigg [ \hat c_a j_2 +\hat l_a \ln p^2 -{f_a\over 2k} j_0 
+{\hat d_a\over k^2} j_{-1}\bigg ]_{k^0}+\cdots , \nonumber \\
(x_R Z_R +x_\Phi Z_\Phi +x_H Z_H  +x_F Z_{F^a} ) 
&\simeq &\bigg [-  \hat h j_2 +{b_a \over 8} \ln p^2 +{\hat f_a\over k} j_0 
+{\hat d_a \over k} j_{-1}\bigg ]_{k^0}+\cdots , \nonumber \\
({1\over g^2_{F^{4a} }} +x_{R^2} Z_{R^2} ) 
& \simeq  &\bigg [-2  \hat h j_2 + {h_2 -2 b_a \over 8} \ln p^2 
+\hat f_a j_0 
+\hat d_a  j_{-1}\bigg ]_{k^0} +\cdots ,
\end{eqnarray}
where the dots  stand for  the corresponding massive contributions 
included in $\G '_a$.
We  use the abbreviations, $[\hat c_a, \hat h]={1\over 2 (4\pi )^2} 
[c_a , h], \ \hat l_a= {l_a\over 16},  \  
\hat d_a ={2\pi \over 3} (1-{4\over k}) d_a, \  \hat f_a = (1-{k\over 2} )
f_a -e_a; \quad  j_n = J_n -K_n.$ 
For the 
supersymmetric  case, where  $c_a = l_a = 0$,  with the other supertraces
$f_a, d_a , \cdots $, non vanishing in general, 
the $O(k^2)$ equation for $Z_\L $ 
seems to  contradict the expectation of a vanishing 
one-loop cosmological constant, $Z_\L =0$. 
A similar mismatch  arises with  the $O(k) $ equation, since we expect
$Z_{\Phi , R, H}=0$. 
As already observed,  these discrepancies  probably originate 
in  our neglect of the higher-derivative gravitational 
interactions.
A detailed analysis  of these matching equations 
is beyond the scope 
of this work.  The $O(k^0)$ equation indicates 
that  the string theory $O(F^{a4})$  amplitude 
involves an unknown combination of 
the quartic gauge  and
quadratic gravitational interactions. 
We shall not attempt here to separate these two  couplings.

\subsection{\bf  $D$-term auxiliary fields}

The  background field approach can also be
applied to perturbations involving the  subset of  conserved 
internal space fermionic currents.
$\bar Q_i, [i=1,2,3]$.
In cases where all of the world sheet fermions are free, as in orbifold models,
all three currents $\bar Q_i, [i=1,2,3]$ are conserved and contribute directly 
to the right-moving sector 
conformal weight operators. Since the conformal vertex operators associated 
to the auxiliary $D$-terms are constructed with the linear combination 
$\sum_{i=1}^3 \bar Q_i(\bar z)$, along with the gauge sector currents, 
$J_a(z)$,  a $D-$term perturbation can be represented as a 
deformation of the extended lattice  for the corresponding charges.
(A similar extension, involving the space-time fermionic currents
$J_{\mu \nu }$ and the internal space currents $i\dh X^i$, can be made for 
the auxiliary $F-$terms.) 

The discussion for 
the auxiliary $D-$term field
was initiated by Petropoulos \cite{petro}, and we review  the main 
arguments here. 
Let us introduce 
the world sheet  fields, $H_i (z), [i=1,2,3]$, corresponding to the 
bosonic counterparts   of the 
internal space  fermion fields,
$\psi^{i,\bar i} =e^{\pm iH_i},$ which describe the 
$SO(6)$ affine algebra, with  the Cartan subalgebra
generators,
$ \bar Q_i(\bar z) =i\psi^i 
\psi {\bar i} =i\bar \dh H_i.$  
In terms of the fields $H_i$, one can express 
the conserved $U(1)$ current, 
$ \bar J(\bar z)$,
of the $ {\cal N}=2$ superconformal algebra of the right-moving sector, 
the   space-time supersymmetry currents, 
$ Q^\pm _{susy}(\bar z) $,
and the anti-holomorphic three-form  field, $\bar \e^\pm (\bar z) $,   as:
$$ \bar J(\bar z)= i\sqrt 3 \bar \dh H, \quad 
 Q_{susy}(\bar z) =e^{\pm i {\sqrt 3\over 2} H} , \quad  
\e^\pm (\bar z)=e^{\pm i (H_1+H_2+H_3)}. \quad 
[H={1\over \sqrt {3} } (H_1+H_2+H_3)] $$ 
The $D-$term   vertex operator 
for a  gauge group factor $G_a$ may then be written as \cite{atick}: 
\begin{equation}
V_D^a(z,\bar z)=-{i\over  \sqrt {6k_a }} 
\bar J(\bar z) J^a (z),
\label{eq:vertop}
\end{equation}
where $J_a (z) $ is the $U(1)$ gauge charge density. 
Starting from the conformal generators known dependence  on the  fermionic 
and gauge charges, 
$\bar L_0=\sum_{i=1}^3 {\bar Q_i^2\over 2} +\cdots  , L_0={J_a^2\over k_a}
+\cdots $, 
a D-term  deformation  of the associated zero-modes lattice can be induced
by   performing an orthogonal transformation to the  basis,
\begin{equation}
\bigg [\bar Q_1,\bar Q_2,\bar Q_3\bigg ]\to 
\bigg [\bar Q_{(0)}={\bar Q_1+\bar Q_2+\bar Q_3\over \sqrt 6}, \quad 
\bar Q_{(8)}={\bar Q_1+\bar Q_2-2 \bar Q_3\over \sqrt {12}}, \quad  
\bar Q_{(3)}={\bar Q_1-\bar Q_2\over 2}\bigg ],
\label{eq:lin}
\end{equation}
with $\bar L_0= \bar Q_{(0)}^2+ \bar Q_{(8)}^2+\bar Q_{(3)}^2+\cdots, $  
followed  by a Lorentz transformation of hyperbolic angle, 
$\o $,  acting on the components,  $[\bar Q_{(0)},
{ J^a\over \sqrt k_a}] $. 
The deformed  theory one-loop vacuum functional
in Hamiltonian formalisn, 
$ Z\propto  Tr ( q^{L_0} \bar q^{\bar L_0} 
e^{-4\pi \tau_2  \d L_0})$, with the conformal weight  operator increment, 
$\d L_0=({J^a\over \sqrt {k_a }} \cosh \o 
+\bar Q_{(0)} \sinh \o )^2 - 
({J^a\over \sqrt {k_a }})^2  $,  is to be compared with that obtained in the 
Lagrangian functional integral formalism by adding the perturbed action, 
\begin{equation}
\d S=
{D^a\over 2\pi } \int d^2\s \sqrt h V_{D^a}= 
{D^a \over 2\pi }  (2\tau_2) \int d z
d\bar z  V_{D^a}  = -4\pi \tau_2 {D^a\over \sqrt {6k_a} } (\bar Q_1 
+\bar Q_2+\bar Q_3) J^a .
\label{eq:dterm}
\end{equation}
Identifying $\d S$  with $-4\pi \tau_2 L_0$,  yields: $D_a =\sinh 2\o $.
Proceeding next to the field theory description, one considers 
the  part of the  four-dimensional supersymmetric  Lagrangian depending on the 
auxiliary field $D_a$, 
\begin{equation}
L_{EFF}={D_a^2\over 2g_a^2} +D_a\phi_i^\star (\hat J^a)_{ij} 
\phi_j +c_a D_a +\cdots ,
\label{eq:dterm1}
\end{equation}
where $\phi_i$ denote the charged matter fields  in the gauge 
group representation 
with generators $\hat J_a$ and  $c_a$ stands for a Fayet-Iliopoulos 
interaction coupling constant.
We can now match $L_{EFF}$   to the 
string theory vacuum functional,
$$Z= {V^{(4)} \over 2 (2\pi )^4} \int {d^2\tau \over \tau_2^3}
{Z_W\over \vert \eta \vert^4} {\sum }' Z_0Z_G e^{-4\pi \tau_2 
[\sinh 2\o  \bar Q_{(0)} {J_a\over \sqrt {k_a}} +\ud (\cosh 2\o 
-1) (\bar Q_{(0)}^2 +{J_a^2 \over k_a } )] } ,$$
after expanding  the exponential factor
in powers of  $D_a$.   The linear and quadratic terms in $D_a$ give us, 
respectively,
\begin{eqnarray}
c_a&=&-{1\over 16\sqrt {3k_a} \pi^3} \int_F {d^2\tau \over \tau_2^2}  
{Z_W \over \vert \eta \vert^4 } {\sum } '  Z_0 Z_G 
(\bar Q_1+\bar Q_2+\bar Q_3)J^a = {1\over 48\pi^2} Tr(\hat J^a),\nonumber \\
{(4\pi )^2\over g_a^2} & =& {4\over 3 }\int_F {d^2\tau \over \tau_2} 
Z_W{\sum } '  
{Z_0Z_G\over \vert \eta (\tau )\vert ^4}
[(\bar Q_1+\bar Q_2+\bar Q_3)^2(J^{a2}-{k_a \over 8\pi \tau_2} )].
\label{eq:dterm2}
\end{eqnarray}
To obtain the second equation for 
the linear term $c_a$, we have used the property 
which identifies the world sheet 
charge $(\bar Q_1+\bar Q_2+\bar Q_3)/\sqrt 3$ as a 
space-time R-charge operator, whose trace 
for massive supermultiplets  gives a net zero and for
massless supermultiplets combines to 
$-1$.  Thus, $c_a$ is proportional   to the trace over the massless 
fermions of the charge generator
$\hat J_a$ (normalized as $Tr(\hat J_a^2 ) =\ud $), 
which is non-vanishing only 
for $U(1)$ gauge group factors.  This expectedly reproduces the 
known result \cite{atick} that an apparently 
anomalous  $U(1)$ can indeed  arise 
in string theory with a one-loop order 
finite, universal coefficient \cite{witten2}. 

The quadratic term in $D_a$  gives us  an equation for the one-loop 
correction to the gauge coupling constant. 
The formula in eq.(\ref{eq:dterm2}) is, of course, 
valid in the supersymmetric case only.
The comparison with the corresponding result obtained  with a 
constant magnetic background field, as given   in eq.(\ref{eq:string}),
would show agreement
if one had the operator identity, 
$ (\bar Q_1+ \bar Q_2+ \bar Q_3)^2 =-3
\bar Q^2$,  noting that
here,  $(\bar Q +\bar I_3 )^2 = \bar Q^2$, 
since the term $\dh_{\bar \tau } \ln \bar \eta $ vanishes for
supersymmetric solutions.
This identity can indeed  be established for orbifold models  by use of
the generalized Riemann-Jacobi identity, as was first shown in \cite{petro}.

\section{\bf  Numerical results}
\label{sec:secgauge}
\subsection{\bf Quadratic order  gauge and gravitational interactions}
\label{sec:gauge1}

The grand desert scenario for the 
minimal supersymmetric standard model 
is known \cite{amaldi} to favor an
unification   scale, 
$M_{GUT}=2\times 10^{16}  $ GeV, with an unified gauge coupling constant, 
$g_{GUT} =(4\pi \a_{GUT} )^\ud \simeq (4\pi /25)^\ud \simeq 
0.71$. Transposed to a string theory framework, where the unification 
scale is determined at tree level 
in terms of  the Planck mass $M_P$ and  string constant
$g_X$ as, $M_X={e^{(1-\g )/2}\over 
4\pi (27)^{1/4}}   g_X M_P \simeq g_X 5.27\times 10^{17} $ GeV, 
the same type of scenario seems
to overestimate the unification scale  $M_{GUT}$ by a factor 
$20$. If one insisted on setting $M_X =M_{GUT}$, this would lead to
an overestimate of Newton constant, $ G_N  ={1\over  M_P^2}$, 
by a factor $400$.  Reasoning in terms of the underlying 
compactified ten-dimensional string  theory, would even turn this estimate   into 
a lower bound,   $G_N \ge {\a _{GUT}^{4/3} \over M_{GUT}^2} $.
This problem has motivated recent proposals 
to examine the alternative  option of 
a strongly coupled string theory \cite{witten,lykken}.
Remaining, however, within  the 
perturbative framework, 
three main known effects could possibly cure 
this  discrepancy: Adjustable 
Kac-Moody levels;  intermediate thresholds;
heavy thresholds.
Of these three
items, the  last one, on which we shall concentrate in this section,
appears as the most
controllable one.
Consider the general splitting of threshold corrections \cite{mayr}, 
$\D_a =-b_a \D +k_a Y +R_a$, involving  two  
components of universal character, $\D  $ and $ Y$, whose contributions 
can  be absorbed into  redefinitions of the 
unification  scale and gauge coupling constant,
\begin{eqnarray}
M_X \to 
M'_X=M_X e^{\D /2}, \quad {1\over g_X^{2}}  \to {1\over {g'}_X^{2}}=
{1\over g_X^{2}}  
+{Y\over ( 4\pi)^2},
\label{eq:mex}
\end{eqnarray}
along with a non-universal residual component $R_a$.
Several studies of threshold corrections using  solvable 
models of string vacua have attempted to justify a decomposition 
of this type \cite{dienes1}. 
In this section, we 
shall pursue the effort started in  our previous paper \cite{chem} 
with the purpose  of  updating the numerical 
results reported there for the  gauge  coupling constants by use of the more 
complete formalism  presented in the previous section.
We perform numerical calculations for the following selection of 
16 abelian orbifold models: (i)  The seven  standard embedding $Z_N$ orbifolds
described for $N=3,4, 6,$ 
 by the internal shift vectors, $ Nv_i= (1,1,-2)$; for  
$N=7,8$, by $N v_i= (1,2,-3)$;  and  for $ N=12$ by $Nv_i=(1,4,-5)$;
(ii) Four non-standard embedding models 
described  by the gauge sector shifts, $ (N V^I) (NV^{'I})=
(1120^5) (1120^5)', (110^6)(20^7)', (1^420^3)(20^7)'$,  for $Z_3$,
and $ (1120^5)(220^6)'$ for $Z_4$;
(iii) Three  non-standard embedding  $Z_3$ models 
with two discrete Wilson lines, due to Ib\'a\~ nez et al.,
\cite{ibanez} and Kim and Kim \cite{kim}.
(iv) Two non-standard embedding  $Z_7$ models 
with one  discrete Wilson line, due to  Katsuki et al., 
\cite{katsuki} and Casas et al., \cite{casas}.
The inputs and gauge group factors for these models 
are described in \cite{chem}. The affine  algebra
levels for the models considered here are $k_a=1$ for non-abelian group 
factors and $k_a= 2\sum_{I=1}^{16} \hat J_a^{I2}$ for the abelian 
$U(1)$ factors.

Let us refer to the two contributions in eq.(\ref{eq:fa2}), which are 
associated with the squared gauge charge  term $J_a^2 $ and the 
modular anomaly compensating 
term $ -{k_a\over 8\pi \tau_2 }$,
as the  zero-modes (or charge)  and anomaly  (or back-reaction) 
contributions,  respectively. 
The  following general trends for the zero-modes  contributions
were found in the results quoted in our  previous paper \cite{chem}.
The $Z_3$ orbifolds show  marked universal features, with 
$\D = 0.068, Y=3.4, R_a=0$. Larger values for the universal 
components appear for 
the $Z_7$ orbifolds, $ \D =0.20\sim 0.40, Y=15$, along with $R_a \ne 0$. 
For the non-prime $Z_N$ orbifolds,  the situation is less clear cut 
since the universal components 
cover wider ranges, 
$\D = -0.2\sim  +0.6, Y= 10\sim 40$. 

\begin{figure} [t]
\centerline{
\psfig{figure=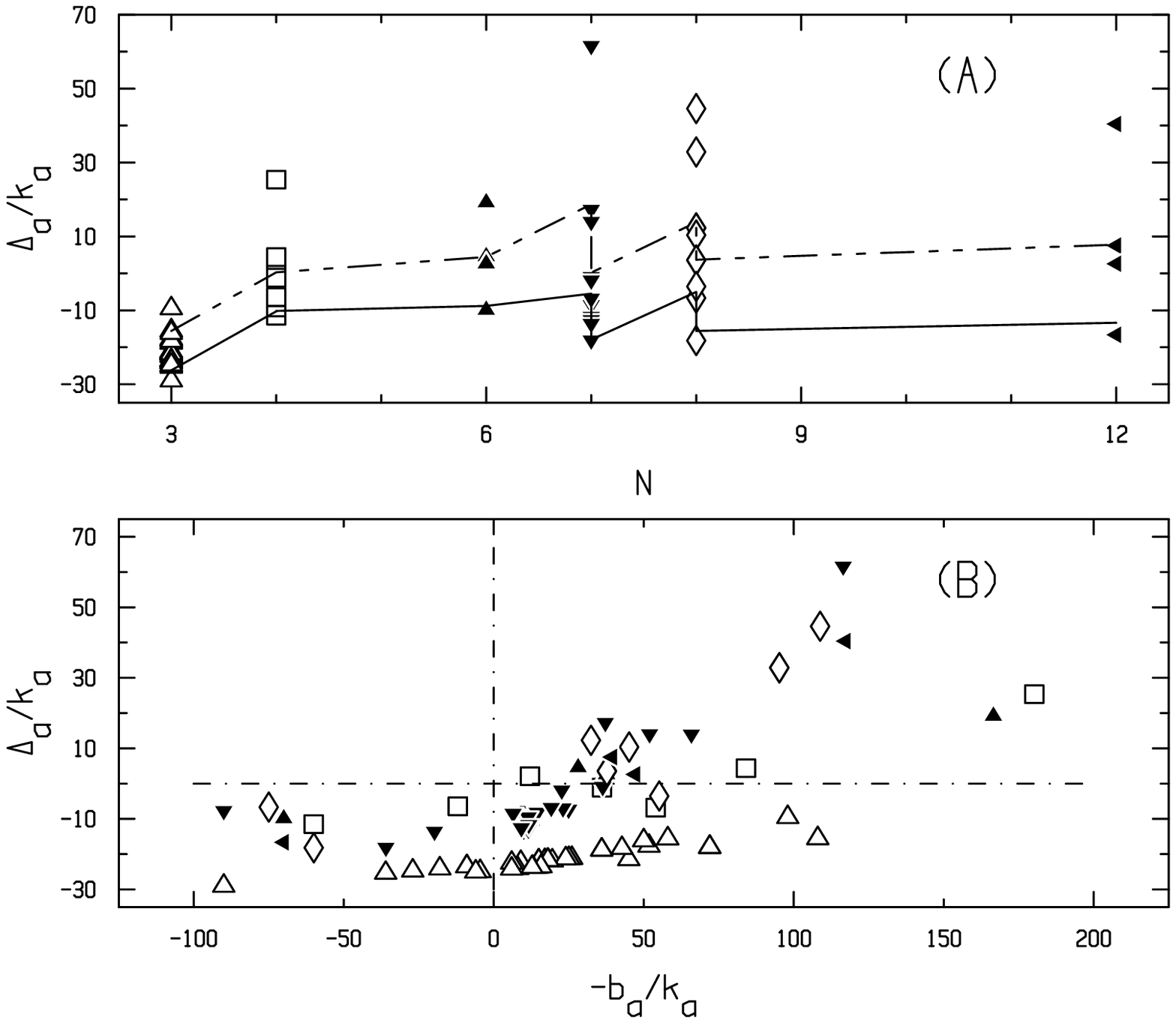,height= 16cm,width=12cm}}
\caption{ {\bf (A)} The one-loop normalized  threshold 
corrections ${\D_a\over k_a} $  to
the gauge coupling constants are
plotted as a function  of the orbifold order $N$  for the sixteen 
orbifold models described in the text. The 
triangles  and squares plotting  symbols  give 
the total ${\D_a\over k_a}$ for the various gauge group factors.
The back-reaction contributions to ${\D_a\over k_a}$ are shown 
by points joined by continuous lines. The full   gravitational 
correction $\D_R$ is shown
by points joined by dashed lines.
{\bf (B)} The  one-loop normalized  threshold corrections are 
plotted as a function  of the normalized 
slope parameter $-{b_a\over k_a}$. 
We denote results for the  gauge  group  factors of 
$Z_3$ orbifolds by open triangles; $Z_4$ by open 
squares; $Z_6$ by filled triangles pointing up; $Z_7$ by 
filled triangles pointing down;  $Z_8$ by open  parallelograms; and
$Z_{12}$ by 
filled triangles pointing leftwards. }
\label{fig1}
\end{figure}

Turning now to the contributions from the  back-reaction component,
we find that this  brings a  large 
negative contribution to $Y$.
(We draw attention to a factor 1/2 discrepancy with the 
results for the back-reaction component reported  in \cite{kkpr3}, 
making our $Y$ twice larger in absolute value.)
The component $\D $ is clearly  unaffected.
The numerical results  for the back-reaction correction 
alone, as well as for the total correction,   
are shown in figure \ref{fig1}.
We see on the plot (A) of figure \ref{fig1}
that the spread between  the various group factors 
is small and that it slightly increases with the orbifold symmetry  order $N$.
The back-reaction  term   dominates and partially 
cancels the zero-modes term. It is  the largest  contribution in  
absolute value for the  $Z_3$ orbifolds, where it
reaches, $Y= -26$,  independently of 
the gauge embedding and Wilson lines  twists. The largest 
contribution to the modular integral here arises from the 
$\tau_2\to \infty $ tail,
which is  determined by the helicity weighted 
massless modes supertrace term, $ -hk_a/12$. 
Since $h $ takes  different signs for chiral and vector supermultiplets, 
the untwisted sector contributes with an opposite sign to that of the
twisted sectors.
One indeed finds large cancellations  between 
the untwisted and twisted 
sectors,  the twisted part being the largest.

The results for the orbifolds other  than $Z_3$ indicate the presence of a 
residual component $R_a$. This appears in a clearer way on 
plot (B) in figure \ref{fig1}.
The signature  for a decomposition of $\D_a$ in terms of 
only two universal components would appear in this  plot 
as a clustering of the points along a single straight line
whose intercept  and slope 
identify with $Y $ and $\D $, respectively.
The conclusions from figure \ref{fig1} is that  no clear
systematic  trend towards such a  universal behavior is visible on the
results.
However, for a fixed orbifold order $N$,  the  deviations $R_a$ 
are  quite small and alignments along straight lines are observed.
The purest case is that of $Z_3$ orbifolds. For the higher-order orbifolds,
common trends do appear, such as a positive $\D $ (negative $Y$) which 
increases (decreases) with increasing $N$. 

It is interesting to  contrast the predictions  for the gauge coupling 
constants against 
the phenomenologically 
favored ones. Naively, a reduction of $M'_X \sim 10^{18} $ 
down to $M_{GUT} \sim 10^{16} $ 
would require $\D \simeq  -10$, while a shift from a dilaton 
vacuum expectation value  set at
a strongly coupled regime, say,   ${4\pi \over g_X^2} \sim O(1)$, 
or at  the self-dual point of  $S-$duality, $ S\to  {(4\pi)^2\over S}$, to
the empirical value, ${4\pi \over {g'}_X^2} \sim  25$, 
would require $Y\simeq 300$.
It appears then  that 
the  predicted moduli-independent  corrections are 
much too small,  and even of wrong sign for $\D $ and 
$Y$,  with respect to a  naive
perturbative string theory 
unification scenario.  Nevertheless, 
viable scenarios  can be found in association
with the  other  expected mechanisms of 
an anomalous $U(1)$ threshold and small 
affine level for $U(1)_Y$ \cite{chem}.

The threshold corrections for the quadratic gravitational   interactions
arising from the term $\tilde E_2 (\tau )$ in eq.(\ref{eq:part4}) are 
also shown in plot (A) of figure \ref{fig1}. 
The zero-mode contribution is again dominated by the  anomaly compensating 
back-reaction contribution from the $-{3\over \pi \tau_2}$ term,
which numerically  coincides with 
that of the gauge interactions case.  
The net correction  to $\D_R$  shows some model dependence 
for orbifold orders $N\ge 6 $ and smoothly increases 
from $-16$  for 
$Z_3 $ to $+8$ for $Z_{12}$ orbifolds.  
Thus, expressed as a shift in the tree  level
coupling constant, $ \d ({4\pi \over g_X^2}) = {\D_R \over 4\pi }$, 
the threshold correction represents 
a tiny few percent effect.

\subsection{\bf Moduli-dependent threshold corrections}
\label{sec:moduli}

In this section we present a comparison  
with the moduli-dependent threshold corrections in 
the quadratic gauge and gravitational  interactions. 
We apply 
the methods initiated in \cite{dkl} and further developed 
in \cite{harvey} to the
curved space-time regularization approach of Section \ref{sec:secin}. 
The one-loop  contributions from the ${\cal N}=2$ supersymmetric 
suborbifolds (with an unrotated two-dimensional torus) 
have a simple representation in terms of a  summation involving the 
right-moving sector massive  BPS states, along with  an unrestricted
sum over the
left-moving sector states. 
The so-called perturbative 
BPS (Bogomolnyi-Prasad-Sommerfield) states are the stable string modes 
which saturate the mass bound, ${\a '\over 4} M_R^2\ge  {p_R^2\over 2}$,
where the ${\cal N}=2$  central charge,   $p_R$, identifies with  the
zero-mode momentum  of the unrotated two-dimensional torus \cite{carlumo}.  
The one-loop contributions in the gauge and gravitational coupling 
constants read:
\begin{eqnarray}
\bigg [{(4\pi )^2 \over k_ag_a^2};
{(4\pi )^2 \over g_R^2} \bigg ]
=  -{1\over 4} \int_F {d^2\tau  \over \tau_2^2} Z_W(\tau , \bar \tau )
(\tau_2 Z_\G (y,\bar y))\bigg [ F_1(q) +{F_2(q)\over \pi  \tau_2};
 F^g_1(q) +{F_2(q)\over \pi  \tau_2} \bigg ],
\label{eq:coupl1}
\end{eqnarray}
where the partition function  for the 
zero-modes  lattice of the internal space
fixed two-dimensional torus,  denoted
$ Z_\G (y,\bar y) =\sum q^{p_L^2/2} \bar q^{p_R^2/2}$,
depends on  the two complex  moduli fields, 
$y= [y_+= T=T_1+iT_2, \ y_-=U= U_1+iU_2]$, parametrizing 
the coset space, $SO(2,2,R)/SO(2,R)\times SO(2,R)$.  
A similar procedure to that of Section \ref{sec:secin} 
(separation of the  massless modes 
at the $\tau_2\to \infty $ degeneration limiti,
subtraction  of the field theory  one-loop contribution and introduction of 
the $\overline {DR}$ renormalized conatant) 
is used to define the 
threshold corrections, $\D_a, \ \D_R $.
The string theory contributions 
involve the following periodic Ramond sector traces 
associated with the so-called 
new supersymmetry index, 
\begin{eqnarray}
[F_1(q),\ F_1^g(q), \ F_2(q)]&=&{1\over \eta^2} {i\over 2}  
Tr_R\bigg (J_0e^{i\pi J_0} 
[ 8\hat J_a^2,\  {E_2(\tau )\over 3},\ 
-1] q^H \bar q^{\bar H }\bigg ), \nonumber \\
&=&\sum_{m\ge -1 }^\infty
[c(m),\  c^g(m), \  \hat c(m) ] q^m,
\label{eq:fonctions1}
\end{eqnarray}
which are meromorphic functions of 
$q$, with at most simple  poles at the cusp point,
$\tau  =i\infty $,  and 
Laurent series expansions  given by the second line of  
eq.(\ref{eq:fonctions1}). 
The zero-mode component, $J_0$,  of the $U(1)$ 
current, $J(\bar z)$, is related to the fermion number operator, $F$,
such that, ${i\over 2} Tr_R(J_0 e^{i\pi J_0}\cdots  ) = Tr_R((-1)^F\cdots )$. 
All three functions, $F_1, \ F_1^g, \ F_2$,  are modular functions for the 
$SL(2,Z)$ modular group, of weight $0$ for $ F_1 , F_1^g$, and  $-2$ 
for  $F_2$, except for modular anomalies of the same form as for 
${1\over \tau_2} $, 
which cancel exactly in the  relevant modular invariant
combinations, $[F_1+{F_2\over \pi \tau_2 },\quad 
F_1^g+{F_2\over \pi \tau_2 }]$.
These functions exhibit simple universality properties
for the  class of non-prime $Z_N$ orbifold models 
associated with decomposable 
six-dimensional tori, $T^6 =T^4/G \oplus T^2$. For the subclass of 
standard gauge embedding orbifolds, we find by explicit calculation, 
\begin{eqnarray}
F_1(q)&=&c(0) \bigg [  1+{11\over 11088} ({E_6^2 - E_2 E_4 E_6\over \D }
+720 ) \bigg ] =c(0)
(1+330 q +26400 q^2 + 881100 q^3 +\cdots ), \nonumber \\
F_2(q)&=&-{4\over \vert G \vert \D } E_6E_4 = {4\over \vert G \vert } 
(-{1\over q}+240  +141444 q + 8529280 q^2 +\cdots ) , \nonumber \\
F_1^g(q)&=&{4\over 3\vert G \vert \D } E_6E_4 E_2={4\over 3 \vert G \vert }
({1\over q}-264  -135756  q  - 5117440 q^2 -\cdots  ), 
\label{eq:mi2p}
\end{eqnarray}
where $  \D  (\tau )= \eta^{24}(\tau ) $ and $E_{2k}(\tau )$ are the Eisenstein 
series functions, normalized as, $E_{2k}(i\infty )=1$.
The model dependence  in eq.(\ref{eq:mi2p})
resides only in the slope parameter  coefficients, 
$b_a  ={c(0)\over 4}, \quad b_R=
 {c^g(0) \over 4} =- {1\over 48} (N_H-N_V+24) $,
 and in the orbifold order, 
$\vert G \vert $. 
($N_H ,\ N_V$ are the numbers of 
${\cal N}=2$ hyper and vector supermultiplets, including the dilaton and graviphoton,
respectively.)
Substituting in eq.(\ref{eq:coupl1}) and carrying 
the modular integrals by means 
of familiar methods \cite{dkl,harvey,kiko}, one finds 
the following formula for the threshold corrections in the  gauge 
coupling constants:

\begin{eqnarray}
\D_a&=& {c(0)\over 4} [\ln (2T_2U_2)-{\pi \over 3} U_2+{\cal K}' ]
-{\pi\over 12} 
T_2(c(0)-24c(-1))+R\sum_{r>0} 
c(kl)\ln (1-e^{2\pi i x })\nonumber  \\
&-&{\hat c(0)\over 4\pi^2 U_2 T_2 } (\zeta (3) +{2\over \pi } 
\zeta (4) U_2^3 ) -{\pi \over 72} 
T_2(\hat c(0)-48\hat c(-1))-{1\over \pi U_2 T_2} 
R\sum_{r>0} \hat c(kl) P(x),
\label{eq:mi3p}
\end{eqnarray}
 where ${\cal K}'=\ln {4\pi e^{3-3\g_E} 
\over 9\sqrt {3}} 
={\cal K} +\ln {e^{2(1-\g_E)}\over 3}, \quad 
\zeta (3)= 1.2020569..., \  \zeta (4)= {\pi^4\over 90}, \ 
\g_E= 0.5772156... $ (Euler-Mascheroni  constant);
$r=(-l, -k)$ is the lattice vector of the fixed two-dimensional torus, 
such that the scalar product, $r\cdot y = ky_+ +ly_-, $ with 
the special definition, $ x= r\hat \cdot y  = \vert R (k T+lU )\vert
+i\vert I(kT+lU)\vert $, where     $R, I $ mean real and imaginary parts;
$r>0 $ means ($ k >0, l\in Z;$) or 
($ k=0, l>0$); 
$P(x) =(Ix) Li_2 (Q) +{1\over 2\pi }Li_3(Q), \ [Q=e^{2\pi ix} ]$ with 
$Li_j(Q)=\sum_{p>0} {Q^p\over p^j} $, the Polylogarithm functions.
To translate  the automorphic fields  notation used here  into 
the string fields notation, one must apply the 
transformation, $ y_\pm \to iy_\pm $ ($T\to iT, \quad U\to iU$).
As functions of $T, U$, 
the $\D_a$ are invariant  under the  modular group,  $O(2,2,Z)\sim 
SL(2,Z)_T \times SL(2,Z)_U$ which includes the interchange 
$T\leftrightarrow  U$.
The representation in eq.(\ref{eq:mi3p}),  which is
only  valid in the domain 
$T_2 > U_2$, is transformed  when passing through the wall at $T_2 =U_2 $,  by
the substitution $T \leftrightarrow U$.

The formula corresponding to (\ref{eq:mi3p}) for  the  gravitational 
correction $\D_R $ is  obtained by simply changing $ c(n)\to c^g (n)$, 
keeping the coefficients $\hat c(m)$ unchanged.

The only discrepancy between our result and those found
in the  approach of \cite{dkl,harvey}
resides in a shift of  the constant term
corresponding to the numerically small  difference,  
${\cal K}'-{\cal K} ={\cal K}' -\ln {4\pi e^{1-\g_E}
\over \sqrt {27} }=  2-2\g_E -\ln 3 \simeq  -0.253$.  
The effective  unification  scale, incorporating the constant term from 
the ${\cal N}=2$  sector,
$M_X^{'2} =M_X^2 e^{-{\cal K}'} ={3\over \a' e^{2(1-\g_E )} 2\pi^2}=
{6M_S^2\over e^{2(1-\g_E )} (4\pi )^2}$,   is then 
a factor 
 $3e^{-2(1-\g_E)}\approx 1.29$ larger 
than that given in \cite{dkl}.
The mismatch originates from the  use there  
of the  simple-minded infrared regularization factor, 
$(1-e^{-N/\tau_2}) ({N\over \tau_2})^\e $,
with the limits,  $N \to \infty , \ 
\e \to 0$. 
In contrast to our curvature infrared  regularization, which is 
realized by the  partition function factor, 
\begin{equation}
Z_W(\tau , \bar \tau )\simeq  [1 +2\phi (\mu )-\phi (2\mu )], \quad 
\bigg [ \phi (\mu ) = (1-\mu {\dh \over \dh \mu })
\sum_{m\ne 0}e^{-{\pi m^2
\over \mu^2 \tau_2 }} \bigg ] 
\label{eq:zw1}
\end{equation}
the  regularization prescription  in \cite{dkl} 
clashes with modular invariance. 

The constant, $O(q^0)$, term in the decomposition of $F_1(q)$
in eq.(\ref{eq:mi2p}), gives rise to the gauge group dependent contribution  involving 
 $c(0)$  in eq.(\ref{eq:mi3p}), which is associated with
the subset  of the BPS states  with 
$kl= 0 $,  namely, $(k=0, l\in Z)$ or  $ (k>0, l=0)$. This 
sums up to, $ (\D_a)_{DKL}  = b_a [\ln (2T_2U_2 \vert \eta (T) \eta (U)\vert ^4 
)+{\cal K }']$, which identifies with the total correction 
initially  discussed by Dixon, Kaplunovsky and Louis \cite{dkl}.
The non-constant terms in $F_1(q)$ of $O(q^n), \ [n>0]$ which arise from 
the combination,
$${E_6^2(\tau )-E_2(\tau )E_4(\tau )E_6(\tau )\over \D  (\tau )} + 720 
= j(\tau ) -{E_2(\tau )E_4(\tau )E_6(\tau )\over \D  (\tau ) } -1008, $$
are associated 
to the contribution from the  subset of states,  $(k>0, \ l\in Z)$ in eq.(\ref{eq:mi3p}). 
Based on the representation, $j(\tau )-744= {1\over q} +196884 q +
21493760 q^2+\cdots = \sum_{m\ge -1}  c^j(m) q^m, $ 
and the Borcherds product formula \cite{harvey}, 
the term  involving $j(\tau )$ yields a  contribution  to $\D_a$ 
proportional to $ \ln q_T^{-1} 
\prod_{k>0, \ l\in Z}  (1-q_T^kq_U^l)^{c^j(kl)}=
\ln [ j(T)-j(U) ]$, where $q_{T,U}=e^{2\pi i [T,U] }$. 
The singularity at the submanifold  $T=U$ mod($SL(2,Z)$ is a reflection of the 
stringy Higgs mechanism responsible for the enhanced symmetry, 
$U(1)_T\times U(1)_U\to SU(3)$, 
in the gauge group associated with the internal space coordinates of the 
fixed two-dimensional torus 
\cite{carlumo}. 
The  absence of  singularities in the threshold corrections 
$\D_a$, associated with gauge symmetries from the gauge sector space,
is due to the 
compensation by a corresponding singular $\ln (T-U)$  contribution from the 
remaining terms in $F_1(q)$.
For the gravitational case, the compensation of the singularity at $T=U$
in $\D _R$ takes place upon adding the contributions from $F_1^g $ and $ F_2(q)$.

A related  discussion of the universality properties for the 
subclass of ${\cal N}=2$ models obtained by toroidal compactification of ${\cal N}=1$ 
models in six dimensions
is given in \cite{peri,kkpr2}.  Another  interesting 
class of ${\cal N}=2$ compactification models
with non-universal behavior is discussed in \cite{kkpr1}.

\begin{figure}[t]
\centerline{\psfig{figure=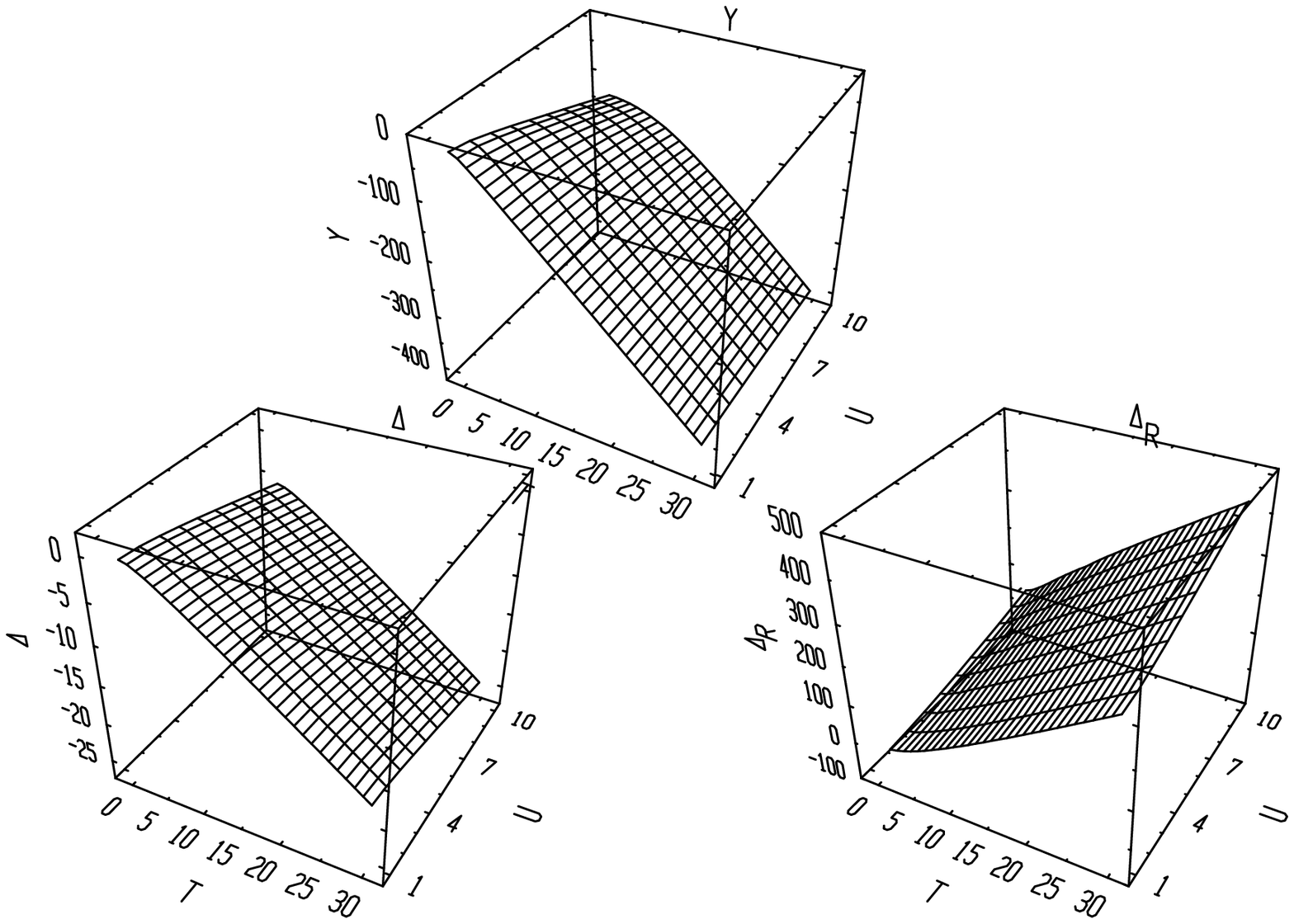,height=15cm,width=14cm} }
\caption{The one-loop threshold corrections for decomposable tori models, 
specialized to the $Z_4$ orbifold case, are 
plotted as a function of $T=T_1+iT_2, U=U_1+iU_2$ at 
$T_1=U_1=0$, with the variables $T_2, \ U_2$ along the  horizontal 
axes. The three plots from 
left to right show the universal components 
$\D , Y$ for the gauge interactions and 
$\D_R$ for the quadratic  gravitational interactions.}
\label{figx2}
\end{figure}

We shall now  present results for the class 
of models described by eq.(\ref{eq:mi2p}).
In terms of the two-component decomposition,
$ \D_a =-b_a \D +k_a Y$, the quantities
$\D , \ Y$ are then identified with the zero-modes and back-reaction 
contributions associated with  $F_1 (q)$ and $F_2(q)$, respectively.
The numerical results
are shown in figure \ref{figx2}. The relationship between $T,U$ and the 
representations of the metric and torsion tensors 
in the zero-modes  lattice basis is given by: 
\begin{eqnarray}
 T&=&2(B_{12} +i\sqrt {det G})= (b+i r_1 r_2 \sin \t ) , \quad  
U= (G_{12} +i \sqrt { det G} )/G_{11}
= {r_2\over r_1} e^{i\t } ,
\end{eqnarray}
 where $ 2 [G_{11},  G_{22}, 
\ G_{12} , B_{12}] = [r_1^2,\  r_2^2, \ 
r_1 r_2 \cos \t , \ b ]$, such that, 
$T, b, r_i^2$ are expressed in units $\a '=1$.
Close to  the self-dual points, $ T=U = i$,
the moduli-dependent corrections 
are comparable in size 
with the moduli-independent  corrections, but with an opposite 
negative  sign for $\D $. For example,  at  $T_2=3, \ U_2=4 $, one has: 
$\D =-2.17, \ Y=-60., \ \D_R= 16$. The variations are rather smooth  and
have a linear power  law increase 
at large $T$ or $ U$. At fixed  $U$, say, $U=4i$ or ${i\over 4}$, 
one  finds: $ \D \simeq -0.66 \vert T\vert , \
Y \simeq -12.7  \vert T\vert , \ \D_R \simeq 10.5 \vert T \vert .$
The dependence on the real 
parts is weak.
Recall that $\D , \  \D_R $ and $ Y$ are 
symmetric   under $T\leftrightarrow  U$ and that 
their  values for  $T , U $  below and above $i$ 
are  connected  by  the duality transformations,
$T\to -{1\over T}, \ U\to -{ 1\over U} $. 
For large compactification volumes, the correction $\D $ has the right 
sign for a reduced unification scale, and the right magnitude  provided that
$T \sim  O(10)$. The correction $Y$, although of sizeable magnitude, 
also has a wrong sign to shift the dilaton vacuum expectation value towards
weak coupling. 
The total  correction  $\D_R$ 
in the gravitational case  is of opposite sign to (and same magnitude 
as) that in the 
gauge case. The associated unification scale, ${M'}^R_X=e^{-\D_R/bR} M_X$,
is enhanced (reduced) for negative (positive) $b_R$.

\begin{figure}[t]
\centerline{\psfig{figure=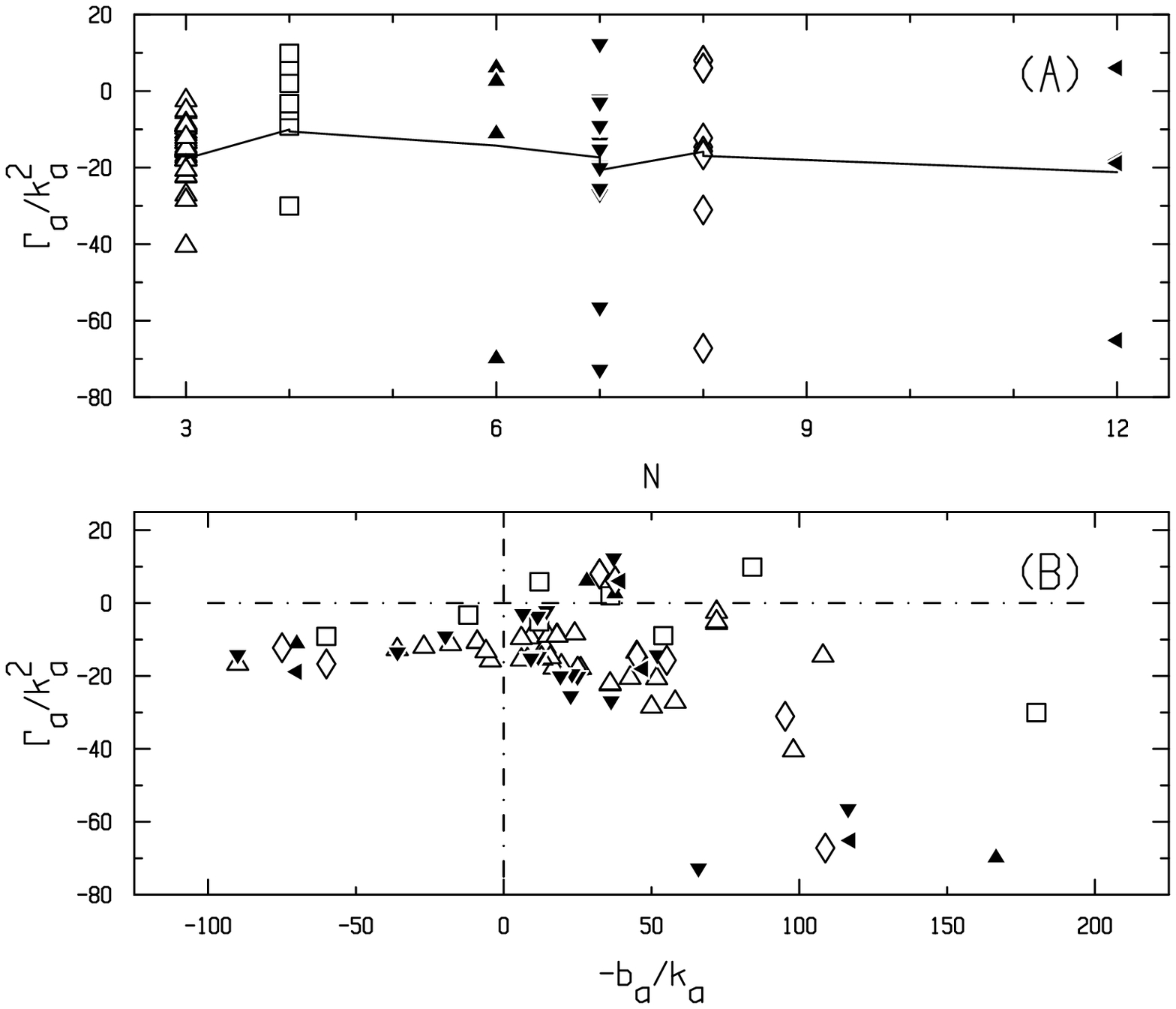,height=16cm,width=12cm} }
\caption{{\bf (A)} The one-loop  threshold corrections to
the quartic $F_a^4$ gauge  interactions coupling constants
are plotted as a function  of the orbifold order $N$  for the sixteen 
orbifolds described in the text. The triangles  and squares plotting 
symbols give 
the normalized quantities ${\G_a\over k^2_a} $
for the various gauge group factors 
in each model. The points joined by continuous lines give the contributions 
of the two gauge group independent terms. 
{\bf (B)} The one-loop normalized  threshold corrections 
are plotted as a function  of the normalized 
slope parameters  $-{b_a\over k_a}$. 
We denote $Z_3$ orbifolds by open triangles; $Z_4$ by open 
squares; $Z_6$ by filled triangles pointing up; $Z_7$ by 
filled triangles pointing down;  $Z_8$ by open  parallelograms; and
$Z_{12}$ by 
filled triangles pointing leftwards.}
\label{fig3}
\end{figure}
For comparison with \cite{kkpr2}, where the $Z_2\times Z_2$ orbifold was 
considered, we observe that,
whereas their function $F_1(q)$  has the same formal structure as in our class of models, 
the decomposition  $ \D_a = (b_a \D -k_a Y)_{KKPR}$ there   identifies 
the gauge-dependent term, $(b_a\D)_{KKPR} $, 
with our term  denoted $(\D_a)_{DKL} $ and  the universal term,
$-(k_a Y)_{KKPR}$,  with the remainder contributions from  $F_1 (q) $ and $ F_2(q)$.
Our total  results 
are compatible with those in \cite{peri,kkpr2}.

\subsection{Quartic order gauge interactions}
\label{sec:quartg}
The numerical calculations for the $F^{a4} $ interactions 
are more  intricate 
than those  of the lower-order  interactions. 
Larger  number of  contributing terms are involved, as can be seen on 
eq.(\ref{eq:jauge4}). The various terms contribute  more or less equally.
We have simply subtracted out the large $\tau_2 \to \infty $ tails for 
all terms, except for the two  $O(1/\tau_2^2)$ terms,  whose 
contribution to the modular integral is finite.

The  results for the moduli-independent threshold corrections
obtained for  our  sample of sixteen orbifold models are
shown in plot (A) of  figure \ref{fig3}.
For a given orbifold order, the corrections 
$\G _a$ are rather  widely spread, somewhat more 
than for $\D_a$. They  cover 
an interval ranging from $10 $ to $-80$.
The contributions from the back-reaction  (gauge group independent) 
terms lie near $-20$  and  significantly 
cancel those  of the zero-modes. 

The corrections $\G _a$ include a sizeable part independent of the 
particular gauge group factor. A dependence on the factor group
is also present, as is seen on  
the plot (B) in figure \ref{fig3}. 
While a correlation between $\G _a$ and 
the slope parameters is not immediately apparent on the figure, 
there appears a systematic trend for an  increase of the corrections   with 
$-{b_a\over k_a}$. For a comparison with  the tree level
coupling constant, we make  the bold approximation that one structure only is
present, say, that associated with $s_1$. 
The total contribution would  read then, 
$S_{EFF}/F^{a4} \sim  s_1(e^{-2\Phi } +Z_{1F^a} ) =
{s_1\over 2\pi } ({4\pi \over g_X^2} +{\G_a \over 2\pi s_1})$. For 
$\G_a \sim O(10)$ and $s_1={3\over 128}$,  it appears that the one-loop  
threshold corrections
correspond to a  large, almost  one
order of magnitude effect.

\section{\bf  Discussion and conclusions}
\label{sec:seconc}

The need for an infrared scale parameter which
separates   the  low  and high  energy  theories mass spectra 
is an  inevitable auxiliary item  in any description of threshold corrections. 
Of course, it is always possible to  circumvent 
the infrared  regularization by  restricting oneself to the
consideration of (moduli or matter fields)
derivatives of  the  coupling constants, 
${\dh  G\over \dh y_\pm }, \  [G= g_a^2, g_{R^2}^2 ] $,
as in the calculations of \cite{anton92,anton93}, or  to 
gauge group and moduli-dependent 
components, as in \cite{dkl}. 
In so doing, however, one gives up useful  information on the
absolute size of threshold corrections.

In  point quantum field theories,
a standard choice   for the infrared parameter 
is the  floating 
off-shell  momentum   scale used in 
the renormalization group method.
The description here proceeds  by equating the summed 
tree and ultraviolet divergent 
loop contributions  to the unrenormalized 
coupling constants in the low and high energy theories,  after 
expressing these in terms of the  renormalized 
running coupling constants at the infrared scale, $M$. 
The  threshold corrections emerge  then 
as the boundary condition terms at the ultraviolet  decoupling  scale $M_X$
\cite{weinberg},
as  is exhibited on the formal equation:
$${1\over G (M_X)} +b \ln {M \over M_X} +\D 
={1\over G(p)} +b \ln {M \over p}.$$
Things differ for string theories because the 
first-quantized  (Polyakov or operator) formalisms  are
on-shell S-matrix approaches, where the  
(tadpoles, mass shifts)  divergences in 
the modular  variables integrals
arise through ${0\over 0}$ ill-defined expressions
(see Chapter  8 in \cite{gsw}). 
What is most specific about 
string theory is that infinities   
are of infrared rather than ultraviolet origin. In particular, 
the notion of  renormalized  string theory  coupling constants is pointless.
Several proposals  of
off-shell  formalisms \cite{bern},
and of extensions of the space-time  dimensional 
continuation procedure \cite{green},
have been  been  made  in the literature. 
No simple satisfactory method has emerged so far. 
Thus, to extract the infrared   scale logarithmic dependence of the 
gauge  coupling constant,
Minahan \cite{minahan}   had to evaluate 
the world sheet correlator for  three gauge bosons, using 
a prescription for going off the energy shell
consistent with conformal and modular 
invariance. 
The background field approach stands as one promising alternative approach.
In the approximate treatment 
of Kaplunovsky \cite{kapl}, 
an infrared cut-off was  introduced  implicitly, as can be seen by writing
his matching formula  in the notations 
of the present paper, 
\begin{eqnarray}
 {(4\pi )^2\over g_a^{2} (p)}-
 {(4\pi )^2k_a\over g_{X}^{2}}&-&
 b_a \ln {p^2\over \L^2} -\D_a  =  
 \sum_i(I_i^0-L_i^0) \nonumber \\ 
& =& -b_a \bigg (\int_F {d^2\tau \over \tau_2}
 -\int_0^\infty  {dt \over t} C_\L (t)\bigg ) =
 -b_a[1-\g_E+\ln {2\over \pi \sqrt {27} \L^2} ].
\end{eqnarray}
The choice of the ultraviolet cut-off function,
$C_\L (t)=(1-e^{-t\L^2} )$, which characterizes  $ g_a (p)$  as 
the  Pauli-Villars renormalized field theory 
coupling constant, yields the same answer as the $\overline {DR}$ prescription.
The infrared divergences   at large values of 
$ \tau_2  $ and  $t=\pi \a ' \tau_2  $,  when regularized
by a sharp cut-off $\tau_2\le 
(M^2\a ')^{-1}$, 
mutually cancel out, so that one is  led to an 
effective unification scale 
$M_X$  which coincides with ours in eq.(\ref{eq:unif}).
On the other hand, as we already pointed out
after eq.(\ref{eq:mi3p}),
for the moduli-dependent threshold corrections,  the use  of a 
similar technically inspired cut-off in the modular integrals \cite{dkl}
leads to a different constant term.

The  crucial advantage of the infrared  regularization by a 
space-time curvature  is that it can be
implemented for both the string and field theories using 
well-defined correspondence  rules.
The infrared cut-off
on the string theory side is 
the  dimensionless Kac-Moody level  parameter,
$\mu ^2  ={1\over k+2}$,  while 
the corresponding parameter on the 
effective field theory side,  $\mu_e^2= {1\over k}$, identifies with 
the  spatial manifold  $S^3$ radial 
scale factor. Since all  
dimensions are  set by the 
string theory  tension, the analog of the dimensional 
floating scale here is  
$  M^2 ={\mu^2 \over \a '}$. In fact, 
by exploiting the representation  of the semiwormhole partition function
in terms of a 
$\Gamma (1,1)$ lattice partition function, the analogy can be 
extended at a deeper level by deriving 
an infrared renormalization group flow equation
for the string theory vacuum functional \cite{kiko}.

The motivations for  the 
finite curvature approach of Kiritsis and Kounnas \cite{kiko}
bear a remote analogy 
with those which inspired earlier proposals in 
field theory \cite{adjac}. These applications
rather aimed at 
formulations of scale invariant theories  exhibiting 
manifest invariance with respect to the conformal  transformations
group, $O(4,1)$  or $O(5)$. 
Negative curvature space-time,  in particular,  
was argued to have beneficial effects on the dynamics, for example, by
suppressing certain non-perturbative effects \cite{callan}.
The full implementation of a  curvature regularized theory  would run into 
significant technical complications  if
one needed an explicit construction of the world sheet correlators, as
this would entail solving an interacting two-dimensional sigma model.
Fortunately, with threshold corrections,    one  is only concerned with the 
spectrum of would-be massless states and their degeneracy, an 
information which is encoded 
in the partition function.

The use of curved space-time string theory solutions formed from 
${\cal N}=4$ superconformal field theory building blocks 
has  desirable features listed in \cite{kiko}: 
A free curvature parameter to monitor  the decompactification  limit;
A well-identified classical  field theory limit;
Preservation of the space-time  supersymmetry properties; 
Solvable marginal deformations to  represent
covariantly constant  gauge and gravitational 
background  fields.
The simplest  models of type,  $M^{(4)} \times K$, 
where  the  ${\cal N}=4$ block, $M^{(4)}$, is
substituted for the  flat space-time, $  R^4$, 
allows one to explore  the class of  phenomenologically viable
orbifold compactification  models for the internal space $K$.
The semiwormhole solution $M^{(4)}=W_k^{(4)}$   presents the enormous 
advantage  of the  partition function 
factorizability.
The consideration of  the  other  known solutions \cite{koun3,koun1,ferrara}, 
 $M^{(4)}= \D_k^{(4)} $ or $ C_k^{(4)} $, 
may involve a less transparent formalism which has not been developed so far.

The combination of a  conformal  algebra structure along with 
a geometrical sigma-model description 
is the main attraction of the approach of \cite{kiko}.
This completes the approach of \cite{kapl} by a consistent 
account for the back-reaction terms, which are essential for ensuring the 
modular invariance. While the form of these terms 
for the gauge coupling constants   could have been guessed
on the basis  of modular invariance, a systematic procedure 
is clearly needed for 
the higher-derivative  interactions.

The reason that only the knowledge of the 
space-time block partition function is needed in calculating the
threshold corrections
rests on the conjectured equivalence between  
the low  energy  limit of the 
string theory  vacuum functional and the effective 
action encoding the world sheet conformal
symmetry constraints \cite{tseyt2}. 
The  dependence on the background fields 
must then  be identical in  the string and field theories 
functionals, up to appropriate 
field redefinitions.
The term by term identification of  powers of the  background 
fields uses in an essential way the connection formulas 
between the vertex operators  and the sigma-model   deformation parameters.
These matching equations provide sets of 
relations between the  string one-loop  contributions and the 
renormalization constants.  They can also be 
interpreted as corrections to the  string theory equations of motion
involving renormalization   contributions  to the  various
interactions. 
The matching assumption can be viewed as an 
implementation of the Fischler-Susskind  mechanism \cite{susskind}, where the  
world sheet infrared regulator associated with 
the  size of the world sheet torus 
handle is   replaced by the space-time curvature parameter.

The connection formulas  between the string theory and sigma-model 
deformation parameters  follow from a consideration of
the mass spectra. However,    
once the  semiwormhole solution is deformed or 
tensored with a non-trivial internal
space, as in $ W_k^{(4)}\times K $, the dependence of the effective action
on the  background fields is only valid in the classical 
limit, corresponding to the leading powers of $ { 1\over k}$ for a given 
interaction.
The semiwormhole would be an  exact quantum solution 
only for the special   model $ W_k^{(4) } \times  R^6$, also 
known as the fivebrane soliton \cite{calhast}. Thus, unlike the 
string theory
functional dependence on $k$,
which is exactly known
thanks to the conformal and modular 
symmetries constraints,
the information  on the $k-$dependence of the 
effective  action   is only  limited to the leading 
powers. This  fact is responsible for 
the restricted  predictive power of the approach. 
The higher orders could possibly be
generated perurbatively  by solving the 
string  theory equations of motion, which 
would then  involve more detailed  information contained 
in the Green's functions. 
To develop the  background fields 
formalism in an  order by order in 
${1\over k}$ identification would also require a 
systematic consideration  of all the 
higher-derivative interactions in  the effective action.
Dealing with these technicalities  would make the analysis quite unwieldy, 
but perhaps not beyond reach.

The implication of our numerical results for the 
moduli-independent threshold corrections is that  these represent
small effects for the quadratic order interactions, corresponding to
a few percent corrections on  the 
improved coupling constant or scale.
Thus, a simple-minded explanation 
for the standard model  gauge coupling constant unification, 
as  being solely  due to heavy threshold corrections is ruled out.
The new revised results  leave the initial 
conclusions \cite{chem,dienes1} unchanged.
The scenarios  for perturbative string theory unification
appealing to the  combined  effects of affine algebra levels,
anomalous  $ U(1)$ factor  and enhanced threshold corrections from 
large compactification volumes
should continue to provide a viable alternative. However, the last 
item here  may prove less effective, since 
considerations from  string duality  indicate
the existence of an upper bound on the compactification  radius \cite{caceres},
$R\le {1\over \a_X M_P}$.

The one-loop threshold corrections  to the higher-derivative $R^2$
gravitational  interactions could possibly inform us about 
the existence of higher-derivative
interactions which are   not reducible to 
the topological  Gauss-Bonnet  combination.  Unfortunately, 
we have been unable to answer this question, because of the
technical difficulties in disentangling the  three 
invariant structures. 
For the  quartic order $F^{a4}$ gauge interactions,
we also had to restrict consideration to 
a linear combination of the invariant couplings which remains unknown. 
Our treatment is  also
qualitative with respect to  the  group theory factors and the 
unresolved mixing with the quadratic gravitational interactions. 
However, the numerical results  here indicate the presence of 
large moduli-independent  threshold corrections
relative to the 
tree level coupling constants.
\appendix
\setcounter{subsection}{0}
\setcounter{equation}{0}
\renewcommand{\theequation}{A.\arabic{equation}}
\renewcommand{\thesubsection}{A.\arabic{subsection}}
\section*{ \centerline {\bf APPENDIX A} }
\subsection{Expansion of partition function in 
background fields}
\label{sec:appa}
We consider the expansion of the deformed 
theory partition function, $Z(F^a, R)$, 
eq.(\ref{eq:defp}), in powers of the 
the gauge and gravitational field strength parameters, $F^a, R$, using 
eq.(\ref{eq:incr}) to describe the 
perturbed conformal weights. The power expansion
up to quartic order is:

\begin{eqnarray}
\label{sec:expand}
  &&e^{-4\pi \tau_2 \delta L_0}  = 1 -  {{I_3\,{\it \bar {Q'}}\,t\,R}\over {2\,{\sqrt{{\it K}}}\,{\sqrt{k} } }} + 
    \left( {{-\left( {I_3^2}\,t \right) }\over {8\,k}} - 
       {{{{{\it \bar {Q'}}}^2}\,t}\over {8\,{\it K}}} + 
       {{{I_3^2}\,{{{\it \bar {Q'}}}^2}\,{t^2}}\over {8\,{\it K}\,k}} \right) \,{R^2}  \nonumber \\
     &+& \left( {{{I_3^3}\,{\it \bar {Q'}}\,{t^2}}\over 
         {16\,{\sqrt{{\it K}}}\,{k^{{3\over 2}}}}} + 
       {{I_3\,{{{\it \bar {Q'}}}^3}\,{t^2}}\over 
         {16\,{{{\it K}}^{{3\over 2}}}\,{\sqrt{k}}}} - 
       {{{I_3^3}\,{{{\it \bar {Q'}}}^3}\,{t^3}}\over 
         {48\,{{{\it K}}^{{3\over 2}}}\,{k^{{3\over 2}}}}} \right) \,{R^3} \nonumber \\  & + & 
    \left( {{{I_3^2}\,t}\over {32\,k}} + 
       {{{{{\it \bar {Q'}}}^2}\,t}\over {32\,{\it K}}} + 
       {{{I_3^4}\,{t^2}}\over {128\,{k^2}}} + 
       {{{I_3^2}\,{{{\it \bar {Q'}}}^2}\,{t^2}}\over {64\,{\it K}\,k}} + 
       {{{{{\it \bar {Q'}}}^4}\,{t^2}}\over {128\,{{{\it K}}^2}}} - 
       {{{I_3^4}\,{{{\it \bar {Q'}}}^2}\,{t^3}}\over {64\,{\it K}\,{k^2}}} - 
       {{{I_3^2}\,{{{\it \bar {Q'}}}^4}\,{t^3}}\over {64\,{{{\it K}}^2}\,k}} + 
       {{{I_3^4}\,{{{\it \bar {Q'}}}^4}\,{t^4}}\over {384\,{{{\it K}}^2}\,{k^2}}}
        \right) \,{R^4} \nonumber \\ & + &  
   \left( {{-\left( {J_a^2}\,t \right) }\over {8\,{\it k_a}}} - 
        {{{{{\it \bar {Q'}}}^2}\,t}\over {8\,{\it K}}} + 
        {{{J_a^2}\,{{{\it \bar {Q'}}}^2}\,{t^2}}\over {8\,{\it K}\,{\it k_a}}} \right ) {F_a^2} + 
   \left( {{{J_a^3}\,{\it \bar {Q'}}\,{t^2}}\over 
          {16\,{\sqrt{{\it K}}}\,{{{\it k_a}}^{{3\over 2}}}}} + 
        {{J_a\,{{{\it \bar {Q'}}}^3}\,{t^2}}\over 
          {16\,{{{\it K}}^{{3\over 2}}}\,{\sqrt{{\it k_a}}}}} - 
        {{{J_a^3}\,{{{\it \bar {Q'}}}^3}\,{t^3}}\over 
          {48\,{{{\it K}}^{{3\over 2}}}\,{{{\it k_a}}^{{3\over 2}}}}}\right ) {F_a^3} \nonumber \\  &+& 
   \left( {{{J_a^2}\,t}\over {32\,{\it k_a}}} + 
        {{{{{\it \bar {Q'}}}^2}\,t}\over {32\,{\it K}}} + 
        {{{J_a^4}\,{t^2}}\over {128\,{{{\it k_a}}^2}}} + 
        {{{J_a^2}\,{{{\it \bar {Q'}}}^2}\,{t^2}}\over {64\,{\it K}\,{\it k_a}}} + 
        {{{{{\it \bar {Q'}}}^4}\,{t^2}}\over {128\,{{{\it K}}^2}}} -
        {{{J_a^4}\,{{{\it \bar {Q'}}}^2}\,{t^3}}\over 
          {64\,{\it K}\,{{{\it k_a}}^2}}} - 
        {{{J_a^2}\,{{{\it \bar {Q'}}}^4}\,{t^3}}\over 
          {64\,{{{\it K}}^2}\,{\it k_a}}} + 
        {{{J_a^4}\,{{{\it \bar {Q'}}}^4}\,{t^4}}\over 
          {384\,{{{\it K}}^2}\,
	  {{{\it k_a}}^2}}} \right ) {F_a^4}  \nonumber \\
        &+&  {{{\rm O}(R)}^5} 
     + {{{\rm O}(F_a)}^5}.
\label{eq:app21}
\end{eqnarray}
We have omitted, for simplicity, the mixed power terms, $O(F^n R^m)$, and 
used the following abbreviations: $ \bar Q'= \bar Q+\bar I_3,
K=k+2, t=8\pi \tau_2$.

\subsection {Zero-mode operators}
\label{sec:appb}
Let us first recall  our  normalization conventions for the 
$SU(2)_k $ and $U(1)_{k_a}$  affine algebras:
\begin{equation}
I_i(z) I_j(0)\simeq { k\d_{ij} \over 2 z^2}, \quad 
J_a(z) J_b(0)\simeq { k_a\d_{ab} \over 2 z^2},\quad 
\bar Q_a(\bar z) \bar Q_b(0)\simeq { \d_{ab} \over \bar z^2},
\label{eq:ap1}
\end{equation}
and the definition of the conformal weights operators in 
eq.(\ref{eq:cfwa}). 
The zero-mode operators for the Cartan subalgebra of commuting  generators 
act essentially 
by inserting the charge  component 
as a factor inside the sum representation of the corresponding 
theta-function or character function. 
This statement can be schematically expressed as, 
$ \bar Q^2 \to \bar  \vt_{\a \b }'' \sim \sum p^2q^{p^2/2} ,\ 
 \bar Q^4 \to \bar  \vt_{\a \b }^{(iv)} \sim 
 \sum p^4 q^{p^2/2}, $ with derivatives operating on 
 $ \vt_{\a \b } (\nu \vert \tau ) \sim 
 \sum q^{p^2/2} e^{2\pi i \nu p} $ as  $\dh /(2\pi i \dh \nu )$. 
More specifically, the  action can be expressed as 
a logarithmic
derivative  with respect to $q $ or $\bar q$ 
acting on the corresponding determinantal factor 
which occurs in the partition function $Tr(q^{L_0} \bar q^{\bar L_0})$. 
For the gauge operator zero-mode,  
using $  L_0^{gauge} = J_a^2/k_a $,
the component decomposition of the world sheet current   operator,
$J_a =\sqrt {k_a\over 2} \hat J_a (z), \ 
\hat J_a (z) =\sum_{I=1}^{16} J_a^I i\dh F^I $,  normalized 
as $Tr \hat J^{2}_a = {\psi^2\over 2} =\ud $, and the schematic 
correspondence:
\begin{eqnarray}
J^2_a \to  k_a L_0 &=&{k_a\over 2} 2q{d\over dq}=
{k_a\over 2}  J_a^{I2} \bigg (2q{d\over dq }\bigg )_I =
{k_a\over 2}  J_a^{I2} \bigg ({ \dh \over 2\pi i \dh \nu_I }\bigg )^2,
\label{eq:ap2}
\end{eqnarray}
one derives the general formula: 
\begin{equation}
 (J_a^2 -{k_a\over 8\pi \tau_2 } )\prod_{I=1}^{16} \vt^I \to
  { k_a\over 2}\big [\sum_I J_a^{I2}   (2q{d\over dq})_I
  +\sum_{I\ne J} J_a^I J_a^J {\dh^2\over (2\pi i)^2 \dh \nu_I \dh \nu_J}
  -{1\over 4\pi \tau_2 }\big ] 
 \prod_I^{16} \vt^I(\nu_I\vert \tau )\vert_{\nu_I=0}.
\label{eq:ap3}
\end{equation}
The fermionic $SO(8)$ Cartan subalgera  helicity generators,  $\bar Q_a=(
\bar Q, \bar Q_i), \ [i=1,2,3] $ 
contribute to the conformal weight  as, 
$\bar L_0=\sum_a{\bar Q_a^2\over 2}$, so that  their action on 
on the spinor fields  theta-functions 
factors reads:
\begin{eqnarray}
 \bar Q^2 &\to & 
  2\bar q {d\over d\bar q} \ln \bar \vt [{\bar \a \atop  \bar  \b }] , \quad
 \bar Q_i^2 \to  
  2\bar q {d\over d\bar q} \ln \bar \vt [{\bar \a +g_i \atop  \bar \b +h_i }]. 
\label{eq:nul2}
\end{eqnarray}
 The space-time orbital  helicity   generators, 
$I_3 , \bar I_3$, for 
the $SU(2)_k$   factor,  enter the conformal weight as
$L_0^{SU(2)_k}= \vec I^2/(k+2) $,  which implies that 
 $[\vec I^2, \vec {\bar I^2}] \to  (k+2)  [L_0,\bar L_0]  =
 {(k+2)\over 2} [2q{d\over dq}, 2\bar q {d\over d\bar q} ]$,
 operating on the corresponding   character function factors,
$\chi_{l,k} ,\bar \chi_{l,k}$, 
 in  the 
partition function $X'(\mu )$, namely, on the factor $X_2 $ in the 
decomposition, $X'(\mu )=X_1 X_2,\quad 
 [X_1=(\sqrt {\tau_2}\eta (\tau ) \bar \eta (\bar \tau ) )^3]$, where 
 $X_1$ comprises the flat space limit contributions from the 
 zero-modes and oscillator excitations. 
 Simultaneously, we must
subtract the free coordinate contributions, 
$I_3^2 \to {k\over 2} 2q{d\over dq} {1\over \eta } $.
It is convenient to 
express the  action  of
$I_3^2 , \bar I^2_3$ directly as derivatives
acting on $X'(\mu )$, so as to exploit the known properties of this 
function at the various limits. 
For this purpose, as just  stated,  we need to correct for the contributions
arising from the explicit action of the 
 derivatives  on the  flat space
and oscillator terms in  $X_1$, 
and  to subtract 
out the action on the free coordinate  oscillator contribution.
 Using the  group symmetry properties, $<I_i^{2p+1}>=0,
 <I_3^2>={1 \over 3}  <\vec I^2>$,  and writing, 
\begin{eqnarray}
I_3^2 [{ X'\over \eta } ]  &=& {k+2\over \eta } \bigg [
{1\over 6} (2q{d\over dq} +{1\over 
8\pi \tau_2 }- {1 \over 2 } (2q{d\over dq}) \ln \eta \bigg ] X' 
+{k\over 2} {X'\over \eta} (2q{d\over dq} ) \ln \eta \nonumber \\
 &=& [{k+2\over 6} 2q{d\over dq} X' +{k+2\over 
8\pi \tau_2 } X']{1\over \eta } - {X'\over \eta } 2q{d\over dq} \ln \eta ,
\label{eq:ap4}
\end{eqnarray}
 we find the explicit  formulas: 
\begin{eqnarray}
\big [ ( \bar Q +\bar I_3)^2 -{k+2\over 8\pi \tau_2} \big ] X'(\mu ) &=&
  \big [2\bar q {d\over d\bar q} \ln {\bar \vt 
  [{\a \choose \b }] \over \bar \eta }  
 +{k+2\over 6} 2\bar q {d\over d\bar q} \big ]X', \nonumber \\
 (I_3^2 -{k\over 8\pi \tau_2 } )X'(\mu )&=& 
 \bigg [ -2q{d\over dq} \ln \eta (q) +{1\over 4\pi \tau_2 } 
 +{k+2\over 6} 2q{d\over dq}  \bigg ] X'(\mu )  \nonumber \\
 &=&{i\over \pi } [\tilde 
 E_2(\tau ) -{k+2\over 6} \dh_\tau ]X'(\mu ).\nonumber \\
\bigg [\tilde E_2(\tau ) & =&  \dh_\tau  \ln \eta (\tau ) -{i\over 4\tau_2 }
= \ud  \dh_\tau  \ln (\eta^2 (\tau ) Im (\tau ))
={i\pi \over 12} (E_2 (\tau ) -{3\over \pi \tau_2} ) , \nonumber  \\
E_2(\tau )&=&{12\over i\pi } \dh_\tau \ln \eta 
= 1-24\sum_{n=1}^\infty  \s_1 (n) q^n \bigg ]
\label{eq:ap5}
\end{eqnarray}
In the sum representation of the zeta-function regularized 
Eisenstein function,  $\s_1(n)=\sum_{d|n} d$  
denotes the sum of the divisors of $n$.
The  antiholomorphic  quantities are deduced by complex conjugation.

The treatment of higher powers of the zero-modes operators follows 
from similar considerations.  One finds:
\begin{eqnarray}
I_3^4 [X']&=& {1\over 5}\bigg  [({k+2\over 2})^2 X_1 (2q{d\over dq})^2
X_2 -({k\over 2})^2 X'(\mu ) (2q{d\over dq} )^2 \ln{1\over  \eta^3 }\bigg ]\nonumber \\ &=&
-{1\over 5\pi^2} \bigg \{ ({k+2\over 2})^2 
\bigg [\dh_\tau ^2 +3({\eta^{'2}\over \eta^2} -
{\eta''\over \eta } -{1\over 8\tau_2^2})\nonumber \\
&+&6({i\over 4\tau_2} 
-{\eta '\over \eta } ) \dh_\tau \bigg ]+{3k^2\over 4 } ({\eta ''\over \eta }
-{\eta^{'2}\over \eta^2 } ) \bigg  \}X' (\mu ) \nonumber \\
&\to & -{3\over 5} X' (2q{d\over dq})^2 
\ln \eta  = {3X'\over 5\pi^2} 
({\eta ''\over \eta } -{\eta '^2\over \eta^2} ),
\label{eq:ap6}
\end{eqnarray}
where $\eta ' = \dh_\tau \eta (\tau )$ and 
we expressed the angle  averaging by means of the classical 
formula, $ I_3^4 \to \vec I^2/5$.  The result appearing on the last line 
in eq.(\ref{eq:ap6}) is obtained by  selecting  the $O(k^0)$ term and 
dropping all terms 
proportional to powers of   $k$
or $(k+2)$.  The corresponding  formula for the 
total angular momentum projection is, 
\begin{equation}
(\bar Q +\bar I_3)^4=    \bar Q^4 +\bar I_3^4+6\bar Q^2
\bar I_3^2
\to  {\bar \vt_{\a \b }^{(iv)} 
\over \bar \vt_{\a \b } } 
-6   {\bar \vt_{\a \b }^{''} \over \bar \vt_{\a \b } }
(2\bar q {d\over d \bar q} \ln \bar \eta )-{3\over 5} 
(2\bar q {d \over d \bar q})^2 \ln \bar \eta . 
\label{eq:ap11}
\end{equation}
The gauge zero-mode operators action is given by:
\begin{eqnarray}
J^{a4} & \to & \sum_I J_I^{a4}\dh_{IIII}^4 +
\sum_{I\ne J} (3J_I^{a2}J_J^{a2}\dh_{II}^2\dh_{JJ}^2 
+4J_I^{a3}J_J^{a}\dh_{III}^3\dh_J) \nonumber \\
&+& \sum_{I\ne J\ne K}6J_I^{a2}J_J^aJ_K^a\dh_{II}^2\dh_{JK}^2+
\sum_{I\ne J\ne K\ne L}J_I^aJ_J^aJ_K^aJ_L^a \dh^4_{IJKL}.
\label{eq:ap12}
\end{eqnarray}
where  the derivatives $\dh_I= {\dh\over 2\pi i \dh \nu_I}$ operate on the 
$E_8\times E_8$ Cartan subalgebra  components of the 
fermionic gauge fields determinantal factors,
$\vt [{\a \atop \b }](\nu_I\vert \tau )$.

\subsection{ Modular integrals}
\label{sec:appc}
To  evaluate the modular integrals in the limit 
$\mu \to 0$, it is convenient to use the representation, 
$X(\mu )= Z_T (\mu ) - Z_T (2\mu )$, while truncating 
the winding modes summation  to the $n=0$ term,
$$Z_T(\mu )= \sqrt {\tau_2} \sum_{(m,n)\in Z^2}
e^{2\pi i mn\tau_1 } e^{-\pi \tau_2 (m^2
\mu^2+{n^2\over \mu^2} )}
\simeq  \sqrt {\tau_2}\vt_3(i\tau_2\mu^2) = {1\over \mu } 
\vt_3({1\over -i\tau_2\mu^2} ) ,$$
where the equation in the last step is deduced by means of the 
duality transformation $\tau \to -1/\tau $.
An important observation is that in the 
difference $Z_T(\mu )-Z_T(2\mu )$, the contribution from the 
$m=0$ term  in the momentum modes sum cancels out,
so that there occurs an  overall 
damping factor,  $e^{-\pi \tau_2 \mu^2 }$, as 
required by the regularization. If one simply factored  out 
the $e^{-\pi \mu^2\tau_2}$ dependence,  this would
prevent one from  simplifying the integrands by using 
the duality transformation.  Nevertheless, a convenient 
approximation for the integrals
$J_1, K_1$,  valid up to small corrections $O(e^{-1/\mu^2 }),
O(e^{-\L^2 /\mu^2 })$,   can be obtained by using the two truncations
detailed in the following steps:
\begin{eqnarray}
X(\mu ) &\simeq  &\sqrt {\tau_2} e^{-\pi \tau_2 \mu^2} \sum_{m\ne 0} e^{-
\pi \tau_2 (m^2-1)} -( \mu \to 2\mu ),  \nonumber \\
&\simeq &\sqrt {\tau_2} e^{-\pi \tau_2 \mu^2}
{1\over \mu \sqrt {\tau_2} } 
[\vt_3({1\over -i \tau_2 \mu^2}) -1  ] -(\mu \to 2\mu ) \nonumber \\
& \simeq & 
{1\over \mu } e^{-\pi \tau_2 \mu^2} -(\mu \to 2\mu ),
\label{eq:ap7}
\end{eqnarray}
where the first truncation  amounts to the substitution, $(m^2-1) \to m^2$,
and the second to retain the leading term in the limit
$\mu \to 0$. One can now evaluate the integral analytically, 
\begin{eqnarray}
J_1 &=&4\pi \mu^2 {\dh \over \dh \mu } 
\bigg [ {1\over \mu } \int_{-\ud }^\ud d\tau_1 
\int_{\sqrt {1-\tau_1^2}}^\infty {d\tau_2 \over \tau_2}
e^{-\pi \tau_2 \mu^2}  -(\mu \to 2\mu )\bigg ] \nonumber \\  
&=& -4\pi \mu^2 {\dh \over \dh  \mu } \bigg [
{1\over \mu }\int_{-\ud }^\ud  d\tau_1 Ei(-\pi \mu^2 (1-\tau_1^2)^\ud -
(\mu \to 2\mu )\bigg ] \nonumber \\ 
&=& 2\pi [\g_E -3 +\ln ( {\pi \sqrt {27} \over 8} \mu^2 ) ],
\label{eq:ap8}
\end{eqnarray}
where $Ei $ is the Exponential integral function and we 
have used: $\int_{-\ud }^\ud dx \ln (1-x^2) = (3\ln 3 -2 -2\ln 2),\quad  
Ei(\infty )\to 0,\quad  Ei(z)\simeq \g_E +\ln z +O(z)$. 
A similar approximation applies 
to the  corresponding field theory integral, 
\begin{equation}
 K_1 \simeq 4\pi \mu_e^2 {\dh \over \dh \mu_e } 
\bigg [ {1\over \mu_e } \int_{1\over \pi \L^2}^\infty {dt\over t}
e^{-\pi t \mu^2}  -(\mu_e \to 2\mu_e )\bigg ] =
2\pi [\g_E -2 +\ln {\mu_e^2 \over 4 \L^2}  ].
\label{eq:ap9}
\end{equation}
A useful approximation for the integrals, $J_n, K_n , $
for $ ( n=2,3)$, is again to truncate the  
winding modes $n\ne 0$, but to retain all the 
momentum modes, $m\in Z$,  while  extending the range of the
modular integral for $m>0$ 
to the entire upper  half-strip. The procedure,  valid up 
to the same exponential accuracy as quoted  above,  can be described as:
\begin{eqnarray}
\int_F {d^2\tau \over \tau_2^n } Z_T(\mu )&\simeq &
{1\over \mu }\bigg [\int_F {d^2\tau \over \tau_2^n} +
\int_{-\ud }^\ud d\tau_1 \int_0^\infty {d\tau_2 \over \tau_2^n} 
(\vt_3({1\over -i\tau_2 \mu^2}) -1)\bigg ] \nonumber \\
&=& {1\over \mu }
\bigg (f_n +2\mu^{2n-2} \pi^{1-n} \G (n-1) \zeta (2n-2)\bigg )
 . \nonumber \\ && \bigg [ f_n = [{\pi \over 3},
 \ud \ln 3 ], \quad [n=2,3]\bigg ]
\label{eq:ap10}
\end{eqnarray}

\end{document}